\documentclass[12pt,a4paper]{article}
\usepackage[utf8]{inputenc}
\usepackage[english]{babel}

\usepackage[left=2.75cm, right=2.75cm, top=3cm, bottom=3cm]{geometry}
\usepackage{url}
\usepackage{natbib,hyperref}
\usepackage{indentfirst,multirow,array,lscape}
\usepackage{amsmath,mathtools,amssymb,amsthm}
\usepackage{subcaption}
\usepackage{lscape}
\usepackage{graphicx}
\usepackage{array}
\usepackage{color,xcolor,fancyhdr}
\usepackage[ruled,vlined,linesnumbered]{algorithm2e}

\newcommand{\ACoala}{{\sc AmoCoala}}
\newcommand{\Coala}{{\sc Coala}}
\newcommand{\lca}{\textsf{mrca}}

\newcommand{\ts}{\tilde{s}}
\newcommand{\tS}{\tilde{S}}

\newcommand{\pvs}{p_{\text{vs}}}
\newcommand{\pjump}[2]{p_{\text{jump}}(#1 \to #2)}
\newcommand{\phs}{p_{\text{hs}}}
\newcommand{\phstar}{p^*}
\newcommand{\pjAcyclic}[3]{1\{E_{#1, #2, #3 } \}}
\newcommand{\pinvade}[3]{p_{\text{invasion}}(#1 \to #2, #3 )}
\newcommand{\pspread}{p_{\text{spread}}}

\newcommand{\ie}{\textit{i.e. }}

\DeclareGraphicsExtensions{.pdf, .jpg, .jpeg, .png, .gif,.eps}

\newtheorem{definition}{Definition}
\newtheorem*{rem}{Remark}
\newtheorem*{lemma}{Lemma}

\newcounter{daggerfootnote}
\newcommand*{\daggerfootnote}[1]{%
    \setcounter{daggerfootnote}{\value{footnote}}%
    \renewcommand*{\thefootnote}{\fnsymbol{footnote}}%
    \footnote[2]{#1}%
    \setcounter{footnote}{\value{daggerfootnote}}%
    \renewcommand*{\thefootnote}{\arabic{footnote}}%
    }

\newcommand{\nocontentsline}[3]{}
\newcommand{\tocless}[2]{\bgroup\let\addcontentsline=\nocontentsline#1{#2}\egroup}
    
\begin{document}

\begin{center}
\noindent{\Large \bf \textsc{
Cophylogeny Reconstruction Allowing for Multiple Associations Through Approximate Bayesian Computation}}
  \bigskip

\noindent {\normalsize \sc{ Blerina Sinaimeri$^{1,2,\dagger}$, Laura Urbini$^{2,}$\daggerfootnote{First co-authors.}, Marie-France Sagot$^2$ and
Catherine Matias$^3$}}\\
\noindent {\small \it 
$^1$ LUISS University, Rome, Italy \\
$^2$ Inria Lyon, 56 Bd Niels Bohr, 69100 Villeurbanne, France, and
	Universit\'e de Lyon, F-69000, Lyon; Universit\'e
        Lyon 1; CNRS, UMR5558; 43
	Boulevard du 11 Novembre 1918, 69622 Villeurbanne cedex, France \\
	$^3$ Sorbonne Universit\'e,  Universit\'e de Paris Cité,
Centre National de la Recherche Scientifique, Laboratoire de Probabilit\'es, Statistique et Mod\'elisation, Paris,
France
}
\end{center}

\medskip
\noindent{\bf Corresponding author:} 
Blerina Sinaimeri, LUISS University, Rome, Italy; E-mail: bsinaimeri@luiss.it.\\

\begin{abstract}
{
{Phylogenetic tree reconciliation is extensively employed for the examination of coevolution between host and symbiont species. An important concern is the requirement for dependable cost values when selecting event-based parsimonious reconciliation. Although certain approaches deduce event probabilities unique to each pair of host and symbiont trees, which can subsequently be converted into cost values, a significant limitation lies in their inability to model the \emph{invasion} of diverse host species by the same symbiont species (termed as a spread event), which is believed to occur in symbiotic relationships. Invasions lead to the observation of multiple associations between symbionts and their hosts (indicating that a symbiont is no longer exclusive to a single host), which are incompatible with the existing methods of coevolution.} 

{Here, we present a method called \ACoala\ (an enhanced version of the tool \Coala) that provides a more realistic estimation of cophylogeny event probabilities for a given pair of host and symbiont trees, even in the presence of spread events. We expand the classical 4-event coevolutionary model to include 2 additional spread events (vertical and horizontal spreads) that lead to multiple associations. In the initial step, we estimate the probabilities of spread events using heuristic frequencies. Subsequently, in the second step, we employ an approximate Bayesian computation (ABC) approach to infer the probabilities of the remaining 4 classical events (cospeciation, duplication, host switch, and loss) based on these values.
}  

{By incorporating spread events, our reconciliation model enables a more accurate consideration of multiple associations. This improvement enhances the precision of estimated cost sets, paving the way to a more reliable reconciliation of host and symbiont trees. To validate our method, we conducted experiments on synthetic datasets and demonstrated its efficacy using real-world examples. Our results showcase that \ACoala\ produces biologically plausible reconciliation scenarios, further emphasizing its effectiveness.
The software is accessible at \url{https://github.com/sinaimeri/AmoCoala} and supplementary material on a Dryad repository at \url{https://datadryad.org/stash/share/SHDH-seLRIznGHCRdQRUNuWE01TnmD5BipocuFrdNUg} with an associated DOI of doi:10.5061/dryad.5x69p8d6v (this last link will only be active upon publication).
} 
}
\end{abstract}

\tocless\section{Introduction}

A powerful framework for modelling host-symbiont coevolution is provided by cophylogeny, a method  which allows to infer combined evolutionary scenarios for a pair of phylogenetic trees of hosts and symbionts. In the following, we refer to symbionts in a wide sense: an organism living in symbiosis, which is not necessarily detrimental nor  beneficial to any of the organisms. The cophylogeny problem is often envisioned as a problem of mapping the phylogenetic tree of the symbionts into the one of the hosts 
\citep[see \emph{e.g.}][]{Charleston2003,MM2005,Page1994,Donati2015}. Such mapping, called a \emph{reconciliation}, allows the
identification of (up to) four types of 
biological events: (a) cospeciation, when the symbiont diverges in correspondence to the divergence of a host species; (b) duplication, when the symbiont diverges but not the host; (c)
host switch, when a symbiont switches from one host species to another independently of any host divergence; and (d) loss, 
{which describes independent extinction of the symbiont lineage while the host lineage survives without an associated symbiont  \citep[also referred to as symbiont extinction, see for instance][]{Dismuskes_review}.}

The reconciliation method is abstract enough that it may actually be applied to different types of data,  of which a
common one is gene-species associations \citep{Bansal2012,DHC2011,HL01,Stolzer2012,THL11}.  In fact, the trees that
are compared do not even need to be representations of phylogenies.  For instance in~\cite{Becerra97}, the phylogenetic
tree of the beetle genus  \emph{Blepharida}  is compared to a tree of the host plants (genus \emph{Bursera}) whose
construction is based on chemical similarity. Such generality may be seen as an advantage since  the methods  developed
for host-symbiont associations \citep{CFOLH2010,MMW2010,Baudet2015,Donati2015} could be applicable to other situations
(such as the gene-species context). However, this also shows that these models do not fully capture the specificity of
the host-symbiont context.  Among the most important aspects that have been only partially addressed is the fact that the 
same symbiotic species can interact, and therefore be associated with more than one host species; we refer to this as a \emph{multiple association}. To mention one example, the same species of insects may pollinate different species of plants \citep[see the example of wasps and figs in][]{Silvieus2008}. This  {has been identified a long time ago \cite[the 'widespread taxon' problem already appears in][]{Page1994} and} 
is in sharp contrast with the gene-species context where a gene (sequence) is naturally associated to only one species 
 \citep[the one it is extracted from, see for instance][]{Stolzer2012,Bansal2018}. We refer to the recent review by \cite{Libes_review} focusing  on the theory of reconciliation in the context of
host-symbiont cophylogenetics.

In host-symbiont systems, a multiple association can result from a combination of biologically different situations. Following \cite{Banks2005}, such association can indeed be explained by: (i) cryptic symbiont species (that is, different symbiont species that are morphologically indistinguishable); (ii) misclassified (over-split) hosts (if the apparently different host species to which the symbiont is related represent in fact a same single species); (iii) recent host switches (when the symbiont  has recently colonised a new host species and in the newly established population, there is very limited genetic diversity compared to the original symbiont population); (iv) failure to speciate by the symbiont population despite the fact that the host diverged (which might happen if the symbiont populations maintain genetic contact despite the host speciation); and  (v) incomplete host switching (if a symbiont colonised a sister taxon of its original host, and maintained genetic contact with the source population). 

While  in the cases  (i)-(ii) the multiple associations are due to errors in defining the real input, in the cases
(iii)-(v)  those are caused by the ability of the symbiont to be associated to more than one host species and hence
require the introduction of an additional biological event that has been called \emph{spread} in the literature. The first use of such term seems to be in~\cite{brooks1991phylogeny}.  Several methods in the literature deal with multiple associations in a more or less ad-hoc way but to the best of our knowledge none of them fully considers spread events. As multiple
associations can be caused by spread events,  any method that deals with multiple associations without considering
spread events is not satisfying.  Below, we briefly review the state of the art of {reconciliation} 
methods that consider multiple associations.

{
Cophylogenetic methods can generally be categorized into three groups: pattern-based statistics, event-scoring methods, and generative model-based approaches \citep{Dismuskes_review}. In this discussion, our focus is on the subset of phylogenetic tree reconciliation methods \citep{Menet_review}, which belong to the latter two categories. 
Event-scoring methods} are based on an optimisation problem where,
given a cost for each of the events, an optimal reconciliation is found by minimising its total cost. 
{These methods allow not only} to estimate the frequencies of each of the events but also to infer the past associations. However, a major problem with these methods is that the solutions
obtained are strongly dependent on the costs that have to be chosen \textit{a priori}. Indeed, costs are inversely proportional to the obtained frequencies: the larger an event cost, the smaller the corresponding frequency of this event. {Statistical approaches based on generative models} 
can then be used in addition to or as an alternative as they remove the subjective step of cost parameter choice and rely instead on a simultaneous inference of parameter values (\ie event probabilities) and events.

To the best of our knowledge, the parsimony-based {reconciliation} methods that address multiple associations are the 
following: {\sc TreeFitter} \citep{Ronquist2003},  {\sc CoRe-Pa} \citep{MMW2010}, {\sc Jane 4} \citep{CFOLH2010} and {\sc WiSPA} (unpublished, see \cite{drinkwater2016wispa}). 
The tool {\sc TreeFitter} \citep{Ronquist2003} treats each multiply associated symbiont as ``an unresolved clade consisting of one lineage for each host in its repertoire. The ancestral host of this terminal clade can then be determined according to one of three separate methods: the ancient, recent and free options \citep{sanmartin02}.'' {These three solutions correspond, respectively, to  scenarios (iii), (iv), and (iii)+(iv) combined.}
{\sc{CoRe-Pa}} \citep{MMW2010}  deals only with the case of cryptic species and solves the multiple associations locally in a parsimonious way. In {\sc Jane 4} \citep{CFOLH2010}   and {\sc WiSPA} \citep{drinkwater2016wispa},  
only parasite tips are permitted to fail to diverge (case (iv) above).

For what concerns the statistical approaches {for reconciliation},  only  \cite{alcala2017host} proposed a method of inference addressing multiple associations. 
The authors develop an approximate Bayesian computation (ABC) method to infer the rates of only two events: host switch and cospeciation. 
Their approach is different from the current literature on tree reconciliation in many ways. First, their method  relies on symbiont genomic sequences to produce sets of dated phylogenies instead of relying on a single symbiont tree. Moreover,  they  pre-estimate extinction and speciation rates from the set of reconstructed symbiont phylogenies. As cospeciation occurs independently from the speciation process in their cophylogeny  model, one might expect that the symbiont trees obtained with this method exhibit more speciations than expected.  Finally, their method outputs only a host-shift rate and a cospeciation probability but no quantification of duplication or loss events. 
{
In their study of figs and wasps, \cite{Satler_19} employ a combination of various approaches. Notably, they propose two ad-hoc methods to address the issue of multiple associations. Firstly, they prune the wasp species that pollinate more than one host taxon, and secondly, they split the shared wasp species into two sister tips. The phylogenetic reconciliation component of their approach is based on the method ALEml by \cite{Szollosi_ALE}, which is designed for phylogenetic reconciliations without multiple associations.
} 
Note that a very recent work addresses multiple associations in 
host-parasite systems, by modelling host repertoire evolution along the branches of a parasite tree \citep{braga_etal_sysbio}. However, this method is far from the reconciliation approach and uses the host tree only through host pairwise distances.

In this paper,  we introduce spread as a fifth
event in the method called {\sc Coala}  (for \emph{COevolution Assessment by a Likelihood-free Approach}) originally proposed in \cite{Baudet2015}  which to our knowledge was the first method to rely on ABC in the context of tree reconciliation.  {\sc Coala}  infers a 
probability for each of the four cophylogeny events: cospeciation, duplication, host switch and loss  but requires that the input has no multiple association.  Introducing a spread event  is a challenge and there is yet no canonical way to do this.  

We choose to  introduce two kinds of spread events, called vertical and horizontal spreads respectively. In this way, we capture the two different situations occurring in the cases (iii)-(v) above. 
The first event, called vertical spread, corresponds to a spread of a symbiont in the entire subtree below a host species. This event could also be called  a \emph{freeze} in the sense that the evolution of the symbiont \emph{freezes} while the symbiont continues to be associated with a host and with the new species that descend from this host.  As will be further detailed in Section Model and Method, this event covers case (iv) above and is related to what is known in the literature as \emph{failure to diverge}  \citep[see for example][]{CFOLH2010}. This also corresponds to the \emph{speciation as a generalist} introduced in~\cite{alcala2017host}. 
Note that there is some abuse of notation in calling this an ``event'' as, from the symbiont lineage's point of view, the diversification of the hosts is not sufficient to be ``noticed'' or to impact on the symbiont lineage diversification.
Thus, this rather corresponds to the absence of an event.
Also, from the biological point of view, 
the term ``freeze'' might be too strong as it suggests  that the symbiont lineage is not able to diversify anymore, while it is just that, once again, the host divergence does not  impact on the symbiont enough to affect the divergence trajectory.
The second event, called horizontal spread, informally corresponds to the combination of
a ``host switch'' with 2 different vertical spreads, one occurring in the initial host subtree and the second in the new host subtree. Thus, this horizontal spread event 
includes both an \emph{invasion} of the symbiont which remains with the initial host but at the same time gets associated  with (\emph{invades}) another host that is not a descendant of the first, plus a \emph{freeze}, actually a double freeze as the evolution of the symbiont \emph{freezes} in relation to the evolution of the host to which it was initially associated and
to the evolution of the second host it \emph{invaded}.  This event is useful to describe the cases (iii) and (v) from above.  It allows to explain the case where two host clades that are phylogenetically distant are associated with the same symbiont species. Notice that a fundamental difference between host switch and horizontal spread is that in the former, the  symbiont that switches hosts will further create 2 different  symbionts, each one associated to the initial and to the new host respectively.
In particular, a host switch never induces a multiple association, in sharp contrast with a horizontal spread. Notice also that cases (i) and (ii) above correspond to input errors rather than real biological events. Nonetheless, these situations are dealt with by our model. Indeed, case (i) is considered as a horizontal spread while case (ii) counts as a vertical spread. Our goal here is not to correct for these potential input errors but to provide a comprehensive framework that handles the diversity of biological situations.

In this article,  we propose a method, called \ACoala, which for a given pair of host and symbiont trees, first pre-estimates the 
probabilities of spread events directly from the input (relying on heuristic frequencies estimates) and second estimates the probabilities of the remaining four classical cophylogeny  events, relying on an ABC approach.  In doing so, we also define a new distance to compare two symbiont trees that are associated with the same host tree in presence of multiple associations. Indeed, ABC methods heavily rely on the ability to compare observations with simulated datasets. In the cophylogeny context, this means comparing trees (as these are the most complete information on the data), a task that is far from trivial.  Our new distance  is an extension of the classical Maximum Agreement SubTree distance (\emph{MAST}) \citep{GGJRW05} to what we call \emph{set-labelled} trees; {we call it \emph{MASST} for Maximum Agreement Set-labeled SubTree}. We believe this new distance can be of independent interest (see Section Model and Method and also Section~B.3 
in the Supplementary Material).

We test \ACoala\ on both synthetic and real datasets and compare the results with \Coala. We could not compare our approach with the tool  \cite{alcala2017host} due to the, previously described, substantial differences both in the model and in the input.  Our tests show that \ACoala\ produces results that seem closer than those of \Coala\ to what is expected from the judgment of a biological expert.


\tocless\section{Model and method}
\tocless\subsection{Reconciliations and cophylogeny events}
Similarly to {\sc Coala},  \ACoala\ 
is built on the event-based model presented in \cite{Charleston2002,THL11}.
The \textit{input} of \ACoala\ consists of a triple $(H,S,\phi)$ where $H$ and $S$ correspond to the phylogenetic trees
of the hosts and symbionts, respectively, and  $\phi$ is a relation from the leaves of the symbiont tree $L(S)$ to the
leaves of the host tree $L(H)$. 
The relation $\phi$ describes  the existing associations between currently living symbiont species and their hosts. More precisely, $\phi$ is a function from the set of symbiont leaves to the set of all subsets of host leaves. Notice that a multiple association will correspond to a leaf in the symbiont tree that is associated to more than one  leaf in the host tree.  
The number of multiple associations is defined as the total number of those supernumerary leaf associations (\emph{e.g.} a symbiont leaf associated {to $k\ge 2$ different host leaves amounts to $k-1$ multiple associations}). 
In {\sc Coala}, as well as in all the models that do not allow for multiple associations, the relation $\phi$ assigns to each $s \in L(S)$ exactly one host leaf in $L(H)$ (notice however that one host can be associated to more than one symbiont). In 
\ACoala, this constraint will be dropped and thus we have that each leaf $s\in L(S)$ in the symbiont tree is associated to $\phi(s)$, a subset of $L(H)$.

A \emph{reconciliation} $\lambda$
is a function from the  vertices of the symbiont species tree 
 to the set of all subsets of vertices of the host tree 
that is an extension of $\phi$, 
 \textit{i.e.} that is the same function as $\phi$ when restricted to the sets of leaves. In the classical 
 setting, a reconciliation can be associated to a set of cospeciations, duplications, host switches and
losses (the four classical cophylogeny events).  For more details about the reconciliation model, we refer to \cite{Charleston2002,THL11,Stolzer2012,Donati2015,Baudet2015} and Section
~A.2 in our Supplementary Material. In this article, we extend the classical reconciliation model to include other biological events. 

Finally notice that here we focus on  models that do not require the host tree to be dated. This is a clear advantage of the method as this information is rarely available and when it is 
available, is often not reliable \citep{Guindon,Brom}.  However, as we do not require the host tree to be dated  some combinations of host switches can introduce an incompatibility due to the temporal constraints imposed by the host and symbiont trees, as well as by the reconciliation itself. We say that a reconciliation is \emph{time-feasible} if it does not violate the time-feasibility constraints. The exact criterion we use to assess time-feasibility is the one defined in \cite{Stolzer2012} and that was already the one used in {\sc Coala}.

\paragraph{Spread events.}
In \ACoala, we introduce two new additional cophylogeny events: vertical and horizontal spreads.  We now define and illustrate both of them.

\begin{figure}[!htb]
  \centering
\begin{minipage}{0.45\linewidth}
\centering
  \includegraphics[width=\textwidth]{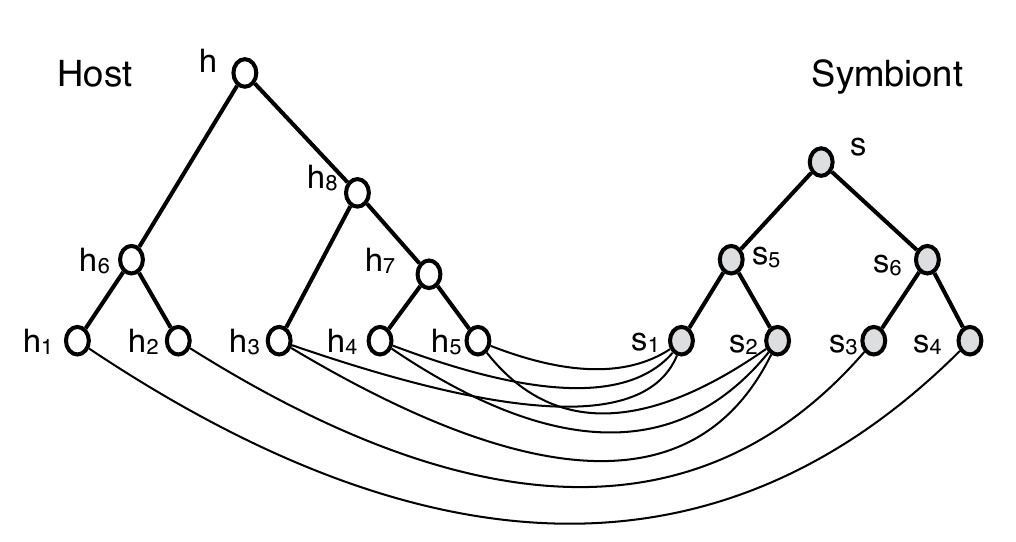}\\ ($a$)
 \end{minipage}
 \begin{minipage}{0.5\linewidth}
   \centering
  \includegraphics[width=\textwidth]{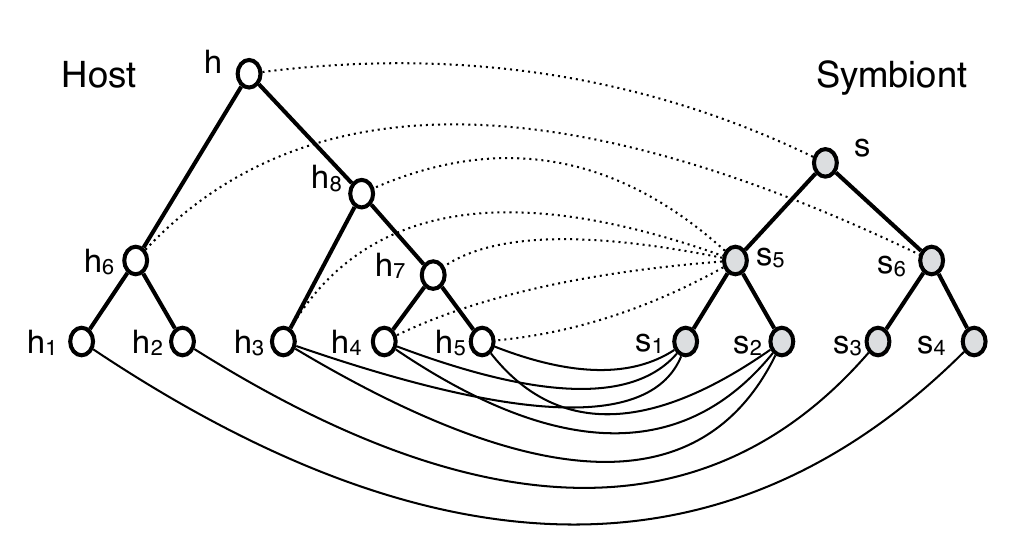}\\ ($b$)
 \end{minipage}
\caption{(a) Example of a dataset with multiple associations.  The leaf associations are represented by plain lines and given by $\phi(s_1)=\{ h_3,h_4,h_5\} ; \phi(s_2)=\{ h_3,h_4,h_5\} ; \phi(s_3)=\{ h_2\} ;\phi(s_4)=\{ h_1\}$.  (b) In dotted lines, a reconciliation involving  2 cospeciations in $s$ and $s_6$ and 1 vertical spread in $s_5$. More precisely, the reconciliation is given by $\lambda(s)=\{h\} ;
\lambda(s_6)=\{h_6\}$ and 
$\lambda(s_5)=\{h_3,h_4,h_5,h_7,h_8\}$ (on the symbiont leaves, we have $\lambda=\phi$). 
}
\label{fig:exampleVerticalSpread}
\end{figure}

\emph{Vertical Spread.} When for a symbiont $s$ that is currently associated to a host $h$, and with  probability $\pvs(h)$, a vertical spread happens at that host $h$,  the evolution of the symbiont $s$ \emph{freezes} in $h$,
\ie $s$ continues to be associated with $h$ and with the new species that descend from $h$. In the toy example depicted in Figure~\ref{fig:exampleVerticalSpread}$($a$)$, we see that the symbionts $s_1,s_2$ are both related to all the hosts $h_3, h_4, h_5$.  One possible explanation is that the symbiont $s_5$ (the most recent common ancestor of $s_1, s_2$) was present in all the 
clade of $h_8$ (which is the most recent common ancestor of $h_3, h_4, h_5$). In that case, we say that $h_8$ is the ancestral host of $s_5$  and the two clades $S_{s_5}$ (which denotes the symbiont clade rooted in $s_5$) and $H_{h_8}$ (the host clade rooted in  $h_8$) are related.  We say that a vertical spread has happened at symbiont $s_5$ and we  associate $s_5$ with all the vertices in the subtree rooted in $h_8$ (see Figure~\ref{fig:exampleVerticalSpread}$($b$)$). 

\begin{figure}[!htb]
\minipage{0.45\textwidth}
\centering
  \includegraphics[width=\linewidth]{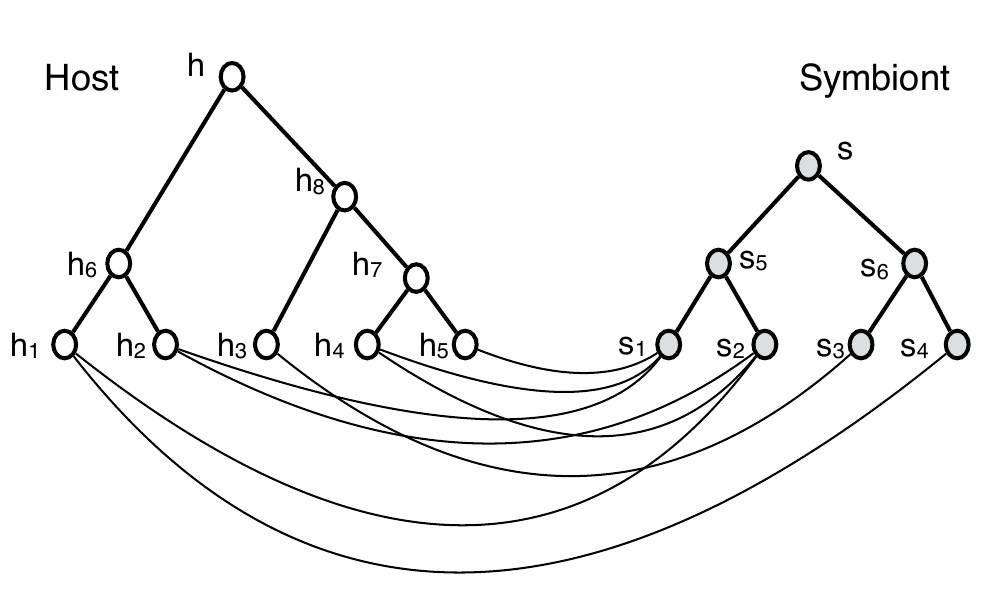}\\ ($a$)
 \endminipage\hfill
\minipage{0.5\textwidth}
\centering
  \includegraphics[width=\linewidth]{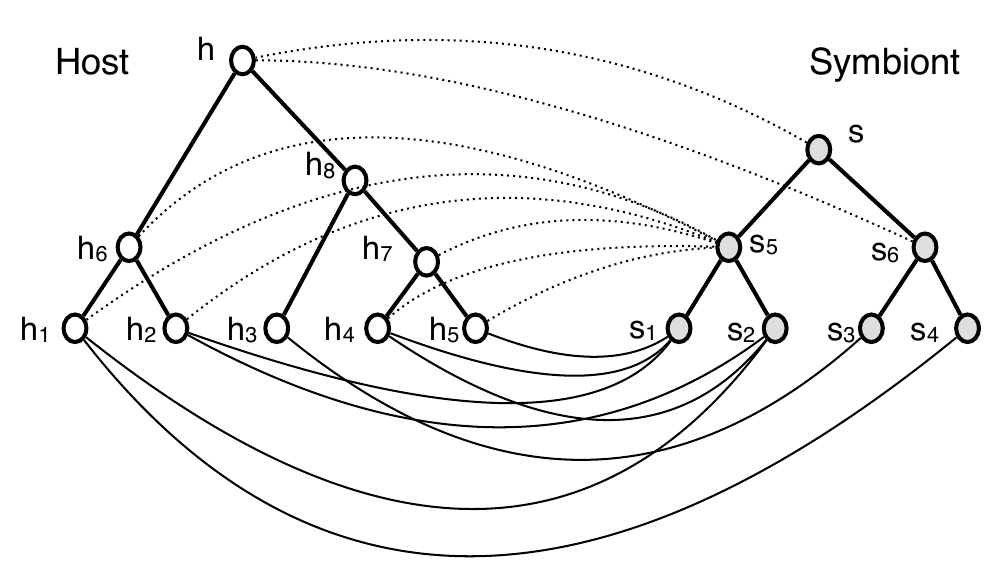}\\ ($b$)
\endminipage\hfill
\caption{(a) Example of a dataset with multiple associations. The leaf associations are represented by plain lines and given by $\phi(s_1)=\{ h_2,h_4,h_5\} ; \phi(s_2)=\{ h_1,h_2,h_4\} ; \phi(s_3)=\{ h_3\} ;\phi(s_4)=\{ h_1\}$.
(b) In dotted lines, a reconciliation involving a 
  horizontal spread event is shown. The symbiont $s_5$ makes a horizontal spread from $h_6$ to $h_7$ (or from $h_7$ to  $h_6$) and thus is associated to the two subtrees $H_{h_6}$ and $H_{h_7}$ (\emph{i.e.} $\lambda(s_5)=H_{h_6}\cup H_{h_7}$). The symbiont $s$ is associated to a duplication (and $\lambda(s)=\{h\}$) and the symbiont $s_6$ to a cospeciation (and $\lambda(s_6)=\{h\}$).}
    \label{fig:exampleHorizontal_Spread_Input}
\end{figure}

\emph{Horizontal Spread.} In some 
datasets, we see the  occurrence of the same symbiont in two different clades of the host tree. Such a  situation cannot occur when relying only on cospeciation, duplication, host switch, loss  or vertical spread events. Indeed, as already underlined, the four initial events never produce multiple associations, while the vertical spread produces them only within clades.
For this reason, we introduce a horizontal spread event. In the horizontal spread event, the symbiont  remains with the initial host but at the same time gets associated with  (\emph{invades}) another host incomparable with the first, and undergoes a freeze, actually a double freeze as the
evolution of the symbiont \emph{freezes} in relation to the evolution of the host to which it was initially associated and in relation to the evolution of the second one it \emph{invaded}. 
A horizontal spread event involves two probabilities: the probability $\phs(h)$ that the horizontal spread occurs at node $h$ of the host tree, and for any other host node $h'$ that is incomparable to $h$, a probability $\pjump{h}{h'}$ (symmetric wrt $h,h'$) that the symbiont jumps from host $h$ to host $h'$ (and then freezes both under $h$ and $h'$). In fact, $\phs(h)$ is deduced from the values $\{\pvs(h),\pvs(h'),\pjump{h}{h'}\}_{h'}$ for all $h'$ incomparable to $h$ (details are given in Section
~A.4 from the Supplementary Material).
For illustrative purposes only,  we show in
Figure~\ref{fig:exampleHorizontal_Spread_Input} an example of a reconciliation involving 
a horizontal spread
event. The horizontal spread event happens in vertex $s_5$ as it is associated to two subtrees of the host tree,  
rooted in $h_6$ and $h_7$, respectively.

{It is worth noting that when a symbiont spreads into a host, it becomes restricted to being present in every descendant of that particular host, and no further events occur. While this restriction may appear limiting, it is crucial to consider that spread events are more likely to occur in the lower part of the tree, specifically among the most recent events (refer to Section
~A.4 in the Supplementary Material). These spread events are introduced to account for situations where ``not enough time has passed yet'' (cases (iii) to (v) listed above). From this perspective, it is reasonable to assume that no subsequent event takes place after the spread event.
This restriction is also driven by concerns regarding identifiability. By introducing two additional events (horizontal and vertical spreads), it is essential to maintain a simple model to prevent the creation of indistinguishable scenarios.
}

\tocless\subsection{General framework of \ACoala}\label{subsec:GeneralFramework}

The method  we propose is based on the approximate Bayesian computation (ABC) method that was already used in \Coala\ \citep{Baudet2015}. We briefly recall it here for the sake of completeness. ABC methods belong to a family of
likelihood-free Bayesian inference algorithms that attempt to estimate posterior densities for problems where the
likelihood is unknown or may not be easily computed. ABC only requires that simulations under the statistical model
at stake are possible. We recall that  the likelihood function expresses the probability of the observed data under a particular statistical model.  More specifically, given a set of observed data $D_0$ (in our case the input $(H,S,\phi)$) and starting with a prior distribution $\pi$ on the space of the  parameters of the model (here, the probabilities {$\theta=\langle p_c,p_d,p_s,p_l\rangle$} of the four classical cophylogeny events), the objective is to estimate the parameter values $\theta$ that could lead to the observed data using a Bayesian framework. 
Formally, we are interested in the posterior distribution $p(\theta | D_0)={p(D_0|\theta) \pi(\theta)}/ {p(D_0)}$.

For simple models, the likelihood function $p(D_0|\theta)$ can typically be derived. However, for more complex models
the likelihood function might be computationally very costly to evaluate. In these  cases, ABC methods  approximate
the posterior distribution by simulations, the outcomes of which are compared with the observed data.  
  First,  a population of $N$  parameter values $\theta^i$ is sampled from the prior distribution. Then, for each sampled
  parameter  $\theta$, a dataset $\tilde{D}_\theta$ is simulated. It consists of a simulated symbiont tree $\tS_\theta$ together with a 
  reconciliation $\tilde \lambda$ from $\tS_\theta$ to $H$. This dataset $\tilde{D}_\theta$ is then compared with the real
  dataset $D_0$ through a summary measure which is used as a \emph{quality measure} to accept or reject the
  candidate parameter value $\theta$. In many cases when it is believed that the prior and posterior densities are very
  different, the acceptance rate is very  low.  To deal with that issue, we can rely more specifically on  a likelihood-free Sequential Monte Carlo (SMC) search  that  involves many iterations of the simulation procedure, each iteration targeting more precisely good candidate parameter values.

Given an input dataset $(H,S,\phi)$, an ABC-SMC method was developed in {\sc{Coala}} \citep{Baudet2015} to infer the posterior density of the probability of each of the four classical events, namely cospeciation, duplication, host switch and loss. {\sc{Coala}} includes two main parts. The first consists in a  simulation algorithm of  
the coevolutionary history of symbionts and their 
hosts. More specifically, given the host tree $H$ and a vector $\theta = \langle p_{c},p_{d},p_{s},p_{l} \rangle$ specifying the probability of each of the classical cophylogeny events, the model generates a symbiont tree $\tS_\theta$ together with a reconciliation from $\tS_\theta$ to $H$ describing the ancient host-symbiont associations. In \ACoala, this first part is improved by introducing spread events whose probabilities of occurrence are fixed throughout all the simulations, while being  heterogeneous along the host tree and specific to the original dataset. More precisely, these probabilities are pre-estimated on each dataset through simple frequency estimates related to the symbiont and host associations. Their values are specific to each node $h$ of the host tree.
The second part concerns a method to select the most likely probability vectors based on an ABC-SMC variant. 
It relies on the main idea that the most likely vectors $\theta$ will generate trees $\tS_\theta$ together with reconciliations $\tilde \lambda_\theta$ from $\tS_\theta$ to $H$ that are similar to the original 
input $(H,S,\phi)$.

In \Coala, the symbiont trees together with their 
leaf associations were summarised through labelled trees and this step thus relied on a phylogenetic distance between labelled trees. In \ACoala, this part is improved by the introduction of a new distance  that accounts for the possibility of multiple associations between $\tS_\theta$ and $H$.
Indeed,  the symbiont trees together with their leaf association may now be summarised through set-labelled trees (\emph{i.e.} trees with leaves labelled by subsets of $L(H)$). 
We thus provide and rely here on  a new phylogenetic distance metric, called $d_{MASST}$ between set-labelled trees. To the best of our knowledge, distances between set-labelled trees have not been considered in the literature and our proposal for such may be of independent interest.

In a nutshell, to deal with multiple associations coming from spread  events, we thus extend {\sc Coala}  as follows: (i) 
we first propose estimators $\pspread$ for all the probabilities needed to define spreads (namely, {$\pspread=\{\pvs(h),\pjump{h}{h'},\phs(h)\}_{h,h'\in H}$ contains vertical and horizontal spreads as well as jumps probabilities}) given the input $(H,S,\phi)$;  (ii) we introduce a new method to simulate the  cophylogeny of the 
symbiont tree, along the host tree and given a candidate probability {$\theta=\langle p_c,p_d,p_s,p_l\rangle$} for each of the four classical cophylogeny events (cospeciation, duplication, host switch and loss) which also takes into account the probabilities of vertical and horizontal spread {(gathered in $\pspread$)};  (iii) we introduce a new distance to compare the simulated to the real symbiont
trees in presence of multiple associations to the host tree.

%

\tocless\subsection{Estimation of the probabilities of the events}\label{sub:computing_spread_prob}

In \ACoala, the probabilities $\langle p_{c},p_{d},p_{s}, p_{l}\rangle$ of the four  classical cophylogeny events (cospeciation, duplication, host switch and loss)  are parameters inferred relying on the ABC-SMC approach, namely they are first sampled from a prior distribution and then later selected according to some criteria that are specified later. \textbf{On the contrary, the probabilities $\pvs(h)$ and $\phs(h)$  for the (vertical and horizontal) spread events at each host node $h$ are not estimated within the ABC-SMC method but rather in a  preliminary step, directly from the input}.  
This choice is mainly driven by the fact that in a  realistic model the spread probabilities are not constant throughout the host tree.  
For instance, a spread event appearing near to the root is less likely to happen than  one close to the leaves. Indeed, spread events were introduced partly to account for recent host switches (see point (iii) in the introduction) and more generally they are motivated by the fact that symbionts may not diversify immediately, which is less likely close to the root. 
Then, as the probability of a spread event is specific to each vertex of the host tree, sampling the spread events will increase significantly the size of the parameter space and thus the size of the space of the generated symbiont trees.  Hence, in this framework the spread probabilities cannot be inferred in the ABC procedure. Nevertheless, these probabilities are clearly related to the shape of the host and symbiont trees and 
to the associations between their leaves. For this reason, we exploit the signal from the input to pre-estimate  the probabilities of the spread events.  These probabilities are used in the generation of the putative symbiont trees and are not inferred through the ABC-SMC method. Details about these estimators as well as an assessment of the robustness of the ABC method with respect to these pre-estimated values  are given in Sections
~A.4 and
~D.3 from the Supplementary Material, respectively.


\tocless\subsection{Simulation of a symbiont tree in \ACoala}\label{subsec:evolution_symbiont_tree}
We now describe the procedure of generation of simulated symbiont trees in \ACoala. 
Similarly to \Coala, our algorithm takes as input  $(H, S, \phi)$ and  the probabilities of each of the events, and simulates the evolution of the symbionts by following the evolution of the hosts, \textit{i.e.}  by traversing $H$ from the root to the leaves, and progressively constructing the phylogenetic tree  $\tilde{S}$ for the symbionts and at the same time mapping them to subsets of vertices of the host tree, \emph{i.e.} constructing $\tilde \lambda$.  In this process, a symbiont vertex can be in two different states: mapped or unmapped. At the moment of its creation, a new vertex $\tilde{s}$ is unmapped and is assigned a temporary position on an arc $a$ of the host tree $H$. We denote this situation by $\langle \tilde{s} : a \rangle$. We let $h(a)$ denote the head of the arc $a$ (\ie the vertex at the endpoint of $a$ that is farthest from the root). Then vertex $\tilde{s}$ is mapped to either vertex $h(a)$ of $H$ (\emph{i.e.} $\tilde \lambda(\ts)=\{h(a)\}$ for cospeciation, duplication and host switch) or to a subset $\cal{H}$ of vertices of $H$ (\emph{i.e.}  $\tilde \lambda(\ts)=\cal{H}$ for vertical and horizontal spread). Notice that for the vertical 
spread, the subset of vertices $\cal{H}$ corresponds to a clade in $H$, while for the horizontal spread it corresponds to the union of two clades in $H$.

In the cases of cospeciation, duplication, and host switch, a speciation has occurred in the symbiont tree and hence  two children are created for $\tS$, denoted by ${\tilde{s}}_1$ and ${\tilde{s}}_2$. Their positioning along the arcs of the host then depends on which of the three events took place. In the case of a loss, no child for $\tS$ is created (at this step) since there is no symbiont speciation, and $\tS$ is just moved to one of the two arcs outgoing from $h(a)$ chosen randomly.

\begin{figure}[!htbp]
\center
\includegraphics[width=0.9\textwidth]{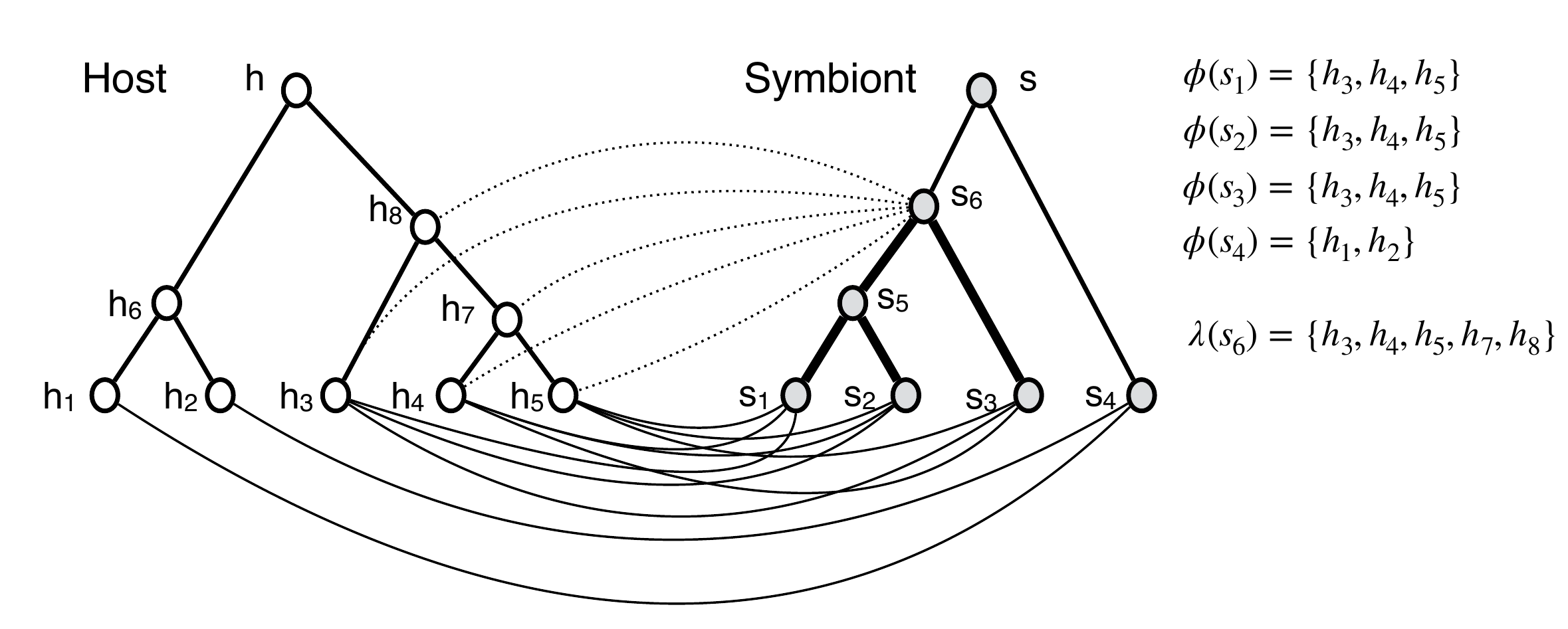}
\caption{The symbiont $s_6$ is associated to a vertical spread on the host $h_8$ and thus is associated to all the subtree $H_{h_8}$. As we do not know exactly how the symbiont $s_5$ is associated, we symbolically associate it to all the vertices in $H_{h_8}$. The subtree of $S$ in bold corresponds to a ghost subtree.
}
\label{fig:ghost_tree} 
\end{figure}

The case of a spread event is different. Consider for instance the example in Figure~\ref{fig:ghost_tree}. A vertical spread occurs at the symbiont $s_6$ on the host $h_8$ and thus $s_6$ is associated to all the subtree $H_{h_8}$ (the host clade rooted in $h_8$).  Moreover, we choose that all the symbionts descendent from $s_6$ are associated to  the same clade as $s_6$ (see Definition
~A.2 in the Supplementary Material). 
We now need to choose a realistic way of continuing the simulation of the symbiont subtree below $s_6$. We call the  subtree of the symbiont tree rooted at a vertex associated to a spread event (vertical or horizontal) a  \emph{ghost subtree}. 
In Figure~\ref{fig:ghost_tree}, the subtree $S_{s_6}$ is a ghost subtree. Then during the generation of the symbiont tree $\tilde{S}$ when a symbiont $\tilde{s}$ undergoes a spread event, we need to simulate the ghost subtree rooted in $\tilde{s}$ up to its leaves, in order to end the simulation in this part of the tree.  After a spread event, with the passing of time, both the host and the symbiont have evolved and in addition, it could be
that some hosts have lost some of their symbionts. Taking into account all the possible evolutions of the symbiont is computationally unfeasible in practice. Therefore, for computational reasons, we decide to promote the simplest situation. In particular, no other event takes place after a spread event and we mimic in this part of the simulated symbiont tree the evolution occurring in the real symbiont tree. Therefore we choose a topology and leaf associations that are identical to those present in $S$.  
More formally, if a vertical spread occurs at $\tilde{s}$ on the host $h$, we consider the set of host leaves descendent from $h$, namely $L=L(H_h)$. Let $L'$ be the set of symbiont leaves that are associated to the leaves in $L$, \ie $L'=\phi^{-1}(L)\cap L(S)$. The ghost subtree $\tilde{S}_{\tilde{s}}$ is then set equal to 
$S_{|L'}$, 
{the smallest subtree of the \emph{real} symbiont tree whose set of leaves is exactly $L'$.}
The case of horizontal spread is analogous, except that the set of leaves $L$ is given by the union of $L(H_{h})$ and 
$L(H_{h'})$ where $h, h'$ are the two host vertices involved in the horizontal spread. Once the ghost tree is set, the simulation ends in this part of the tree. 
Notice that as already mentioned, the spread events are more likely to occur far from the root, so that the loss of variability in the simulated tree $\tS$ induced by this choice is counterbalanced by the fact that it should affect a small part of the tree. More details are given in Section
~B.1 from the Supplementary Material.

The symbiont tree simulation algorithm is summarised in Algorithm~\ref{algo:simu_spreads}. It relies on the following notation.  
 A generic arc from the host tree $H$ is denoted by $a$, its head (end node farthest from the root) is $h(a)$ and the arcs  outgoing from its head are $a_1,a_2$. {A root is denoted $v_{root}$,} $L(T)$ is the set of leaves of $T$, while the subtree of $T$ rooted at node $h$ is denoted by $T_h$. For any set of leaves $L$, we let $T_{|L}$ denote the   subtree  of $T$ whose set of leaves is exactly $L$. This $T_{|L}$ is also a subtree of $T$ rooted at the most recent common ancestor of the elements in $L$ {and whose set of leaves is restricted to $L$. For any  node $v\in T$, we let $\mathcal{I}_T(v)$ be the set of  nodes $v'\in T$ that are incomparable to $v$.} 
 During the algorithm, before defining the simulated association $\tilde \lambda(\ts)$ of a simulated symbiont node $\ts$, the node is temporarily positioned on an arc $a$, which is denoted by $\langle \ts :a \rangle$. 
 { For a switch of symbiont $\ts$ located on arc $a$ (i.e., $\langle \ts :a \rangle$) to be possible, two conditions must be met. Firstly, there must be another host vertex that is incomparable to the head vertex $h(a)$ (i.e., $|\mathcal{I}_H(h(a))|\ge 1$). Secondly, there must exist an arc $a'$ in the host tree $H$ where placing a children node $\ts_2$ on this arc (i.e., $\langle \ts_2 :a' \rangle$) would not violate the time feasibility condition.
During the simulation procedure, a filtering step is executed at the final stage. Any simulated symbiont tree with a size larger than twice that of the observed symbiont tree is discarded. This filtering step, which is already employed in \Coala, is vital in further assessing the similarity between the simulated trees and the observed one.
}


\begin{algorithm}[!htbp]
 \DontPrintSemicolon
 \SetAlgoLined 
 \SetKwInOut{Input}{Input}\SetKwInOut{Output}{Output}
{
\Input{$(H,S,\phi)$ and  event probabilities 
$\theta=\langle p_c,p_d,p_s,p_l\rangle$,  
$\pspread=\{\pvs(h),\pjump{h}{h'},\phs(h)\}_{h,h'\in H}$.} 
\Output{Simulated symbiont tree $\tilde{S}$ and reconciliation $\tilde \lambda$ to host $H$.}
Initialization:  Create root  $\tilde{s}_{root}$ and position  $\langle \ts_{root}:a\rangle$, where $a$ s.t.  $h(a)=h_{root}$  \;
Add $\tilde{s}_{root}$ to the set $U$ of unmapped nodes of $\tS$ \;
\SetKwProg{While}{While}{}{}
\SetKwBlock{Else}{else}{}
 \While {$U$ not empty}{
Pick $\ts \in  U$,  its position is $\langle \ts:a\rangle$ \;
\SetAlgoVlined 
Horizontal spread: Sample $HS\sim Bern(\phs(h(a))$ \;
\SetAlgoVlined 
\If{$HS=1$ \textbf{and} $|\mathcal{I}_H(h(a))|\ge 1$}{ 
Sample $h'\in \mathcal{I}_H(h(a))$ with probability $\pjump{h(a)}{h'}$  \;
Map $\tilde{\lambda}(\tilde{s})=H_{h(a)}\cup H_{h'}$ and remove  $\ts$ from $U$ }  
\SetAlgoLined 
 \Else{Vertical spread: Sample $VS\sim Bern(\pvs(h(a))$ \;
\SetAlgoVlined 
  \If {$VS=1$} {
Map $\tilde{\lambda}(\tilde{s})=H_{h(a)}$  and remove  $\ts$ from $U$ \; 
 For $L'=\phi^{-1}(L(H_{h(a)}))\cap L(S)$, paste $S_{|L'}$ in $\tS$ below $\tilde {s}$ \;  
 For all $s' \in \tilde {S}_{\ts}$, map $\tilde{\lambda}(s')=H_{h(a)}$  
 }
 \SetAlgoLined 
 \Else{Classical event: Sample $E\sim \mathcal{M}(1,\theta)$ multinomial in $\{\mathbb{C,D,S,L}\}$ \;
\SetKwProg{If}{if}{}{}
\SetAlgoVlined 
 \If {$E=\mathbb S$ (switch) \textbf{and} 'switch possible'} 
  {Map $\tilde{\lambda}(\tilde{s})=\{h(a)\}$ and remove  $\ts$ from $U$ \; 
  Create $\ts_1,\ts_2 $ children of $\ts$ in $\tilde S$ \;
  Position $\langle \ts_1:a\rangle$   and add $\ts_1$ to $U$ \;
   Randomly choose arc $a'$ (among those that do not violate time feasibility condition) in $H$ and    position $\langle \ts_2:a'\rangle$ \;
\SetAlgoVlined 
   \If{$h(a')$ is a leaf of $H$}{Map $\tilde{\lambda}(\ts_2)=h(a')$  }
   \Else{Add $\ts_2$ to $U$}
   }
  } } }	
}
   \caption{Generation of a symbiont tree under model with spreads.}
  \label{algo:simu_spreads}
\end{algorithm}

\begin{algorithm}
\setcounter{AlgoLine}{22}
\DontPrintSemicolon
\SetKwProg{While}{}{}{}
\SetKwProg{If}{}{}{}
{
 \While{}{
\If{}{
\If{}{
 \SetKwProg{If}{if}{ then}{}
 \If{$E=\mathbb S$ (switch) \textbf{and} 'switch impossible'}
 {Sample $E\sim \mathcal{M}(1,\theta')$ multinomial in $\{\mathbb{C,D,L}\}$ with $\theta'=\langle p_c,p_d,p_l\rangle /(p_c+p_d+p_l)$  }
\If {$E\in \{\mathbb{C,D}\}$} 
 {Map $\tilde{\lambda}(\tilde{s})=\{h(a)\}$ 
 and remove $\ts$ from $U$ \;
 Create $\ts_1,\ts_2 $ children of $\ts$ in $\tilde S$ \;
 \If {$E=\mathbb C$ (cospeciation)} 
 {Position $\langle \ts_1:a_1\rangle$ and $\langle \ts_2:a_2\rangle$ ($a_1,a_2$ arcs outgoing from $h(a)$)\;
 \For{$i=1,2$}{\If{$h(a_i)$ is a leaf of $H$} 
 {Map $\tilde{\lambda}(\ts_i)=\{h(a_i)\}$ }
 \Else{Add $\ts_i$ in $U$}
 }
 }
 \If {$E=\mathbb D$ (duplication)} 
 {Position $\langle \ts_1:a\rangle$ and $\langle \ts_2:a\rangle$ \;
 Add $\ts_1, \ts_2$ in $U$
 }
 }
\If {$E=\mathbb L$ (loss)} {Randomly choose $a'\in \{a_1,a_2\}$ and position $\langle \tilde{s}: a' \rangle$\;
\If{$h(a')$ is a leaf of $H$} 
 {Map $\tilde{\lambda}(\ts)=\{h(a')\}$ and remove $\ts$ from $U$}
} 
}}}
\tcc{ Filtering step} 
\SetKwProg{If}{if}{ then}{}
\If{$|\tilde S|\ge 2 |S|$} {Discard the tree and restart the algorithm}
}
\end{algorithm}

\begin{figure}[!htbp]
\center
\includegraphics[scale=0.5]{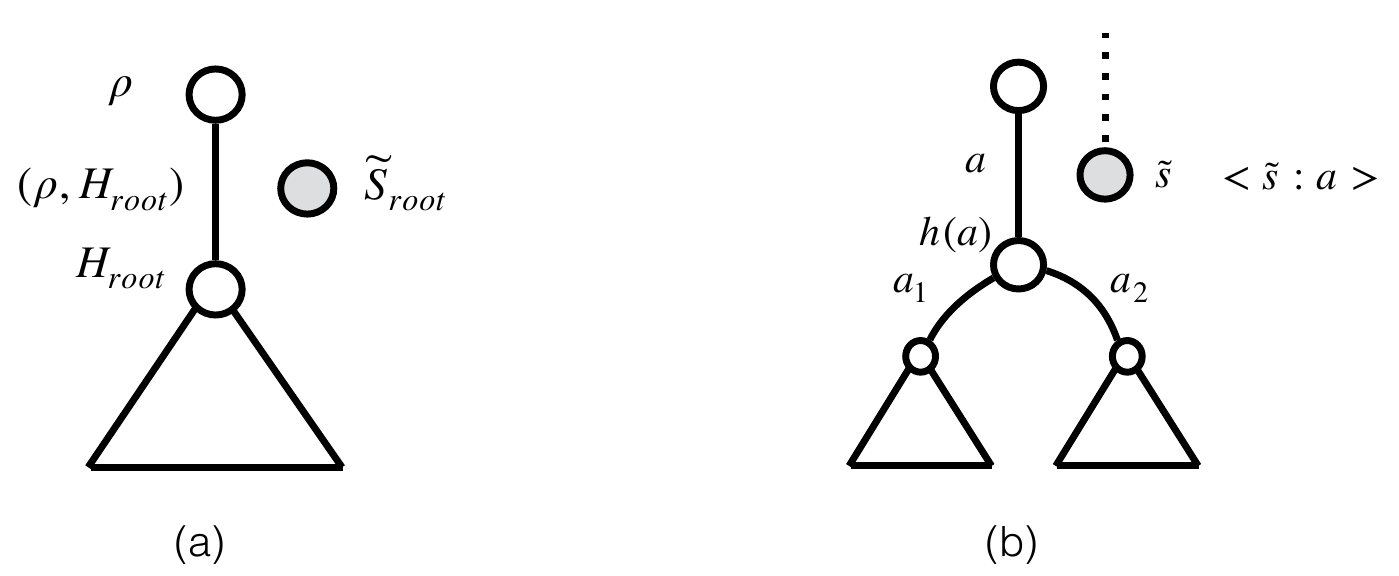}
\caption{
Simulation algorithm. (a) Starting configuration. (b) Unmapped vertex $\ts$.}
\label{fig:initial_config} 
\end{figure}
\begin{figure}[!htbp]
\center
\includegraphics[width=\textwidth]{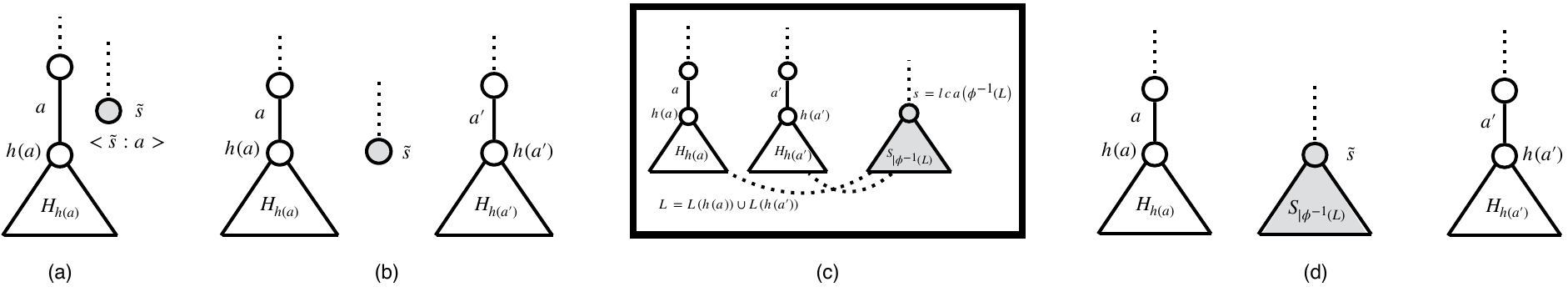}
\caption{
Simulation of a horizontal spread. 
(a) Initial configuration. (b) Mapping of vertex $\ts$: we let $\tilde\lambda(\ts)=H_{h(a)}\cup H_{h(a')}$.  (c) Looking in the real symbiont tree for the ghost subtree to be used in the next step. (d)  Creating the ghost subtree in $\ts$ and stopping the evolution of the leaves of $\tilde{S}_{\ts}$.}
\label{fig:horizontal_spread_simulation} 
\end{figure}
\begin{figure}[!htbp]
\center
\includegraphics[width=\textwidth]{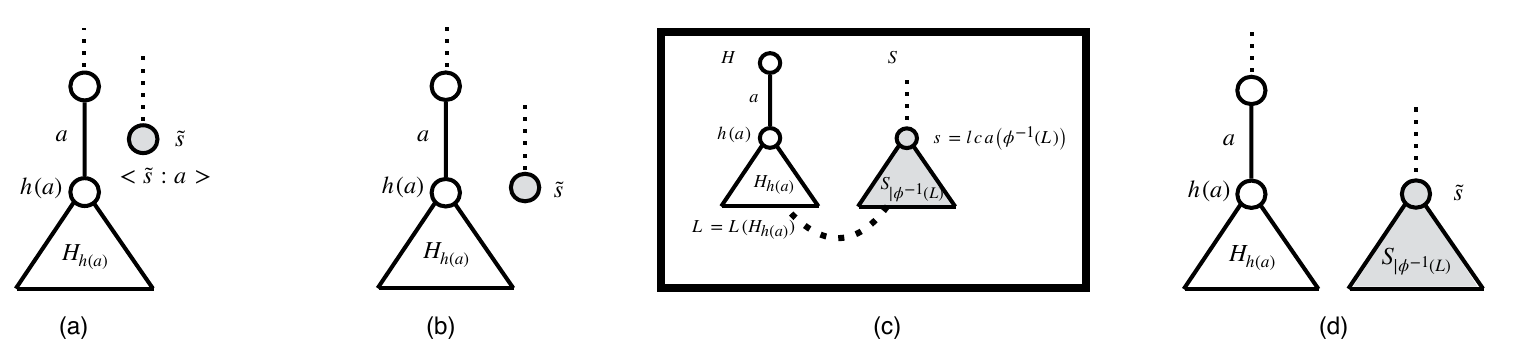}
\caption{
Simulation of a vertical spread. 
(a) Initial configuration. (b) Mapping of vertex $\ts$: we let $\tilde\lambda(\ts)=H_{h(a)}$. (c) Looking in the real symbiont tree for the ghost subtree to be used in the next step. (d)  Creating the ghost subtree in $\ts$ and stopping the evolution of the leaves of $\tilde{S}_{\ts}$.}
\label{fig:vertical_Spread_simulation} 
\end{figure}
\begin{figure}[!htbp]
\center
\includegraphics[width=\textwidth]{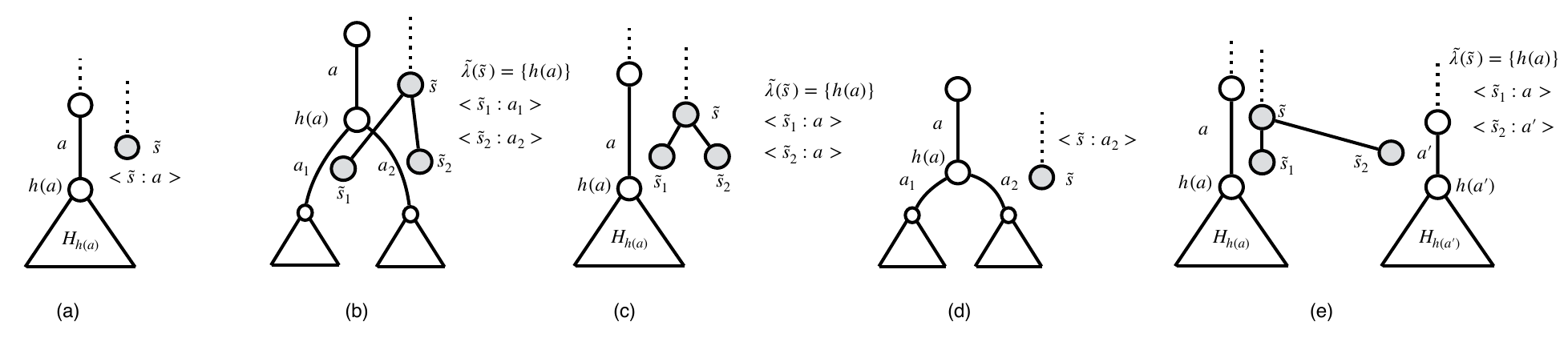}
\caption{
Simulation of a classical event. 
(a) Initial configuration. (b) Cospeciation. (c) Duplication. (d) Host switch. (e) Loss.}
\label{fig:classical_event_simulation} 
\end{figure}

Finally, Figures~\ref{fig:initial_config} to~\ref{fig:classical_event_simulation} illustrate the different steps of the symbiont tree generation procedure.

\tocless\subsection{ABC-SMC inference method}\label{subsec:parameter_estimation}

\ACoala\ is based on the same ABC-SMC method presented 
in \Coala. It is an iterative method with many rounds, and it involves a summary discrepancy  that describes the quality of any candidate vector $\theta$ (\emph{i.e.} how much it is susceptible to have generated the observed dataset).
We first present Algorithm~\ref{algo:simu} that describes how we rely on simulated trees in a reconciliation model with spreads produced through Algorithm~\ref{algo:simu_spreads}, to characterize the quality of a candidate vector $\theta = \langle p_c,p_d,p_s,p_l\rangle$ with respect to the observed dataset $(H,S,\phi)$. In particular, for each candidate vector $\theta$, we produce many different trees and summarise them into a discrepancy $d_\theta$ that characterizes the quality of $\theta$ as a candidate to produce the observed data. The structure of this  procedure is unchanged from \Coala, except for the way we compute the discrepancy $d_\theta$.

\begin{algorithm}[!htbp]
\DontPrintSemicolon
 \SetKwInOut{Input}{Input} 
 \SetKwInOut{Param}{Parameters}
 \SetKwInOut{Output}{Output}
\Input{$(H,S,\phi)$, probabilities $ \langle \theta,\pspread\rangle$}
\Param{$M$ number of simulated symbiont trees} 
\Output{Distance $d_\theta$} 
\For{m=1 to M}{ 
Apply Algorithm~\ref{algo:simu_spreads} with input $(H,S,\phi)$ and  $\langle \theta,\pspread\rangle$ and output $(\tilde S_{\theta,m},\tilde{\lambda}_{\theta,m})$ \;
Compute $d_{\theta,m}$ discrepancy between $(\tS_{\theta,m},\tilde{\lambda}_{\theta,m})$ and $(S,\phi)$
}

Compute  $d_\theta$ as the average value of $\{d_{\theta,m}\}_{1\le m \le M}$
   \caption{Symbiont tree simulation algorithm overview}
  \label{algo:simu}
\end{algorithm}

We then present a general overview of the ABC-SMC procedure in Algorithm~\ref{algo:ABC_SMC}.
We include all the details of the method in Section
~B.2 from the Supplementary Material. Moreover, we report below the differences between this procedure and the one at stake in \Coala.

\begin{algorithm}[!htbp]
\DontPrintSemicolon
 \SetKwInOut{Input}{Input} 
 \SetKwInOut{Param}{Parameters}
 \SetKwInOut{Output}{Output}
\Input{$(H,S,\phi,\pspread)$}
 \Param{ 
 $R$ rounds, 
 $N$ initial number of vectors, 
 $\{\tau_i\}_{1\le i \le R}$ tolerance values at each round, 
 $M$ simulated symbiont trees for each vector
 }
\Output{Selected vectors $\theta$}
Sample $N$  vectors $\theta=\langle p_c,p_d,p_s, p_l \rangle \sim \mathcal{D}(1,1,1,1)$,  
 store them in $A_0$ \;
\tcc{Simulation at first round:} 
 \ForAll{$\theta$ in $\Theta$}{
Apply Algo~\ref{algo:simu} with input  $(H,S,\phi,\theta,\pspread)$, parameter $M$. Output  $d_\theta$ \;
Select $Q_1=\tau_1 \times N$ values in $A_0$ with smallest $d_\theta$, store them in  $A_1$
 and set $\epsilon_1=\textrm{Argmax}_{\theta \in A_1} d_\theta$ 
 }
\tcc{Simulation at other rounds:} 
  \For{r = $2$ to $R$}{
  $\mathcal{Q}_r\longleftarrow \emptyset$\;
  \While{$|\mathcal{Q}_r|\le Q_{r-1}$}{
Sample $\theta^\star$ uniformly in $A_{r-1}$
and  create $\theta^{\star \star}$ by perturbing $\theta^\star$ \;
Apply Algo~\ref{algo:simu} with input  $(H,S,\phi,\theta^{\star\star},\pspread)$, parameter $M$. Output  $d_{\theta^{\star\star}}$ \;
 \If{$d_{\theta^{\star\star}}\leq \epsilon_{r-1}$}{
 add $\theta^{\star\star}$ to quantile set $\mathcal{Q}_r$}  
}
Select $Q_r=\tau_r \times Q_{r-1}$ values in $\mathcal{Q}_r$ with smallest $d_{\theta^{\star\star}}$, store them in  $A_r$ and set $\epsilon_r=\textrm{Argmax}_{\theta \in A_r} d_\theta$ 
}
    \caption{ABC-SMC procedure}
  \label{algo:ABC_SMC}
\end{algorithm}


The main difference between Algorithms~\ref{algo:simu} and \ref{algo:ABC_SMC} and their respective  corresponding versions in \ACoala\ lies in the summary discrepancy $d_\theta$ used to quantify the quality of the vector $\theta$.  The summary discrepancy  between a simulated  dataset (the generated symbiont tree and its host associations) and the observed one (the real symbiont tree and its host associations) is measured through a distance between phylogenetic trees which can be calculated in polynomial time.  Similarly as in \Coala, this discrepancy is built from two components: (i) $d_1$, 
that describes how much the simulated tree  $\tS_{\theta}$ is representative  of the vector  $\theta$,  and ii) $d_2$ that measures how much  is $\tS_{\theta}$ (and its labels) topologically similar to $S$ (and its labels). The value of $d_1$ is computed identically as in \Coala.  As concerns point (ii), the distance used here is  different from the one used in \Coala\ and we detail its definition and motivation in the next paragraph.

\tocless\subsection{A distance between set-labelled trees}
There are many distances between tree topologies, though not all are simple to compute.  However, the topology of a simulated tree is not sufficient to characterize its similarity in the reconciliation context. Here, we want to consider, on top of the topology, the leaf labels of the tree. Indeed, the sets that label the leaves of the (simulated) symbiont tree contain information on the associations given by the coevolution of symbionts with their hosts. 
In \ACoala, the leaves of both the observed and the simulated symbiont trees ($S$ and $\tS$ respectively) are labelled by the host leaves to which they are associated.  Thus, due to possible multiple associations in \ACoala, those symbiont trees are what we call \emph{set-labelled} trees, that is, their leaves are labelled with sets and not with singletons. To the best of our knowledge, distances for set-labelled trees have not been considered in the literature and we believe our proposal for such is thus of independent interest.  

We first recall that the MAST distance of two phylogenetic trees $T_1$ and $T_2$ corresponds to the number of leaves in the largest isomorphic subtree that is common to the two  trees (subtrees common to the two trees are called agreement subtrees and we look for the one with the largest number of leaves).
Clearly this isomorphism takes into account the labels of the trees. The MAST distance can be calculated in $O(n^2)$ time where $n$ is the size of the largest input tree \citep{Ganapathy2006}.  For set-labelled trees, we need to take into account the sizes of the sets of labels in the possible agreement subtrees. 

\begin{figure}[!htbp]
\center
\includegraphics[width=0.9\textwidth]{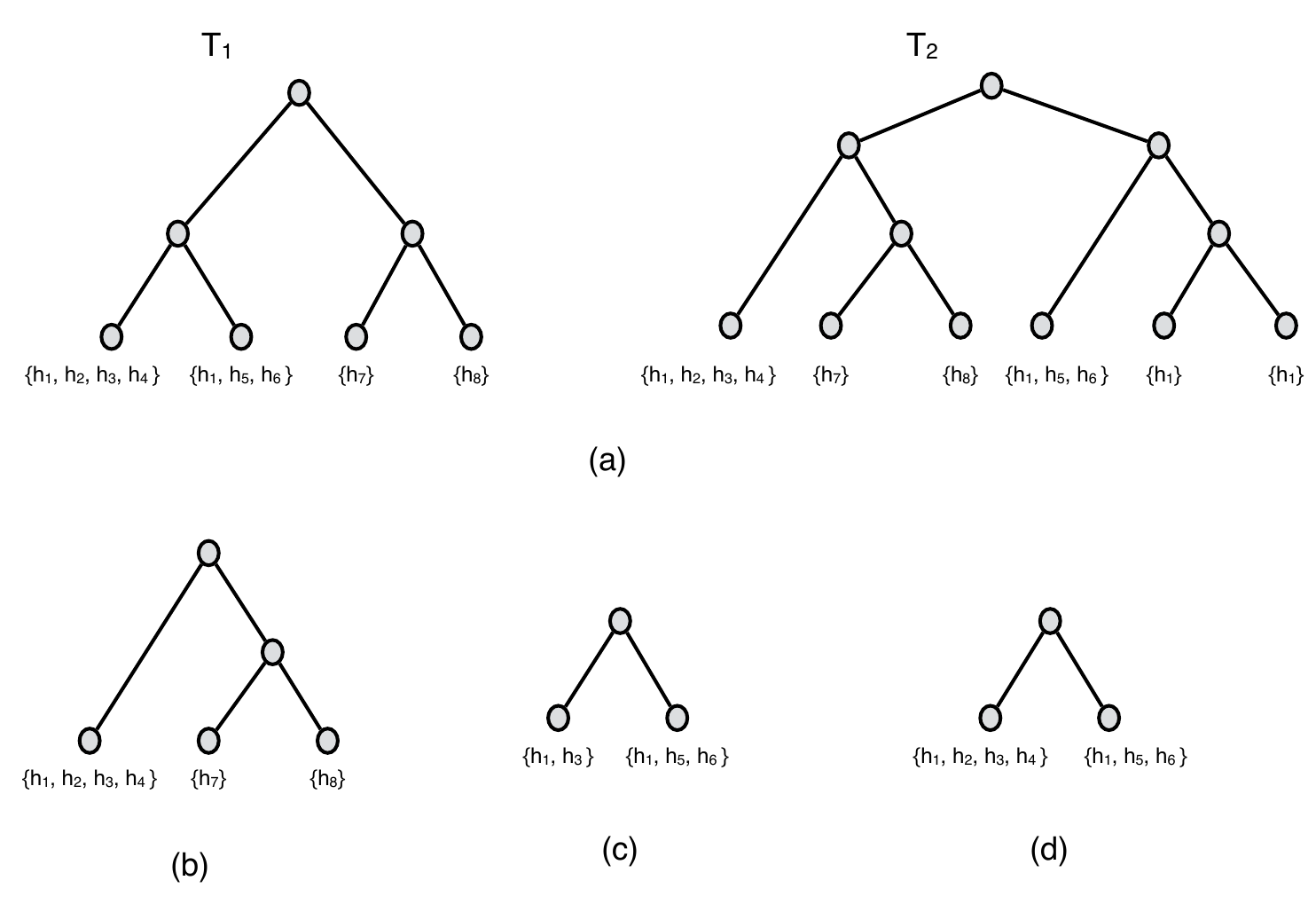}
\caption{(a) Two set-labelled phylogenetic trees. $T_1$ has weight 9 and $T_2$ has weight 11. In $(b)$, $(c)$, $(d)$, three different agreement set-labelled subtrees of weights 6, 5 and 7 respectively. The maximum agreement set-labelled subtree is the one depicted in $(d)$ and notice that it does not have the maximum number of leaves. }
\label{fig:MASST}
\end{figure}

Thus, given a set-labelled tree $T$, we denote its \emph{weight} by
$w(T)= \sum_{v\in L(T)} |l(v)|$, where $l(v)$ is the set of labels associated to the leaf $v$.  Now, a   \emph{maximum agreement set-labelled subtree}, denoted 
by $MASST(T_1,T_2)$, is a set-labelled subtree that is common to the two trees $T_1,T_2$ and which has largest weight.  
Notice that a common subtree may have leaf labels that are subsets of the original ones. As a consequence, 
the maximum agreement subtree of two trees does not  necessarily have the maximum number of leaves among the set-labelled agreement subtrees, as shown in Figure~\ref{fig:MASST}.  
In the same way 
as the MAST distance is defined, we introduce the \emph{maximum agreement set-labelled subtree} distance, denoted by $d_{MASST}$, between two set-labelled phylogenetic trees 
$T_1$,$T_2$ as well as a normalized related quantity $d_2$, respectively defined as
\begin{align*}
d_{MASST}(T_1,T_2)&= \max\{w(T_1),w(T_2)\} - w(MASST(T_1,T_2))\\
d_2(T_1,T_2)&= \frac{d_{MASST}(T_1,T_2)}{\max\{w(T_1),w(T_2)\}}=1- \frac{w(MASST(T_1,T_2))}{ \max\{w(T_1),w(T_2)\}}.
\end{align*}
We can prove that  $d_{MASST}$ is a distance metric and that it can be calculated in polynomial time using a dynamic programming algorithm. Note that the normalized quantity $d_2$ has the advantage of lying in $[0,1]$ and is computed with the same complexity as $d_{MASST}$. 
It is only a pseudo-distance (as it does not satisfy the triangular inequality).  The resulting  $d_\theta$  defined relying on $d_2$ is a \emph{summary discrepancy}
(see details in Section
~B.3 from  the Supplementary Material).

\tocless\subsection{Summary of \ACoala}
Algorithm~\ref{algo:ACoala} presents a final summary of the algorithm at stake in \ACoala. The output of the ABC-SMC procedure is a specified number of selected vectors $\theta$. Similarly as \Coala, \ACoala\ further performs a hierarchical clustering procedure, with an automatic selection of the number of groups, to cluster the final list of accepted parameter vectors.
The clusters and their number are automatically selected through the \texttt{R} package \texttt{dynamicTreeCut} \citep[for details, see][]{LZH2007}. 
 {Each cluster can be summarized by a ``representative'' parameter vector, which is computed as follows: for each coordinate, the ``consensus'' parameter vector is determined by taking the mean value of the respective coordinate across all parameter vectors within the cluster. Subsequently, the ``consensus'' coordinates are normalized to ensure their sum is equal to one, resulting in a representative parameter for the cluster.
 }

\begin{algorithm}[!htbp]
\DontPrintSemicolon
 \SetKwInOut{Input}{Input} 
 \SetKwInOut{Param}{Parameters}
 \SetKwInOut{Output}{Output}
\Input{$(H,S,\phi)$}
 \Param{ 
 $R$ rounds, 
 $N$ initial number of vectors, 
 $\{\tau_i\}_{1\le i \le R}$ tolerance values at each round,  
 $M$ simulated symbiont trees for each vector
 }
 \Output{Selected vectors $\theta$,  
and (optional) clusters of these vectors} 
 Compute: Vertical spread $\pvs(h)$ and jump $\pjump{h}{h'}$ probabilities   for any incomparable nodes $h,h'\in H$; deduce horizontal spread $\phs(h)$ probabilities. Gather these quantities in vector $\pspread$ \;
{\bfseries ABC-SMC procedure:} Apply Algo~\ref{algo:ABC_SMC} with input $(H,S,\phi,\pspread)$ and parameters $(R,N,\{\tau_i\}_{1\le i \le R},M)$
  \caption{\ACoala\ general structure}
  \label{algo:ACoala}
\end{algorithm}

\tocless\section{Experimental results and discussion}\label{sec:Experimental_Risults_Discussion}

\tocless\subsection{Experimental settings}
\paragraph{Parameter settings.} 
For each (synthetic or biological) dataset $(H,S,\phi)$, we ran \ACoala\ with the following parameter values. We simulated $N = 2000$  vectors $\theta^i$, ($1\leq i \leq 2000$) in the first round of simulation. For each vector $\theta^i$, we simulated $M = 1000$ symbiont trees.  
The tolerance value used in the first round was $\tau_1 = 0.1$. We ran $R=3$ rounds  and we defined $\tau_i = 0.25$. Notice that $ \tau_1 \times  N = 200$ defines the size $Q$ of the quantile set which must be produced in each new round. Thus, after the last round, we have $\tau_3 \times Q = 50$ accepted vectors.

\paragraph{Synthetic datasets generation.} 
Synthetic datasets are obtained in a similar way as in \Coala\  \citep[see][for more details]{Baudet2015}, the only difference lying on the fact that the simulation algorithm now includes spread events. 
In particular, we use the real symbiont tree and its (multiple) associations to the host tree to derive the spread probabilities. 
To obtain realistic datasets, we started from a real biological tree and chose the  dataset SFC described in the next section. This host tree $H$ (and associated spread probabilities) is combined with 8 different parameter values. We thus  simulated 8 datasets  $(H,S_{\theta^\star_j},\phi_j)$ for $1\leq j\leq 8$ associated with the following 8 probability vectors, in the form $\theta=\langle p_c,p_d,p_s,p_l \rangle$. We used $\theta^\star_1 = \langle 0.70,0.10,0.10,0.10\rangle$, $\theta^\star_2 = \langle 0.80,0.15,0.01,0.04\rangle$, $\theta^\star_3 = \langle 0.75,0.01,0.16,0.08\rangle$, $\theta^\star_4 = \langle 0.70,0.05,0.02,0.23\rangle$, $\theta^\star_5 = \langle 0.60,0.20,0.00,0.20\rangle$, $\theta^\star_6 = \langle 0.55,0.00,0.20,0.25\rangle$, $\theta^\star_7 = \langle 0.45,0.10,0.15,0.30\rangle$ and $\theta^\star_8 = \langle 0.40,0.20,0.10,0.30\rangle$. 
The choice of these vectors was done with the aim to cover some typical coevolution patterns of probability. Indeed, vectors with very low probability of cospeciation correspond to situations where there is almost no signal of the coevolution of the  species at a macroevolutionary level. In these cases, the cophylogeny reconciliation methods are not appropriate \citep{Baudet2015,Althoff2013}.
Moreover, a high probability of host switches or duplications is not appropriate to produce synthetic datasets due to the variability of the simulated trees. Note that for these reasons, a ninth parameter value used in \Coala\ was discarded here.

\tocless\subsection{Results of the self-test}
The objective of this test is to check whether \ACoala\  produces the correct results for synthetic datasets where we know the truth.  To this purpose, we  ran \ACoala\ 50 times on each of the 8 synthetic datasets generated as explained in the previous subsection with true parameter value $\theta^\star_i$.  We expected to find a vector ``very close'' to $\theta$ among the vectors accepted on the last round of \ACoala.  
Note that contrarily to what we did in \Coala, we here rely on
an Euclidean distance over the parameter vectors. 
At the end of the third round, we therefore took note of the cluster whose representative parameter vector had the smallest Euclidean distance to the true value $\theta^\star_i$ (we call it the ``best'' cluster).
We also stress again that parameter  vectors do not include the probabilities of spread events, which are pre-estimated before applying the ABC-SMC approach.

\begin{figure}[!ht]
    \centering
\includegraphics[scale=0.27]{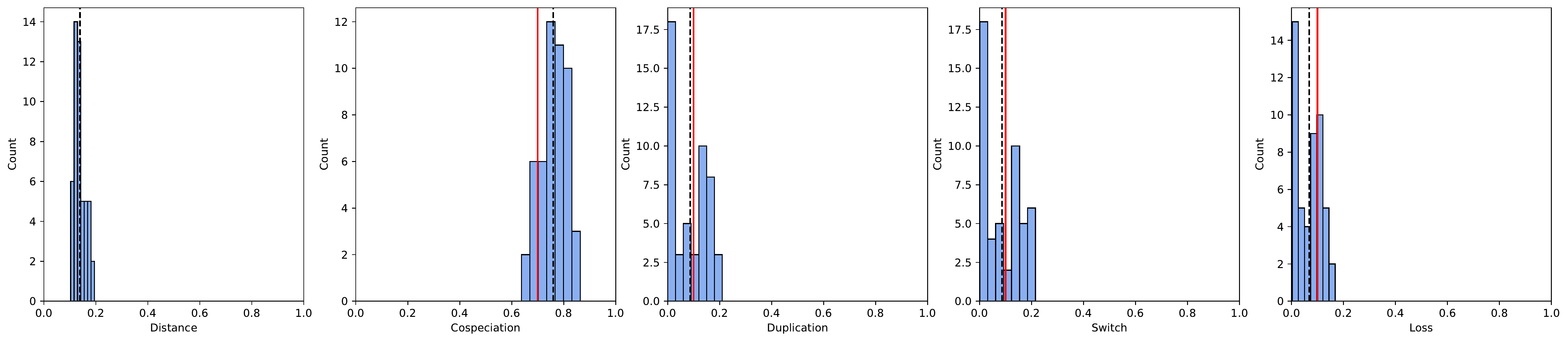} 
\caption{For each simulated dataset with true parameter value $\theta^\star_1 = \langle 0.70,0.10,0.10,0.10\rangle$, we ran \ACoala\ 50 times and, at the end of the third round, we took note of the cluster whose representative parameter vector had the smallest euclidean distance (histograms shown in the first column) to $\theta^\star_1$. Columns 2 to 5 show the histograms of the distributions of the event probabilities in these ``best'' clusters. The dashed vertical black line indicates the mean value. The solid vertical red line indicates the true parameter value. }
\label{fig:theta1}
\end{figure}

The results for the first parameter value  $\theta^\star_1$ are presented in Figure~\ref{fig:theta1}. The results for the other vectors are similar and given in Figures
~Ba to Cd from the Supplementary Material.

The first column shows the histograms of the distances between the true value $\theta^\star_i$ and the representative parameter in the best cluster. Then, columns 2 to 5 show the histograms of the distribution of the event probabilities in these best clusters. The solid vertical red line indicates the true parameter value. The dashed vertical black line indicates the mean value.
 Overall the distances (first columns) are rather small and the parameters are correctly estimated (columns 2 to 5).  In some specific cases, the slightly lower performance of the method may often be explained by the difficulty of the problem. For instance, for true parameter vector $\theta^\star_8$, the low cospeciation level makes the reconciliation problem less relevant. It results in over-estimation of the cospeciation and underestimation of the loss probabilities. 
Overall, these simulations show that \ACoala\ is able to select parameter vectors that are close to the true ones. 

\tocless\subsection{Biological datasets} \label{par: biodata}
To test our method, we selected 4 biological datasets from the literature. The choice of these datasets was dictated by: (1) the availability of the data in public databases, (2) the desire to cover for situations as widely different as possible in terms of the topology of the trees and the presence of multiple associations. The phylogenetic trees of each  dataset can be found in Figures
~D to G from the Supplementary Material.
As already mentioned, any dataset $D$ containing multiple associations cannot be analysed with  \Coala. 
Thus, in order to compare the results with those obtained by  \Coala\ \citep{Baudet2015}, for each real dataset $D$ we generated a dataset $D_{Coala}$ which is obtained from $D$ by randomly choosing exactly one 
association (among existing ones and whenever there are more than 2 such associations) for each symbiont leaf.  Notice that this is what is usually done in the literature when analysing such  datasets with a method that does not allow for multiple associations.  We detail here the results obtained for only two datasets, the reader can find the remaining ones in Section
~D.1 from the Supplementary Material. Computing times are also presented in Section
~D.2 from the Supplementary Material.

\emph{Dataset $1$: AP - \textit{Acacia} \& \textit{Pseudomyrmex}}. This dataset was extracted from 
\cite{gomez2010neotropical} and displays the interaction between \textit{Acacia} plants and 
\textit{Pseudomyrmex}, a genus of ants. Although the authors did not use a cophylogeny reconstruction tool to analyse the dataset, this is  considered as a typical example of mutualism between ants and plants, and the authors show that their relationship
originated in Mesoamerica between the late Miocene to the middle Pliocene, with eventual diversification of both groups in Mexico. 
The host and symbiont trees include 9 and 7 leaves, respectively. The dataset has 22 multiple-associations.
The corresponding dataset with no multiple association is called AP$_{Coala}$. 

\emph{Dataset $2$: SFC - Smut Fungi \& Caryophillaceus plants}. This dataset was extracted from 
\cite{RGJ2008}. 
The host and symbiont trees include 15 and 16 leaves, respectively. 
The dataset has 4 multiple associations. The corresponding dataset with no multiple association is called SFC$_{Coala}$.  Notice that this is the same dataset used in \cite{Baudet2015}.

\begin{figure}[!ht]
	\centering
\begin{subfigure}[b]{1\textwidth}
		\centering
		\includegraphics[width=\linewidth]{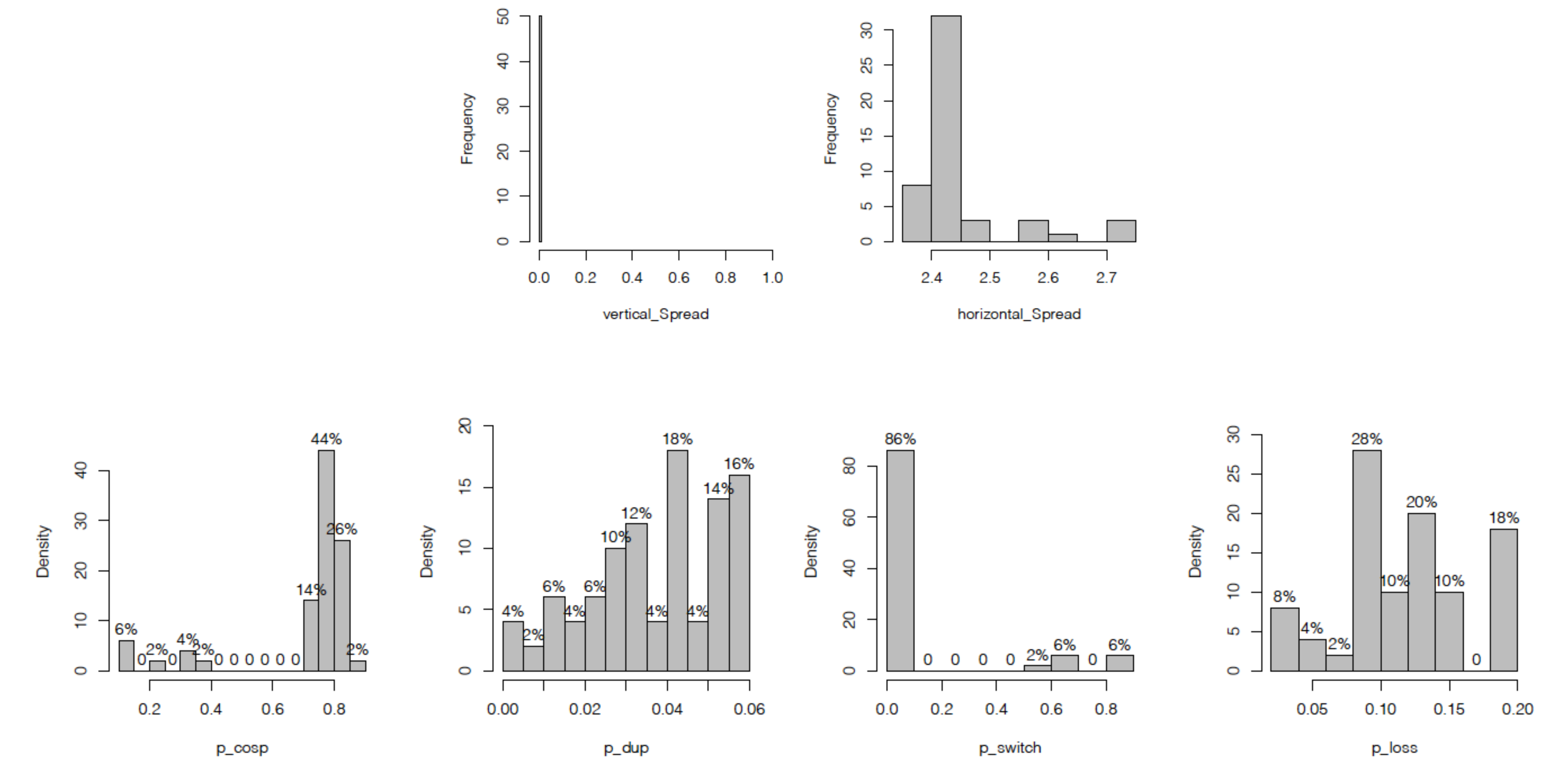}
	\end{subfigure}%
	\\ \medskip
\begin{subfigure}[b]{1\textwidth}
		\centering
		\includegraphics[width=\linewidth]{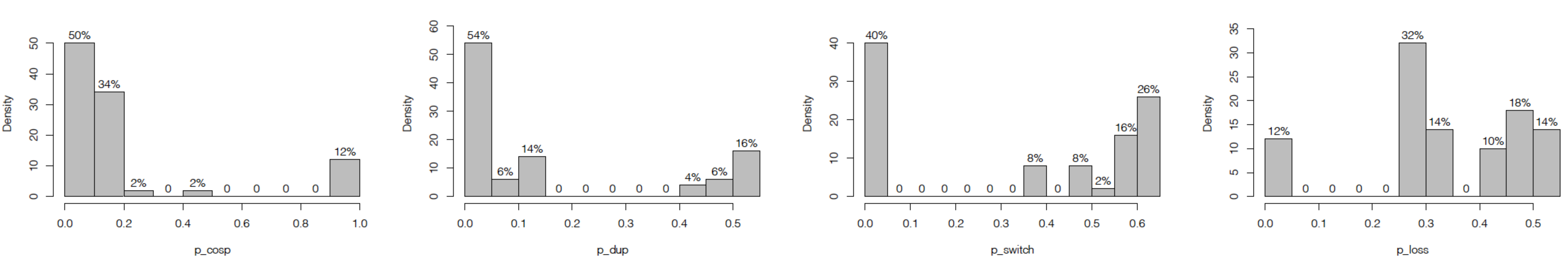}
	\end{subfigure}
	\caption{Comparison of the results obtained with \ACoala\ and \Coala\ for the dataset AP. 
	In each graphic, we show for each event type, the distribution of the parameter
values. In the first two rows, the results provided by \ACoala\ and in the third row, the ones provided by \Coala.}
	\label{fig:AP_datasets}
\end{figure}

\begin{figure}[!ht]
	\centering
\begin{subfigure}[b]{1\textwidth}
		\includegraphics[width=1.03\linewidth]{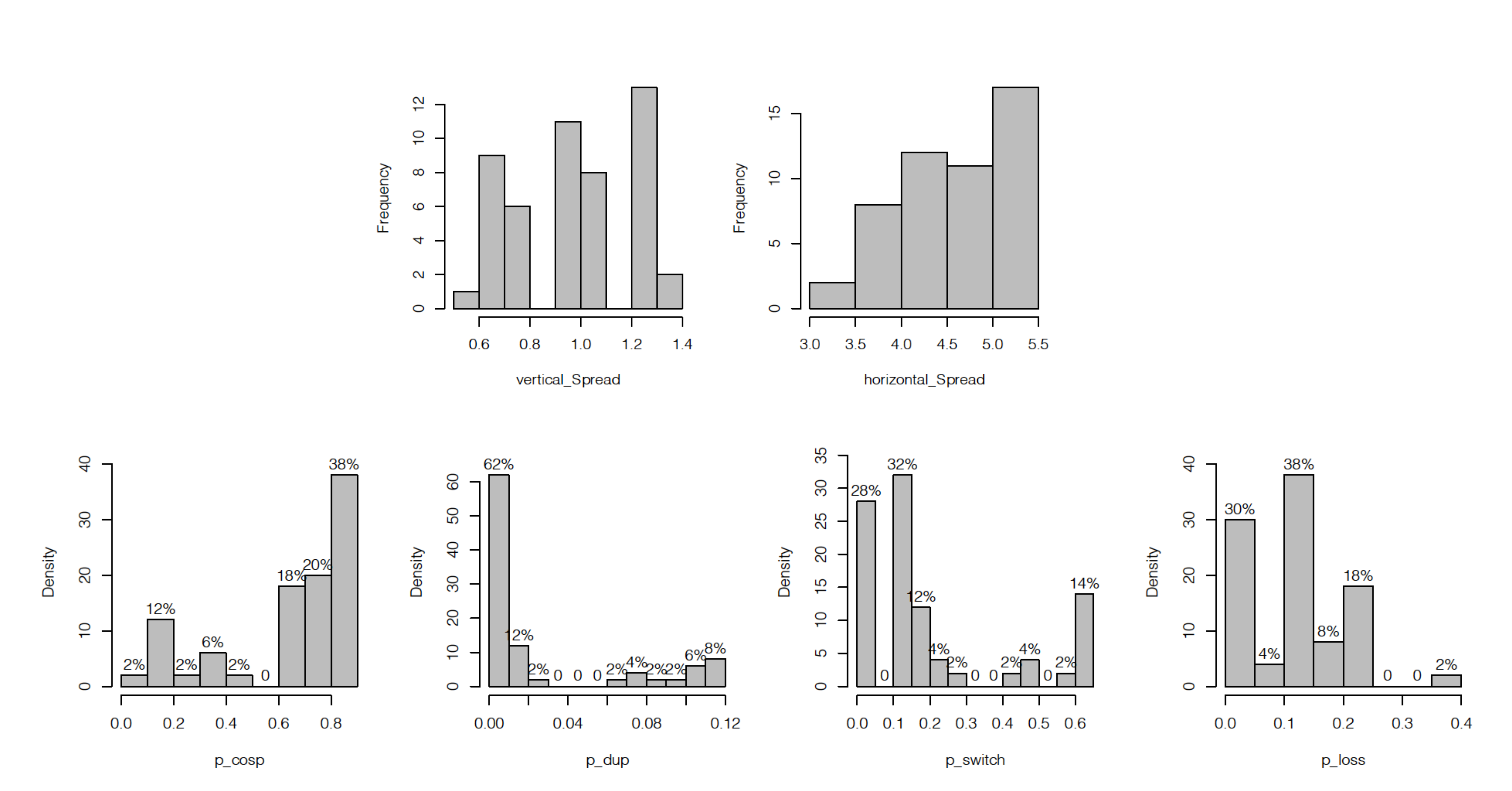}
	\end{subfigure}%
	\\ \medskip
\begin{subfigure}[b]{1\textwidth}
		\centering
		\includegraphics[width=\linewidth]{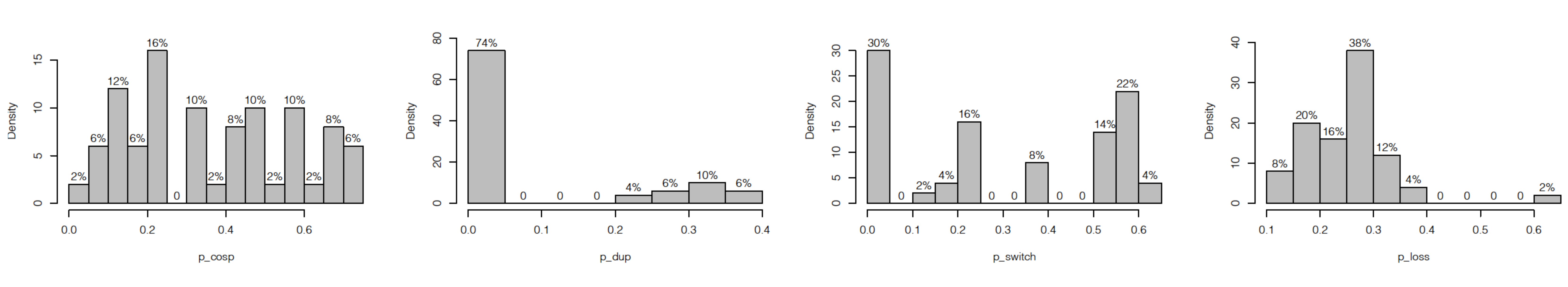}
	\end{subfigure}
	\caption{
Comparison of the results obtained with \ACoala\ and \Coala\ for the dataset SFC. 
In each graphic, we show for each event type, the distribution of the parameter
values.  In the first two rows, the results provided by \ACoala\ and in the third row, the ones provided by \Coala.}
	\label{fig:SFC_datasets}
\end{figure}

In Figures~\ref{fig:AP_datasets} and \ref{fig:SFC_datasets}, we present for each of the cophylogeny events, the distribution of the inferred probabilities  obtained by running \ACoala\ and \Coala.
First notice that the results change substantially when we consider the complete dataset instead of the one obtained by removing the multiple associations. Indeed, from the graphics in the third row of Figure~\ref{fig:AP_datasets}, we see that if we ignore 
multiple associations, then \Coala\ explains the dataset using a very low 
cospeciation frequency and a high number of switches and losses. In general, we can say that \Coala\ detects a high incongruence between the trees which cannot be explained by 
cospeciations. However, if the complete dataset is considered, \ie the one including all the multiple associations, we see from the first two rows of Figure~\ref{fig:AP_datasets} that the dataset can be explained by only 2-3 horizontal spreads, a high 
number of cospeciations, a very low number of duplications and switches and also a significantly lower number of losses.  Thus, the incongruence between the two phylogenetic trees can be explained by approximately 3 horizontal spreads and then most of the events correspond to cospeciations, which is an indication of coevolution. This is in accordance with what is expected for this dataset, which, as already mentioned in the previous paragraph, is  considered as a typical example of mutualism between ants and plants.


Next, we considered the dataset SFC with multiple associations proposed in \cite{RGJ2008}. From 
Figure~\ref{fig:SFC_datasets}, we can see that both methods show similar results concerning 
cospeciations, duplications and host switches while \ACoala\ outputs a smaller number of losses (less then $25\%$)  compared to \Coala\ (less then  $40\%$). 
In \cite{RGJ2008}, the different analyses performed indicated that the most plausible reconciliations presented for the SFC dataset have from 0 to 3 cospeciations, no duplication, 12 to 15 host switches and 0 to 2 losses.  It is impossible for us to calculate the number of events in a parsimony framework because there is no parsimonious algorithm for computing optimal  reconciliations in 
the presence of vertical and horizontal spreads.
Nonetheless, we have access to estimated frequencies of the reconstructed events. Moreover, 
from the definition of the model (see 
Sections
~A.2 and A.3 from  the Supplementary Material) we know that the sum of the classical events (cospeciation, duplication and host switch), excluding the loss event, is equal to the number of internal vertices of the symbiont tree. The symbiont tree (that is the same for SFC and SFC$_{Coala}$) has 15 internal vertices. Based on the analyses 
presented in \cite{RGJ2008}, we expect to have events with the following frequencies: between 0\% and 20\% for cospeciations (from 0 to 3 events), 0\% for duplications (no duplications), between 80\% and 100\% for 
host switches (from 12 to 15 events) and between 0\% and 13\% for losses (from 0 to 2 events). To compare the results output by the two methods (\Coala\ and \ACoala) with those expected from the analyses of \cite{RGJ2008}, we cluster the parameter vectors output by the methods. 
Indeed, both \Coala\ and \ACoala\ perform a hierarchical clustering procedure to group the final list of accepted parameter vectors. We then compared the cluster patterns  found by the two methods.  Table~\ref{tab:probabilities_round3} shows the representative vectors of each of the clusters output by \ACoala\ (for the SFC dataset) and by \Coala\ (for the SFC$_{Coala}$ dataset).  Notice that 
as already mentioned in 
\cite{Baudet2015}, a vector with a high frequency of host switches can generate  a large space of simulated trees, many of which can have a high distance from the real symbiont tree. Thus, it is clear that such vectors are more difficult to be output by both \Coala\ and \ACoala.

\setlength{\tabcolsep}{8pt}
\setlength{\extrarowheight}{2pt}
\begin{table}[htbp]
\begin{center}
\caption{Representative vectors of the clusters produced by \ACoala\ (for the SFC dataset) and by \Coala\ (for the SFC$_{Coala}$ dataset). The column $\# vectors$ indicates the number of vectors in the cluster. }
\label{tab:probabilities_round3}
\begin{tabular}{ccccccc} \hline
$Dataset$ & $Cluster$ & $p_c$ & $p_d$ & $p_s$ & $p_l$ & $\# vectors$ \\ \hline
\multirow{ 4}{*}{ SFC} &  1 & 0.531 &	0.004 &	0.282 &	0.183 & 19  \\ \cline{2-7}
                     &  2 & 0.226	 & 0.004 &	0.543	& 0.228  & 14\\ \cline{2-7}
                     &  3 & 0.898	 & 0.020 &	0.040	 & 0.042 & 12\\ \cline{2-7}
                     &  4 &  0.859 & 0.062	& 0.002	& 0.077 & 5 \\ \hline
\multirow{ 5}{*} {SFC$_{Coala}$} & 1 & 0.437	& 0.002	& 0.357	& 0.204 & 20 \\ \cline{2-7}
                     &  2 & 0.417 &	0.274 &	0.003 &	0.306 & 19 \\ \cline{2-7}
                     &  3 & 0.850	& 0.002 &	0.005 &	0.144 & 5 \\ \cline{2-7}
                     &  4 & 0.005	& 0.418 &	0.003 &	0.575  & 4 \\ \cline{2-7}
                     &  5 & 0.144 & 0.001 & 0.548 & 0.308 & 2 \\ \hline
\end{tabular}
\end{center}
\end{table}

From the results in  Table~\ref{tab:probabilities_round3}, we have that 
the event vector that is most similar to the expected one according to~\cite{RGJ2008}  is 
Cluster 2 for \ACoala\ run on the SFC dataset ($22.6\%$ for cospeciations, $0.4\%$ for duplications, $54.3\%$ for host switches and $22.8\%$ for losses). It is also important to note that the number of vectors that are part of this cluster is high (14 out of 50 vectors accepted in the third round).  
Notice that Cluster 5 of \Coala\ run on SFC$_{Coala}$ is also close to these values, 
however this cluster is supported by only 2 of the accepted vectors. Moreover, all the representative vectors of the clusters output by \ACoala\ have
a frequency of duplication close to $0$, which is in agreement with what is expected from \cite{RGJ2008}.

Overall the results obtained with \ACoala\ are  closer to the result presented in \cite{RGJ2008} than those that were obtained by \Coala\ which ignores such multiple associations. This shows again the importance of taking into account the latter. 

\tocless\subsection{Comments on the algorithm complexity and running time}
\ACoala\ has basically the same algorithmic complexity as \Coala. It first  requires a pre-computation of 
the spread probabilities which scales with the number of pairs of incomparable nodes in the host tree. So this step  has an $O(|L(H)|^2)$ time complexity, 
 which will be negligible 
 compared to the main term. Next, the time complexity depends on the hyper-parameters of the algorithm: the number of rounds $R$ (which in general will be less than 5); the numbers $N_r$ of vectors to be generated at each round (these numbers may also be obtained as the combination of an initial number of vectors and tolerance values, as introduced in Algorithm~\ref{algo:ABC_SMC}) and the number $M$ of symbiont trees to be generated for each parameter vector. 
 First, Algorithm~\ref{algo:simu_spreads} is an iterative process of simulating a tree, whose total number of simulation steps is $O(|L(H)|)$. However, when sampling a ``switch'' event, the time feasibility condition requires at most $O(|L(H)|^2)$ operations to be checked. Thus, the generation of a symbiont tree (namely Algorithm~\ref{algo:simu_spreads} except for its final filtering step) has a time complexity of $O(|L(H)|^3)$. Then, as a default value, Algorithm~\ref{algo:simu} may  simulate up to $5M$  symbiont trees for each parameter vector (to account for the filtering step in Algorithm~\ref{algo:simu_spreads}) and this constant $5$ does not impact on the time complexity of this algorithm. Also, computing the distance between the 2 trees has complexity $O(|L(S)|\times |L(\tilde S)|)=O(|L(S)|^2)$, because the filtering step ensures that the size of simulated tree $\tilde S$ is no more than twice that of $S$ and thus $|L(\tilde S)|=O(|L(S)|)$. 
 Finally, the complexity of Algorithm~\ref{algo:simu}  
 is $O(M \times (|L(H)|^3+|L(S)|^2))$. 
  Thus, \ACoala\ has a global complexity 
  of $O(M\times(|L(H)|^3+|L(S)|^2)\times (\sum_{r=1}^R N_r) )$, which can be quite large. 

Examples of running times are given in Section~\ref{sec:SM-time} of the Supplementary Material; see also the section \emph{Running times} in \cite{Baudet2015}. 
In the experiments of this manuscript, default values were given for all hyper-parameters. In the case of dealing with large trees, it might be wise to modify these values, especially the number of trees $M$ to be simulated. 
However, this will be at the cost of potentially 
losing in accuracy. 
We also mention that the code's  implementation is parallelized for the simulation of the symbiont trees.

\tocless\subsection{{Using \ACoala\ to analyse coevolution}}
{
It is important to emphasize that neither \Coala\ nor \ACoala\ provide a direct reconciliation of the two trees, but instead offer a set of estimated probabilities for coevolutionary events. This is also the case for other algorithms, such as the one proposed in \cite{alcala2017host}.}

{Let us begin by recalling that in datasets without multiple associations, \ACoala\  implements our previous tool, \Coala, and its outputs can be utilized as input costs in a parsimonious reconciliation method. The procedure is briefly described here, with more details available in the work by \cite{Baudet2015}.
\Coala\ provides a comprehensive set of estimated parameter values $\langle p_c,p_d,p_s,p_l\rangle$, organized into clusters, where each cluster is summarized by a representative parameter that includes probabilities for each event.
To proceed, the probabilities need to be transformed into costs. While the choice of the transformation function from probabilities $p$ to costs $c$ requires further research (beyond the scope of this study), a common approach is to employ the classical method of $c=-\log(p)$.
As a result, a parsimonious reconciliation method can be employed with cost values obtained by taking the negative logarithmic transforms of the representative parameter probabilities for each cluster (or for clusters with a sufficiently large relative size).}

{
Currently, there is no existing method to compute a most parsimonious reconciliation under a model that incorporates spreads. Consequently, it is not straightforward to directly utilize the outputs of \ACoala\ and provide them as input costs for a reconciliation method based on the same coevolution model that allows for spreads. 
Therefore, a significant future direction for this research is to develop and design reconciliation procedures that incorporate spread events and can effectively utilize the outputs of \ACoala\ as realistic costs for those events. This would enable a more comprehensive and accurate analysis of coevolutionary relationships.} 

{In the meantime, \Coala\ can be utilized in at least two different ways.
The first approach is to conduct qualitative analysis of datasets, as demonstrated in the four biological datasets mentioned above. In datasets with multiple associations, \ACoala\ enables us to handle the data without arbitrary modifications that would remove those multiple associations. It provides estimated probabilities in the form of representative vectors from the largest clusters for the four classical coevolutionary events. This allows us to estimate the expected numbers (or at least bounds) of cospeciations, duplications, switches, and losses in a reconciliation of the two trees.
The second possibility is to use \ACoala\ in a similar manner as \Coala, namely, by taking the negative logarithmic transformation of the probabilities for the four classical events obtained from representative parameters and inputting them as costs into a parsimonious reconciliation method, even if the method does not handle spread events. We believe that our estimated values are more accurate than those produced by methods that simply remove multiple associations in ad-hoc ways. Although we do not expect a dramatic improvement in this scenario, we anticipate that this approach will provide a more accurate reconciliation scenario.
}

\tocless\section{Concluding comments}\label{Discussion}
In this paper, we propose a method, called \ACoala, which for a given pair of host and symbiont trees, estimates the probabilities of the cophylogeny events, in presence of spread  events, relying on an approximate Bayesian computation (ABC) 
approach. In \ACoala, it is possible to estimate the probabilities
of the classical cophylogeny events (cospeciation, duplication, host switch and loss) and also the probabilities of horizontal and vertical spreads {(heterogeneous along the host tree)}. These two latter events allow to study datasets that contain multiple associations. The model uses set-labelled trees and to compare them we introduced  a new distance, called $d_{MASTT}$, which we believe can be of independent interest.

{\ACoala\ can effectively handle datasets with multiple associations, avoiding arbitrary treatment of such associations. The method leverages the information present in these multiple associations to deliver more precise estimates for the probabilities of the four classical coevolutionary events.} 
{We demonstrate the ability  of our method to produce more accurate results both on synthetic and real datasets.}

This work leads to different research directions. First, it would be interesting to define better distances for set-labelled trees. To the best of our 
knowledge, these types of trees have not been considered in the literature and it would be interesting to generalise (if possible) some of the well-known phylogenetic distances to set-labelled trees. 
Another direction is to include the vertical and horizontal spreads in a parsimonious reconciliation framework. Thus, a perspective to this work is to design a reconciliation procedure that includes these switches. 

\tocless\section{Acknowledgments}
{The authors would like to thank 2 anonymous referees as well as associate and editor of the journal for their helpful comments on previous versions of this work.}

\tocless\section{Software and Supplementary Material}
The software, datasets and Supplementary Material are available at \url{https://github.com/sinaimeri/AmoCoala} and supplementary material is accessible on a Dryad repository at \url{https://datadryad.org/stash/share/SHDH-seLRIznGHCRdQRUNuWE01TnmD5BipocuFrdNUg} with an associated DOI of doi:10.5061/dryad.5x69p8d6v (this last link will only be active upon publication).

\tocless\section{Disclosure statement}
The authors state they have no conflicts of interest to declare.

\newpage
\appendix
\setcounter{equation}{0} 
\setcounter{figure}{0} 
\setcounter{table}{0} 
\renewcommand\theequation{S.\arabic{equation}}
\renewcommand\thefigure{S.\arabic{figure}}
\renewcommand\thetable{S.\arabic{table}}

\pagestyle{fancy}
\rhead{Supplementary Material}
\lhead{\ACoala}
\thispagestyle{empty}

\begin{center}
\noindent{\large \bf
\textsc{Supplementary Material for: Cophylogeny reconstruction allowing for multiple associations through approximate Bayesian computation}}
  \bigskip

\noindent {\normalsize \sc{ Blerina Sinaimeri$^{1,2,\dagger}$, Laura Urbini$^{2,}$\daggerfootnote{First co-authors.}, Marie-France Sagot$^2$ and
Catherine Matias$^3$}}\\
\noindent {\small \it 
$^1$ LUISS University, Rome, Italy \\
$^2$ Inria Lyon, 56 Bd Niels Bohr, 69100 Villeurbanne, France, and
	Universit\'e de Lyon, F-69000, Lyon; Universit\'e
        Lyon 1; CNRS, UMR5558; 43
	Boulevard du 11 Novembre 1918, 69622 Villeurbanne cedex, France \\
	$^3$ Sorbonne Universit\'e,  Universit\'e de Paris Cité,
Centre National de la Recherche Scientifique, Laboratoire de Probabilit\'es, Statistique et Mod\'elisation, Paris,
France
}
\end{center}

\medskip
\noindent{\bf Corresponding author:} 
Blerina Sinaimeri, LUISS University, Rome, Italy; E-mail: bsinaimeri@luiss.it.\\

\tableofcontents


\section{The event-based model}\label{sec:event_model}

\ACoala\ relies on the event-based model presented in \cite{Charleston2002,THL11}. For the sake of completeness, we
detail the model here. We first start with some basic definitions related to phylogenetic trees.

\subsection{Tree-related basic definitions}\label{subsec:preliminaries}
A rooted phylogenetic tree is a leaf-labelled tree that models the evolution of a set of taxa from their most recent
common ancestor (placed at the root). The internal vertices of the tree correspond to the speciation events.  In a
rooted phylogenetic tree, a direction is assumed from the root to the leaves that corresponds to the direction of
evolutionary time. Specifically,  a  phylogenetic tree is a rooted tree with labelled leaves where
the root has in-degree 0 and out-degree 2, the leaves have in-degree 1 and out-degree 0 and every internal vertex has in-degree 1 and out-degree 2. For such a tree $T$, the set of vertices is denoted by $V(T)$, the set of arcs by $A(T)$, and the set of leaves by $L(T)$. The cardinality of set $A$ is denoted 
by $|A|$. The root of $T$ is denoted by $r(T)$. For a vertex $v$ in a tree $T$, we denote by $T_v$ the subtree of $T$ rooted in $v$ (often 
referred to as a \emph{clade}), and we write $L(v)$ for the set $L(T_v)$.  For a vertex  $v \in V(T)$, we denote by $Des(v)$ the set of \emph{descendants} of $v$, \textit{i.e.} the set of vertices in the subtree of $T_v$.  Similarly,  we denote by $Anc(v)$ the set of \emph{ancestors} of $v$,  that is the set of vertices in the unique path from $r(T)$  to $v$ (including the end points). For a vertex $v \in V(T)$ different from the root, we call its \emph{parent}, denoted by $par(v)$, the vertex $x$ for which there is the arc $(x, v) \in A(T)$. We denote by $\lca(v,w)$ the most recent common ancestor of $v$ and $w$ in $T$. Finally, we denote by $\leq$ the partial order induced by the ancestry
relation in the tree. Formally, for $x, y \in V (T)$, we say that $x \leq  y$ if $x \in Anc(y)$. If neither
$x \in Anc(y)$ nor $y \in Anc(x)$, the vertices $x$ and $y$ are said to be \emph{incomparable}. 

For any tree $T$ and any set of leaves $t_1,\dots , t_n$, we denote by $T_{ | \{t_1,\dots, t_n\} }$  the phylogenetic
subtree of $T$ induced by the leaves $t_1,\dots , t_n$ and eventually suppressing the vertices of out-degree 1. When a vertex $u$ with parent vertex $v$ and child vertex $w$ is 
suppressed, both vertex $u$ and arcs $(v,u), (u,w)$ are removed and the arc $(v,w)$ is added to the tree.

\subsection{Reconciliation model from 
Tofigh et al.}
\label{sec:SM-DTLmodel}
In this section, we describe the classical reconciliation model, where  4 coevolutionary events are allowed, producing no multiple associations.
Let $H$ and $S$ be respectively the rooted phylogenetic trees of the host and symbiont species, both binary and
full (\textit{i.e.} each internal vertex has exactly two children). Let $\phi$ be a function 
from $L(S)$ to $L(H),$ representing the symbiont/host associations between extant species.
A reconciliation is a function $\lambda$ that assigns, for each symbiont vertex $s\in V(S)$,  a host vertex $\lambda(h)\in V(H)$,  and satisfies the conditions stated in Definition~\ref{defn:recon}. 

In its classical form, a  reconciliation associates to each vertex $s$ in $V(S)$  an event
$E(\lambda(s))$ among cospeciation ($\mathbb{C}$), duplication
($\mathbb{D}$) and  host switch ($\mathbb{S}$).

\begin{definition}\label{defn:recon}
Given two phylogenetic trees $S$ and $H$, and a function  $\phi : L(S)\to L(H)$, a reconciliation of $(S, H, \phi)$ is a function $\lambda: V(S)\to V(H)$ satisfying the following:
\begin{enumerate}
\item For every leaf vertex $s\in L(S),$ we have $\lambda(s)=\phi(s)$.
\item For every internal vertex $s\in V(S)\setminus L(S)$ with children $s_1,\,s_2,$  exactly one of the following applies:
\begin{enumerate}
\item $E\left(\lambda(s)\right)=\mathbb{S}$, that is, either $\lambda(s_1)$ and $\lambda(s)$ are incomparable and $\lambda(s_2)$ is a descendant of $\lambda(s)$, or $\lambda(s_2)$ and $\lambda(s)$ are incomparable and $\lambda(s_1)$ is a descendant of $\lambda(s)$,
\item $E\left(\lambda(s)\right)=\mathbb{C}$, that is, $\lca(\lambda(s_1),\lambda(s_2)) = \lambda(s)$, and $\lambda(s_1)$ and $\lambda(s_2)$ are incomparable,
\item $E\left(\lambda(s)\right)=\mathbb{D},$ that is, $\lambda(s_1)$ and $\lambda(s_2)$ are both descendants of $\lambda(s)$, and the previous two cases do not apply.
\end{enumerate}
\end{enumerate}
\end{definition}

The loss event is denoted by $\mathbb{L}$ and is identified by a multiset 
(generalisation of a set where the elements
are allowed to appear more than once)  whose elements are in $V(H)$ containing all the vertices $h \in V(H)$ that are in
the path between the image of a vertex $s \in  V(S)$ and the image of one of its children. The images themselves are not included in the count, except for the duplication event, where one of the images is included.

The function $\lambda$ partitions the set of internal symbiont tree vertices into three disjoint subsets
according to the coevolutionary event occurring at that vertex.
The number of occurrences of each of 
the three events and the number of losses make up the \textit{event vector} of the reconciliation.
The \textit{event vector} of a reconciliation is a vector of integers consisting of the total number of each type of events $\mathbb{C},\,\mathbb{D},\,\mathbb{S},\,\mathbb{L}$. 

 We say that a reconciliation is \emph{time-feasible} if it does not violate the time-feasibility constraints. The exact criterion we
use to assess time-feasibility is the one defined in \cite{Stolzer2012} and  that was already in force in {\sc Coala}.

\subsection{Reconciliation model allowing for spreads}
\label{sec:SM-DTL_spread}
The introduction of spread events modifies the previous setting in the following way. 
Let again $H$ and $S$ be respectively the rooted phylogenetic trees of the host and symbiont species, both binary and
full (\textit{i.e.} every internal vertex has exactly two children). 
Now, let $\phi$ be a relation 
between $L(S)$ and $L(H),$ representing the symbiont/host associations between extant species. More precisely, let us denote $\mathcal{P}(L(H))$ the set of all subsets of $L(H)$. Then $\phi$ is now a function from
  $L(S)$ to $\mathcal{P}(L(H))$.
  For any extant symbiont species $s\in L(S)$, whenever the cardinality $|\phi(s)|\ge 2$ (\emph{i.e.} whenever the symbiont is associated to more than one host),  we say that this symbiont has   multiple associations and we count the total number of multiple associations in the dataset as:
  \[
\text{  Nb of multiple associations } = \sum_{s \in L(S)} (|\phi(s)|-1 ).
  \]
A reconciliation is now a function  $\lambda$  from $ V(S)$ to $\mathcal{P}(V(H))$ that assigns, for each symbiont vertex $s\in V(S)$, a set of host vertices $\lambda(s)\subset V(H)$, and satisfies the conditions stated in Definition~\ref{defn:recon_spread}. 
A  reconciliation now associates to each vertex $s$ in $V(S)$  an event
$E(\lambda(s))$ among cospeciation ($\mathbb{C}$), duplication
($\mathbb{D}$), host switch ($\mathbb{S}$), vertical spread ($\mathbb{VS}$) and horizontal spread  ($\mathbb{HS}$). 

\begin{definition}\label{defn:recon_spread}
Given two phylogenetic trees $S$ and $H$, and a function  $\phi : L(S)\to \mathcal{P}(L(H))$, a reconciliation of $(S, H, \phi)$ is a function $\lambda: V(S)\to \mathcal{P}(V(H))$ satisfying the following:
\begin{enumerate}
\item For every leaf vertex $s\in L(S),$ we have $\lambda(s)=\phi(s)$.
\item For every internal vertex $s\in V(S)\setminus L(S)$ with children $s_1,\,s_2,$ such that $\lambda(s)$ is a singleton, exactly one of the following applies:
\begin{enumerate}
\item $E\left(\lambda(s)\right)=\mathbb{S}$, that is, either $\lambda(s)$ and one element of $\lambda(s_1)$ are incomparable and $\lambda(s_2)$ contains a descendant of $\lambda(s)$, or $\lambda(s)$ and one element of $\lambda(s_2)$ are incomparable and $\lambda(s_2)$ contains a descendant of $\lambda(s)$,
\item $E\left(\lambda(s)\right)=\mathbb{C}$, that is, there is some $h_1 \in \lambda(s_1)$ (resp. $h_2\in \lambda(s_2)$)  such that  $\lca(h_1,h_2) = \lambda(s)$, and $h_1$ and $h_2$ are incomparable,
\item $E\left(\lambda(s)\right)=\mathbb{D},$ that is, there is some $h_1 \in \lambda(s_1)$ (resp. $h_2\in \lambda(s_2)$)  such that  both $h_1,h_2$ are descendants of $\lambda(s)$, and the previous two cases do not apply.
\end{enumerate}
\item  For every internal vertex $s\in V(S)\setminus L(S)$     such that $\lambda(s)$ is not a singleton, 
    exactly one of the following applies:
  \begin{enumerate}
  \item  $E\left(\lambda(s)\right)=\mathbb{VS},$ that is $\lambda(s)$ is a clade in $H$, and  all the descendants $s'$ of $s$ are also associated to the same clade, \emph{i.e.} $\lambda(s')=\lambda(s)$.
    \item  $E\left(\lambda(s)\right)=\mathbb{HS},$ that is $\lambda(s)$ is the union of two clades in $H$ whose respective roots are incomparable. Moreover, all the descendants $s'$ of $s$ are also associated to the same clades, \emph{i.e.} $\lambda(s')=\lambda(s)$.
    \item $s$ is the descendant of a node $s'$ where a spread (either vertical or horizontal) occurred (cases (3a) and (3b)). Then $\lambda(s)=\lambda(s')$. In that case, no additional coevolutionary event is recorded at that vertex.
  \end{enumerate}
\end{enumerate}
\end{definition}

The loss event denoted by $\mathbb{L}$ 
is identified by a multiset 
(generalisation of a set where the elements
are allowed to appear more than once)  whose elements are in $V(H)$ containing all the vertices $h \in V(H)$ that are in
the path between the image of a vertex $s \in  V(S)$ which is a singleton and the image of one of its children.
Note that no other event and thus no losses can happen below spread events.

Now, the function $\lambda$ partitions the set of internal symbiont tree vertices into five disjoint subsets according to the coevolutionary event occurring at that vertex, 
plus an additional subset of all internal symbiont vertices that descend from a vertex where a spread occurred.
 The number of occurrences of each of 
 the five events and the number of losses make up the \textit{event vector} of the reconciliation.
The \textit{event vector} of a reconciliation is a vector of integers consisting of the total number of each type of events $\mathbb{C},\,\mathbb{D},\,\mathbb{S},\,\mathbb{L},\,\mathbb{VS},\,\mathbb{HS}$. Note that in the case of spread events (either vertical or horizontal) occurring at internal vertex $s\in V(S)\setminus L(S)$, the event is counted only once and the internal vertices $s'$ descendants of $s$ have no coevolutionary event associated to them. 

The time feasibility condition is unchanged when adding spreads in the list of coevolutionary events.

\subsection{Pre-estimating probabilities for the spread events}
\label{sec:SM-spread}
Given an input dataset $(H,S,\phi)$, we rely on frequency estimators for the spread probabilities that will be used in our algorithm. 
Note that the ``classical events''  (cospeciation, duplication, host switch and loss) have the same probability to occur everywhere in the tree, while the probability of a vertical or horizontal spread is specific to each vertex of the host tree. These probabilities are pre-estimated based on the input $(H,S,\phi)$ as described below rather than in the full ABC procedure. They are estimated through heuristic frequencies observed in the associations of the two trees. In Section~\ref{sec:SM-robust}, we explore the robustness of our results with respect to these pre-computed estimators.

\paragraph{Probability  that a vertical spread occurs at host  $h$. } 
A probability $\pvs(h)$ is associated to a
vertical spread event at host $h$ as follows. If $h \in L(H)$, then $\pvs(h)$ is estimated to 1.  Otherwise, for any internal
vertex $h$ of the host tree $H$, the probability $\pvs(h)$ is estimated to 
\begin{equation}
\pvs(h)=\left( \frac{1}{| S^{L(h)}|} \right)  \frac{\sum_{s \in S^{L(h)} } | \phi(s)\cap L(h) | -1}{|L(h)| -1} 
\label{eq:pvs}
\end{equation}
where $L(h)$ is the set of leaves in $H_h$ (the subtree of $H$ rooted in $h$),  $S^{L(h)}$ is the set of leaves in the symbiont tree $S$ that are associated with at least one leaf of $H_h$ (formally $S^{L(h)}=\{ s\in L(S):\phi(s)\cap L(h) \neq \emptyset\}$),  and $|\phi(s)\cap L(h)|$ is the number of host leaves in $H_h$ associated with a symbiont $s$. 

Intuitively, the probability $\pvs(h)$ is large whenever 
 a large proportion  of the symbionts in $S^{L(h)}$ are associated to  a large proportion of the hosts $L(h)$ (\ie\ most of the symbionts are generalists) and is low  when most of those symbionts are associated only with 
 a few hosts of $L(h)$ (\ie\ most of the symbionts are 
 specialists). Notice that for a host $h$ that is high in the tree, \ie\ that is near to the root of $H$, the set $L(h)$ is large. Thus, a vertical spread to occur at $h$ with high probability
 requires that some symbiont leaves are associated to an unrealistically large set of hosts $L(h)$. 
 Hence usually the probability of  a vertical spread is lower in hosts that are high in the tree. As explained in the next paragraph, the same holds for the horizontal spread event.

\paragraph{Probability that a symbiont present in $h$ invades an incomparable host $h'$.} 
For two incomparable vertices $h$ and $h'$, a probability  $\pjump{h}{ h'}$ is estimated  as follows 
\begin{equation}
\pjump{h}{ h'} = \frac{| S^{L(h) }  \cap S^{L(h') }|}{| S^{L(h) }  \cup S^{L(h') }|}. 
\label{eq:jump}
\end{equation}
The notion of ``jump'' does not refer to a coevolutionary event and should not be confused with a host switch. The jump probability is specific to each pair of vertices of the host tree. It is a symmetric quantity, \ie\  $\pjump{h}{ h'}=\pjump{h'}{ h}$. 
It is high whenever the leaves of the subtrees  $H_{h}$ and $H_{h'}$ 
share 
a large proportion of associated symbionts.  
In particular, it is zero when they do not share any associated symbiont, and 1 when they have exactly the same set of associated symbionts.

\paragraph{Probability that a horizontal spread occurs at host  $h$.} 
From the probabilities $\pjump{h}{ h'}$, we
estimate a probability of horizontal spread at each vertex $h$. The associated probability depends on all the vertices $h'$
that are incomparable with $h$. Indeed, such vertices are all those that may be reached from $h$ through a horizontal
spread event. In fact, a horizontal spread corresponds to a jump  combined with two vertical spreads. We thus associate a
probability of horizontal spread $\phs(h)$ to each vertex $h$ of the host tree that takes into account both a jump and two 
vertical spreads and is set as 
\begin{equation}
\phs(h) = \min \{ 1,  \phstar(h) \},
\label{eq:phs}
\end{equation} 
where
 \[ 
 \phstar(h) =  \pvs(h)   \sum_{\substack{h'\in V(H) \\ h,h'\text{ incomparable}}} \pvs(h')  \pjump{h}{ h'} .
 \]
The probability of 
a horizontal spread $\phs(h)$ is high whenever $\pvs(h)$ is high and there exist vertices $h'$ incomparable to $h$ with large $\pvs(h)$ and large value $\pjump{h}{ h'}$ (so that the leaves below $h$ and $h'$ share many  symbionts).  Observe that $ \phstar(h)$ is not a probability but a positive value, that in particular may be larger than 1.

\paragraph{Probability for sampling a horizontal spread to some specific host $h'$.}
In the simulation process,  
once a horizontal spread is sampled for symbiont $s$ at vertex $h$, we need to choose an incomparable vertex $h'$ where the symbiont $s$ has to jump to. In this case, we need to guarantee that the jump  satisfies the time-feasibility constraints as given in \cite{Stolzer2012} and \cite{Baudet2015}. This constraint depends on the symbionts mapped so far (see 
Section \emph{Simulation algorithm in \ACoala} below).
For a current  partial  mapping $\lambda$ from the vertices of $S$ to the subsets of vertices of $H$, the probability $\pinvade{h}{h'}{\lambda}$ of a vertex $h'$ to be invaded by  a symbiont $s$ mapped in $h$ is estimated as
\begin{align}
\pinvade{h}{h'}{\lambda} &= \frac{ \pjump{h}{ h'}\pjAcyclic{h}{ h'}{\lambda} \pvs(h) \pvs(h')}{ \pvs(h)   \sum_{h''} \pvs(h'') \pjump{h}{ h''}\pjAcyclic{h}{ h''}{\lambda} },\nonumber \\
&= \frac{ \pjump{h}{ h'}\pjAcyclic{h}{ h'}{\lambda}  \pvs(h')}{    \sum_{h'' } \pvs(h'') \pjump{h}{ h''}\pjAcyclic{h}{ h''}{\lambda} },
\label{eq:invade}
\end{align}
where $\pjAcyclic{h}{ h'}{\lambda}  = 1$ whenever 
the horizontal spread of the symbiont mapped in $h$ to the new host $h'$
induces a time
feasible reconciliation, and the sum in the denominator is restricted to the vertices $h''$ that are incomparable to
$h$. If no vertex induces a time feasible reconciliation (namely $\pinvade{h}{h'}{\lambda} =0$ for any $h'$ incomparable
  to $h$), the horizontal spread is not applied and another event is sampled. Otherwise, as 
the probabilities $\pinvade{h}{h'}{\lambda}$ sum up to one,  a vertex $h'$ is necessarily chosen. 

\paragraph{Computing the pre-estimated spread probabilities.} 
The estimated spread probabilities are calculated at the beginning of the algorithm. These values depend only on the host tree $H$, the symbiont tree $S$ and the associations between the leaves $\phi$. In a first step, we start by setting to 1 the probabilities $\pvs$  for the leaves. Then, for the internal vertices $h$, these probabilities are computed as in Equation~\eqref{eq:pvs}. In a second step, the probabilities of a jump are calculated for each pair of incomparable vertices $h$ and $h'$ as in Equation~\eqref{eq:jump}. In the last step, the probabilities of a horizontal spread for vertex $h$ are computed as in Equation~\eqref{eq:phs}.
Observe that the probabilities of invasion (Equation~\eqref{eq:invade}) depend on the current simulation. Indeed, one has to take into account the  time-feasibility in order to choose the target $h'$ of a horizontal spread. Therefore, it may happen that the invasion $\pinvade{h}{h'}{\lambda}>0$ for the current partial mapping $\lambda$ but after some steps $\pinvade{h}{h'}{\lambda'}=0$ for the new mapping $\lambda'$. These probabilities are then  updated, during the simulation algorithm, each time a horizontal spread is selected.

\section{\ACoala\ algorithm}
\subsection{Simulation algorithm in \ACoala} 
\label{sec:SM-algo_simu}

The simulation of a symbiont tree $\tilde{S}$ together with its reconciliation $\tilde \lambda$ starts with the creation of its root vertex $\tilde{s}_{root}$.  This vertex is positioned before the root of $H$ on the arc $a = (\rho, H_{root})$. We add the arc $(\rho,H_{root})$ to allow the simulation of events that happened in the symbiont tree before the most recent common ancestor of all host species in $H$. Figure~\ref{fig:initial_config} in main text depicts this starting configuration.

For any vertex $\tilde{s}$ of $\tilde{S}$ that is not yet mapped and whose position is $\langle \tilde{s} : a \rangle$ (see Figure~\ref{fig:initial_config} in main text), \ACoala\ successively considers the six allowed operations, and chooses one depending on the probability of each event (once an event is picked, the others are not considered). In what follows, we denote by $a_1, a_2$ the arcs outgoing from the head $h(a)$ of the arc $a$. 

\begin{itemize}

\item[I.]  If $h(a)$ is a leaf, we \emph{STOP} the evolution of $\tilde s$.

\item[II.]  We first sample a horizontal spread according to the probability $\phs\bigl(h(a)\bigr)$. When a horizontal spread occurs (Figure~\ref{fig:horizontal_spread_simulation} in main text), we apply the 
mapping  $\tilde{\lambda}(\ts)= H_{h(a)} \cup {H_{h(a')}}$.
The choice of the
  incomparable vertex $h(a')$  varies in order to preserve
  time feasibility   \citep{Stolzer2012,Baudet2015}, thus the probabilities described in Equation \eqref{eq:invade} are updated according to the new set
  of incomparable vertices. If there is no incomparable vertex, it is not possible for a horizontal spread to occur and we go to   Step III. To select the ghost subtree rooted in 
  $\tilde s$, we mimic the real symbiont tree as shown in Figure~\ref{fig:horizontal_spread_simulation} in main text.
 
\item[III.] If a horizontal spread did not occur,  we sample a vertical spread according to the probability $\pvs\bigl(h(a)\bigr)$. When a vertical spread occurs (Figure~\ref{fig:vertical_Spread_simulation} in main text), we apply the mapping $\tilde{\lambda}(\ts)= H_{h(a)}$. To select the ghost subtree rooted in $\tilde s$, we mimic the real symbiont tree as shown in Figure~\ref{fig:vertical_Spread_simulation} from main text.

In both cases of vertical and horizontal spreads, the evolution of $\ts$ stops after the creation of the ghost subtree and its descendants are not processed anymore. 

\item[IV.] If a spread was not sampled, then we sample  with a multinomial distribution a classical event according to the probabilities $\theta=\langle  p_{c},p_{d},p_{s}, p_{l}\rangle$.  Notice that $ p_{c}+p_{d}+p_{s}+ p_{l}=1$ so that one of the four events is selected. This case is handled identically as in \Coala\ and the symbiont is associated to a single host.  We briefly recall the procedure below.

\begin{itemize}

\item Cospeciation (Figure~\ref{fig:classical_event_simulation}(b) in main text): We apply the mapping $\tilde\lambda(\ts)=\{h(a)\}$  and we create the vertices $\ts_1$ and $\ts_2$ as children of $\ts$. We position them as follows: $\langle \ts_1 : a_1 \rangle$ and $\langle \ts_2 : a_2 \rangle$. This operation is executed with probability $p_c$.

\item Duplication (Figure~\ref{fig:classical_event_simulation}(c) in main text): We apply the mapping 
$\tilde\lambda(\ts)=\{h(a)\}$
and we create the vertices $\ts_1$ and $\ts_2$ as children of $\ts$. Both $\ts_1$ and $\ts_2$ are positioned on $a$. This operation is executed with probability $p_d$.

\item Host switch (Figure~\ref{fig:classical_event_simulation}(e) in main text): We apply the mapping $\tilde\lambda(\ts)=\{h(a)\}$ and we create the vertices $\ts_1$ and $\ts_2$ as children of $\ts$. We then randomly choose one of the two children and position it on $a$. Finally, we randomly choose an arc $a'$ that does not violate the time feasibility of the reconstruction so far \citep{Stolzer2012,Baudet2015}. If such an arc does not exist, it is not possible for a host switch to take place. In this case, we choose between the three remaining events with probability $p_i / (p_c + p_d + p_l)$ with $i \in \{c,d,l\}$. Otherwise, we position $\ts_2$ on $a'$. This operation is executed with probability $p_s$.

\item Loss (Figure~\ref{fig:classical_event_simulation}(e) in main text): This operation consists of randomly choosing an arc outgoing from the head $h(a)$ of $a$ and positioning $\ts$ on it. This operation is executed with probability $p_l$.
\end{itemize}
In any of these four cases, the simulation process recursively continues with the new vertices created (back to Step I).

\end{itemize}

Note that in our modelling, losses never occur after a spread event. Indeed, in the case of a vertical spread, a symbiont and its entire clade are associated to one host clade, while in the case of a horizontal spread, they are  then associated to two host clades. This might appear unrealistic. However, this choice is made for computational reasons. Indeed, as mentioned in the Main Manuscript, there is no simple way of simulating the symbiont tree below a symbiont where a spread occurs.

\subsection{ABC-SMC inference method in \ACoala}
\label{sec:SM-ABC_method}
\ACoala\ is based on the same ABC-SMC method as the one developed in \Coala\  \citep{Baudet2015}. 
For the sake of completeness, we now recall the procedure. 

The ABC-SMC procedure is composed of a sequence of $R > 1$ rounds. At each round, parameter vectors $\theta$ are sampled in a specific way, symbiont trees $\tilde{S}_\theta$ are generated under the reconciliation model allowing for spreads with parameter values given by $\theta$ (and relying on the simulation algorithm described in the previous section). Then, these symbiont trees are compared to the original dataset through a summary distance $d$ whose details are given in the next section.  The parameters with the smallest discrepancies  are selected.

For each of these rounds, we define a tolerance value ${\tau_r}$ ($1 \le r \le R$) which determines the percentage of parameter vectors to be accepted. Associated with a tolerance value $\tau_r$, we have a threshold $\epsilon_r$ which is the largest value of the summary distance associated with the accepted parameter vectors.

\begin{itemize}

	\item Initial round ($r=1$): 
	\begin{itemize}
		\item Draw an initial set of $N$ parameter vectors $\{\theta_1^i\}_{(1 \le i \le N)}$ from the prior $\pi$. 
		
		\item Then, for each $\theta_1^i$, simulate $M$ trees $\{\tilde{S}_j(\theta_1^i)\}_{(1 \le j \le M)}$. Compute the corresponding discrepancies $\{d_j(\theta_1^i)\}_{(1\le j \le M)}$ and summarise them into the summary discrepancy  $d_{\theta^i_1}$ through the mean value.

		\item Select $Q_1 = \tau_1 \times N$ parameter vectors $\theta_1$ that have the smallest value $d_{\theta_1}$, thus defining the threshold $\epsilon_1$ and the set $A_1$ of accepted parameter vectors.
	\end{itemize}
	\medskip
	\item Following rounds ($2 \le r \le R$):
	\begin{enumerate}
		\item Sample a parameter vector $\theta^\star$ from the set $A_{(r-1)}$. 
		
		\item Create a parameter vector $\theta^{\star\star}$ by perturbing $\theta^\star$ (through a kernel proposal). 
		
		\item Simulate $M$ trees relying on the parameter value $\theta^{\star\star}$ and compute $d_{\theta^{\star\star}}$. If $d_{\theta^{\star\star}} \le \epsilon_{(r-1)}$, add $\theta^{\star\star}$ into the quantile set $\mathcal{Q}_{r}$. If $ |\mathcal{Q}_{r}| < Q_{r-1}$, return to Step 1.
		
		\item Based on the set $\mathcal{Q}_{r}$, select $Q_r= \tau_r \times Q$ parameter vectors $\theta_r$ that have the smallest $d_{\theta_r}$, thus defining the threshold $\epsilon_r$ and the set $A_r$ of accepted parameters.
	\end{enumerate}

\end{itemize}

\paragraph{Prior distribution.} 
We sample from  a uniform distribution on the simplex $\mathcal{S}_3=\{(p_1,p_2,p_3,p_4) ; p_i \ge 0 \text{ and } \sum_i p_i =1\}$ (we recall that  $ p_{c}+p_{d}+p_{s}+ p_{l}=1$).

\paragraph{Kernel proposal.} We add to each coordinate of $\theta$ a randomly chosen value in $[-0.01, +0.01]$ and 
normalise the result.  The final set of accepted parameter vectors is the result of the ABC-SMC procedure and characterises the list of vectors that may explain the evolution of the pair of host and symbiont  trees given as input. Observe that, since in all experiments a uniform prior distribution is assumed and also the perturbations are performed in a uniform way, the weights induced by the proposals will also appear to be uniform \citep{Beaumont09}. However, in the case of a different prior, weights should be used in the process in order to correct the posterior distribution according to the perturbation made.

\paragraph{Clustering of the vectors.} The final list of accepted vectors are clustered using a hierarchical clustering procedure implemented in \Coala\ \citep{Baudet2015}. 
As final result, we therefore obtain a list of clusters to each one of which a representative vector is associated.

\subsection{Distance measure in \ACoala}
\label{sec:SM-dist}
The discrepancy between the simulated and the original datasets is measured through a distance  between set-labelled  phylogenetic trees which can be calculated in polynomial time. 
Similarly as in \Coala, this distance contains two components: (i) $d_1$, that describes how much the simulated tree  $\tS_{\theta}$ is representative  of the vector  $\theta$, and (ii) $d_2$ 
that measures how much  is $\tS_{\theta}$ (and its labels) topologically similar to $S$ (and its labels).

Let us recall the definition of this first component. 
For a given vector $\theta=\langle p_c, p_d, p_s, p_l\rangle$ and for each simulated tree $\tS_\theta$ that was simulated according to this vector, we 
keep track of the vector of the number of classical cophylogeny events $\langle o_c,o_d,o_s,o_l\rangle$ associated to this simulation. We compute the corresponding expected vector
$\langle e_c,e_d,e_s,e_l\rangle$ as follows
\[ \forall \text{event} \in \{c,d,s,l\}, \quad e_{\text{event}} = |S|\times \theta_{\text{event}} =  |S|\times p_{\text{event}}, 
\] 
where $|S|$ is the size of the symbiont tree, \emph{i.e.} its number of internal leaves. 
Then by comparing the observed and expected vectors, we define a measure $d_1(S,\tS_\theta)$ as follows:
\[
d_1(S,\tS_\theta)= \frac{1}{4} \times \sum_{\text{event} \in \{c,d,s,l\}} {\frac{|e_{\text{event}}-o_{\text{event}}|}{\max\{e_{\text{event}},o_{\text{event}}\}}}.
\]
Note that we did not consider the number of observed spread events, which does not depend on the choice of $\theta$ as the corresponding probabilities are pre-estimated before applying the ABC-SMC approach.

As concerns point (ii), we extend the well-known \emph{maximum agreement subtree} (MAST) 
distance \citep{Finden1985,Fiarach-Coton1995} 
to handle set-labelled trees. This part is the novelty with respect to the proposal in \Coala\  and details were given in the Main Manuscript. We establish in the next sections that $d_{MASST}$ is a distance and that 
it can be computed in polynomial time. 

We use a normalised version of $d_{MASST}$ and define the distance $d_2$ (see Main Manuscript).
The two components are then combined to form the following distance  
\[
d_\theta= \alpha_1 d_1(S,\tS_\theta) + \alpha_2 d_2(S,\tS_\theta).
\]
According to our experiments and also the ones presented in \Coala, the most appropriate values are $\alpha_1 = 0.7$ and $\alpha_2 = 0.3$.

\subsection{A proof that 
$d_{MASST}$ is a distance}
We show that the distance $d_{MASST}$ is a metric. For this, we check that $d_{MASST}$ satisfies the following properties:
\begin{enumerate}
\item $d_{MASST}(T_1,T_2)\geq 0$ for all $T_1$,$T_2$: this is trivial. 
\item $d_{MASST}(T_1,T_2) = 0$ if and only if $T_1 = T_2$. Clearly if $T_1 = T_2$ then $d_{MASST}(T_1,T_2) = 0$. Otherwise, let $d_{MASST}(T_1,T_2) = 0$. Then $\max\{w(T_1),w(T_2)\} = MASST(T_1,T_2)$. The proof follows by observing that  if $T^*$ is a subtree of $T$ such that $w(T^*)=w(T)$ then $T^*=T$.
\item $d_{MASST}(T_1,T_2) = d_{MASST}(T_2,T_1)$: this is trivial.
\item For any triplet of trees $T_1$, $T_2$, $T_3$, it holds that $d_{MASST}(T_1,T_2) + d_{MASST}(T_2,T_3) \geq d_{MASST}(T_1,T_3)$. For simplicity, we set $w_i = w(T_i)$ and $w_{i,j}=w(MASST(T_i,T_j))$. Hence $d_{MASST}(T_i,T_j) = \max\{w_i,w_j\}-w_{i,j}$. Furthermore, we denote by $w_{1,2,3}$ the  weight of the maximum agreement subtree that is common to the three trees $T_1$, $T_2$, $T_3$. We then have:
\begin{align*}
 \label{eq:proof3}
& d_{MASST}(T_1,T_2) + d_{MASST}(T_2,T_3) \\
 &= \max\{w_1, w_2\} - w_{1,2} + \max\{w_2, w_3\} - w_{2,3} \\ 
 &= \max\{w_1, w_2\} + \max\{w_2, w_3\} - (w_{1,2} + w_{2,3} - w_{1,2,3} + w_{1,2,3}) \\  &\geq   \max\{w_1,w_2, w_3\} + w_2 - (w_2 + w_{1,2,3}) \\ &\geq \max\{w_1, w_3\} - w_{1,3},
\end{align*}
where for the first inequality, we use the 
fact that $\max\{w_1,w_2\} + \max\{w_2,w_3\} \geq  \max\{w_1,w_2,w_3\}+w_2$ and we show in the next Lemma that $w_{1,2}+w_{2,3}-w_{1,2,3}$ is at most $w_2$. The last inequality uses $w_{1,2,3}\leq w_{1,3}$. 
\end{enumerate}
This concludes the proof. 

\begin{lemma}
For any three set-labelled trees $T_1, T_2, T_3$ (using the notation from the above proof) it holds that $w_{1,2}+w_{2,3}-w_{1,2,3}\leq w_2$.
\end{lemma}

\begin{proof}
Let $T_{1,2}$ and $T_{2,3}$ be  maximum agreement set-labelled subtrees (MASST) of $T_1, T_2$ and $T_2, T_3$, respectively. Consider any pair of leaf, label that belongs to $T_2$, \textit{i.e.} $(l,lab) \in T_2$. There are only four possibilities: (i)  $(l,lab) \in T_{1,2}$ and $(l,lab) \not\in T_{2,3}$ (we call these leaves of type $A$), (ii)  $(l,lab) \not\in T_{1,2}$ and $(l,lab) \in T_{2,3}$ (we call these leaves of type $B$), (iii)  $(l,lab) \in T_{1,2}$ and $(l,lab)\in T_{2,3}$ (we call these leaves of type $C$), (iv)  $(l,lab) \not\in T_{1,2}$ and $(l,lab) \not\in T_{2,3}$ (we call these leaves of type $D$).
Then we have
\begin{align*}
w_2 &= |A|+|B|+|C|+|D| \\
&= w_{12}-|C| + w_{23}-|C| +|C|+|D|\\
&= w_{12}+w_{23} -|C|+|D|.
\end{align*}
Or equivalently
\begin{equation}
w_{12}+w_{23} = w_2 +|C|-|D|.
\label{eq:weights}    
\end{equation}
Moreover, we define the tree $\tilde T$ as the subtree obtained from $T_2$ by taking all the pairs of leaf, label that belong to $T_{12}$ and $T_{23}$. Notice that $\tilde T$ is also a subtree of $T_1$ and of $T_3$. Thus, $\tilde T$ is included in $T_{123}$. This implies that $|C|\leq w_{123}$.
Going back to~\eqref{eq:weights}, we thus obtain
\begin{align*}
    w_{12}+w_{23} &= w_2 +|C|-|D|\\
    &\le w_2 +|C|\\
    &\le w_2+w_{123}.
\end{align*}
This concludes the proof of the lemma.
\end{proof}

\begin{figure}
    \centering
    \includegraphics[width=\textwidth]{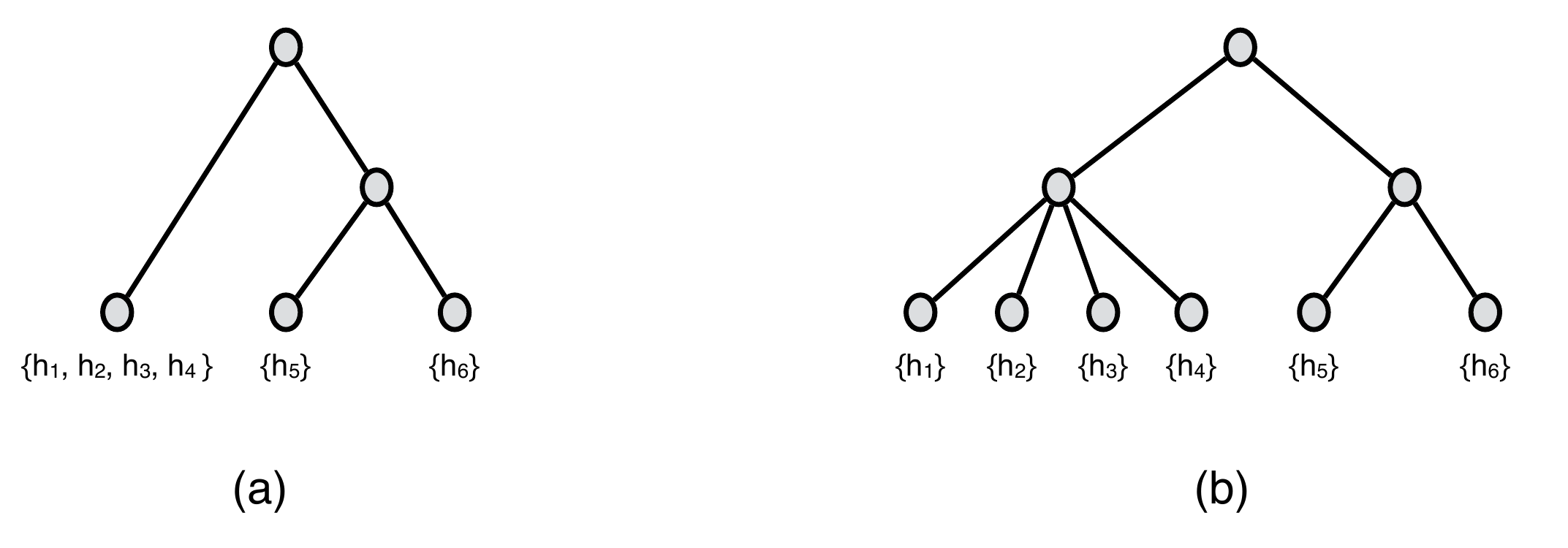}
    \caption{The two phylogenetic trees will be considered at distance 0 if we substitute the vertex labelled by the set ${h_1, h_2, h_3, h_4}$ by a multifurcated vertex.}
    \label{fig:MASST_not_MAAC}
\end{figure}

\begin{rem}
The previous proof and comments show that the MASST distance $d_{MASST}$ is very similar to the MAAC one \citep{GGJRW05} for multi-labelled trees. Thus, it is natural to ask whether comparing two set-labelled trees can be reduced to comparing two multi-labelled trees. One idea is to transform a set-labelled tree into a multi-labelled tree. However, the straightforward transformation seems not to work well for our purpose. For instance, we can transform each set-labelled tree into a multi-labelled tree by substituting each set-labelled leaf by a subtree with a fixed topology (say a complete binary tree, or a multifurcating vertex) as in Figure~\ref{fig:MASST_not_MAAC}. However, in these cases the two trees in Figure~\ref{fig:MASST_not_MAAC} would be considered equivalent, but in our context they are different. In fact, the set-labelled tree in Figure~\ref{fig:MASST_not_MAAC}(a) indicates that there is a symbiont that infects 4 different hosts ${h_1,h_2,h_3,h_4}$, while in Figure~\ref{fig:MASST_not_MAAC}(b), we will have 4 different symbionts infecting each a different host.
\end{rem}

\subsection{Polynomial time algorithm for computing the 
$d_{MASST}$ distance}
We show that it is possible to calculate the distance $d_{MASST}(T_1,T_2)$ in polynomial time with respect to the size of the trees. This boils down to computing the weight of the maximum agreement subtree $w(MASST(T_1,T_2))$ in polynomial time. The algorithm is based on dynamic programming and extends quite straightforwardly the algorithm for calculating the MAAC distance \citep{GGJRW05}. We abbreviate to  $w(v_1,v_2)$ the weight of the  maximum agreement subtree between the two trees $T_1$ and $T_2$ rooted in $v_1$ and $v_2$, respectively. For a leaf $v$, we denote by $l(v)$ the set of labels associated with it. Finally, for an internal vertex $v$, we denote by $ch_1(v)$ and $ch_2(v)$ the two children of $v$.

The dynamic programming algorithm starts from the leaves and ends in the roots of $T_1$ and $T_2$ following a recursion. We have that $w(v_1,v_2)$ is given by:
\begin{itemize}
\item If $v_1$ and $v_2$ are both leaves then $w(v_1,v_2) = | l(v_1) \cup l(v_2)| $
\item If $v_1$ 
or $v_2$ (could be both) are internal vertices, $w(v_1,v_2)$ is the maximum value among the following three quantities
\begin{enumerate}
\item $\max\{w(ch_1(v_1),v_2), w(ch_2(v_1),v_2)\}$ ; 
\item $\max\{w(v_1,ch_1(v_2)), w(v_1,ch_2(v_2)\}$ ; 
\item $\max\{w(ch_1(v_1),ch_1(v_2))+ w(ch_2(v_1),ch_2(v_2))$, $w(ch_1(v_1),ch_2(v_2))$ \\
$+ w(ch_2(v_1),ch_1(v_2))\}$. 
\end{enumerate}
\end{itemize}

\section{Additional results for the self-test}
\label{sec:SM-simus}

\begin{figure}[!ht]

\begin{minipage}{\textwidth}
\includegraphics[scale=0.27]{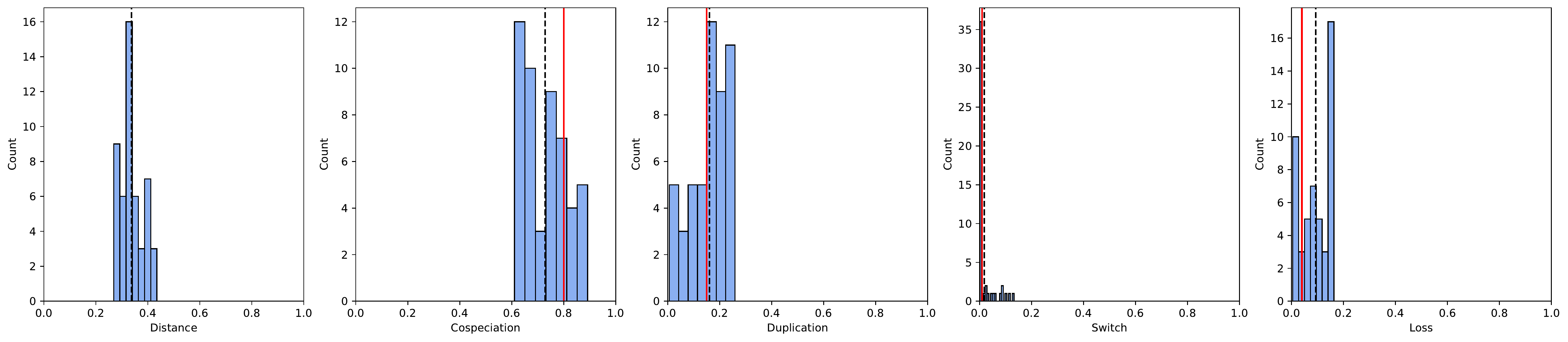} 
    \subcaption{\footnotesize Results for $\theta^\star_2 = \langle 0.80,0.15,0.01,0.04\rangle$.}
\label{fig:theta2}
\end{minipage}

\begin{minipage}{\textwidth}
\includegraphics[scale=0.27]{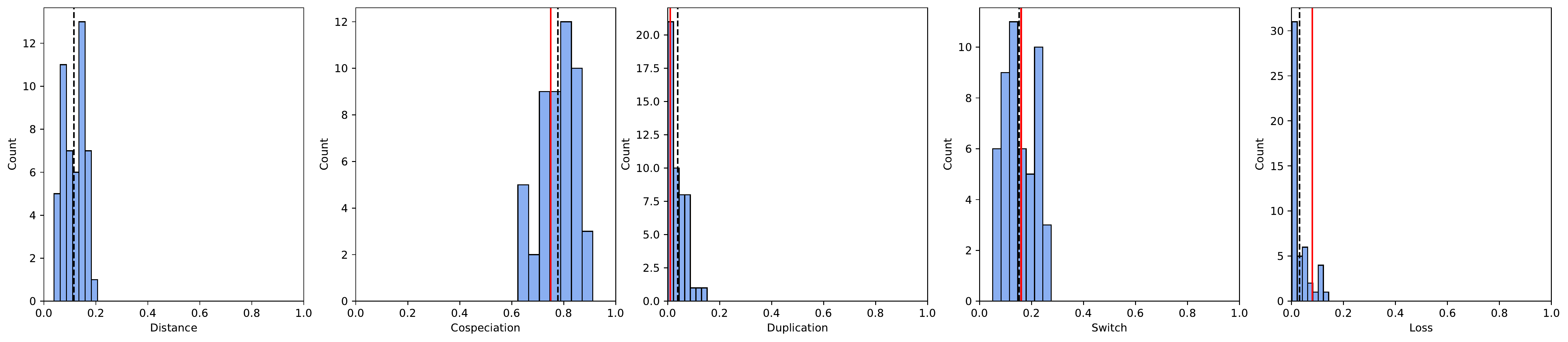} 
    \subcaption{\footnotesize Results for $\theta^\star_3 = \langle 0.75,0.01,0.16,0.08\rangle$}
\label{fig:theta3}
\end{minipage}

\begin{minipage}{\textwidth}
\includegraphics[scale=0.27]{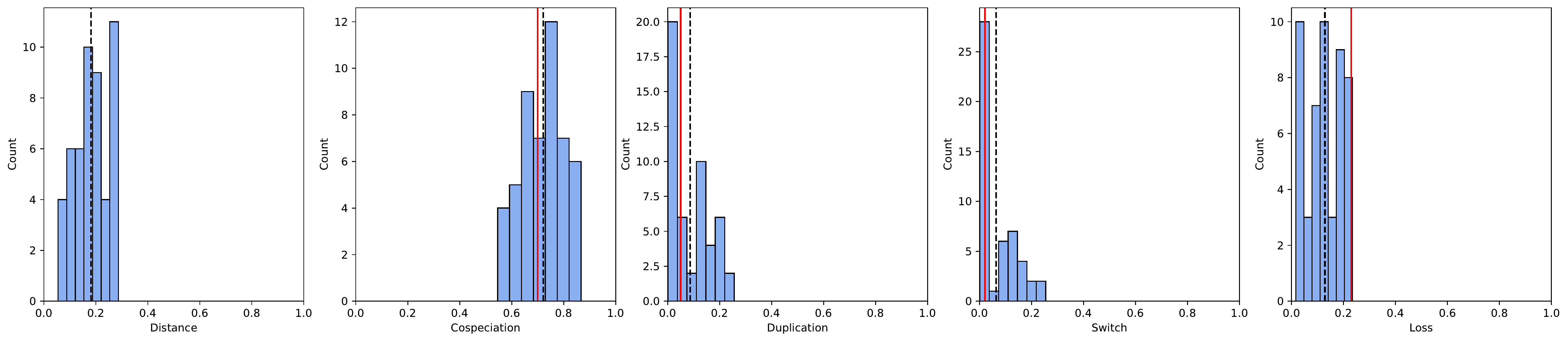} 
    \subcaption{\footnotesize Results for $\theta^\star_4 = \langle 0.70,0.05,0.02,0.23\rangle$.}
\label{fig:theta4}
\end{minipage}

\caption{For each simulated dataset with true parameter value $\theta^\star_i$ and $2\leq i \leq 8$, we ran \ACoala\ 50 times and, at the end of the third round, we took note of the cluster whose representative parameter vector had the smallest euclidean distance (histograms shown in the first column) to $\theta^\star_i$. Columns 2 to 5 show the histograms of the distributions of the event probabilities in these ``best'' clusters. The dashed vertical black line indicates the mean value. The solid vertical red line indicates the true parameter value. }
\end{figure}

\begin{figure}[!ht]
\begin{minipage}{\textwidth}
\includegraphics[scale=0.27]{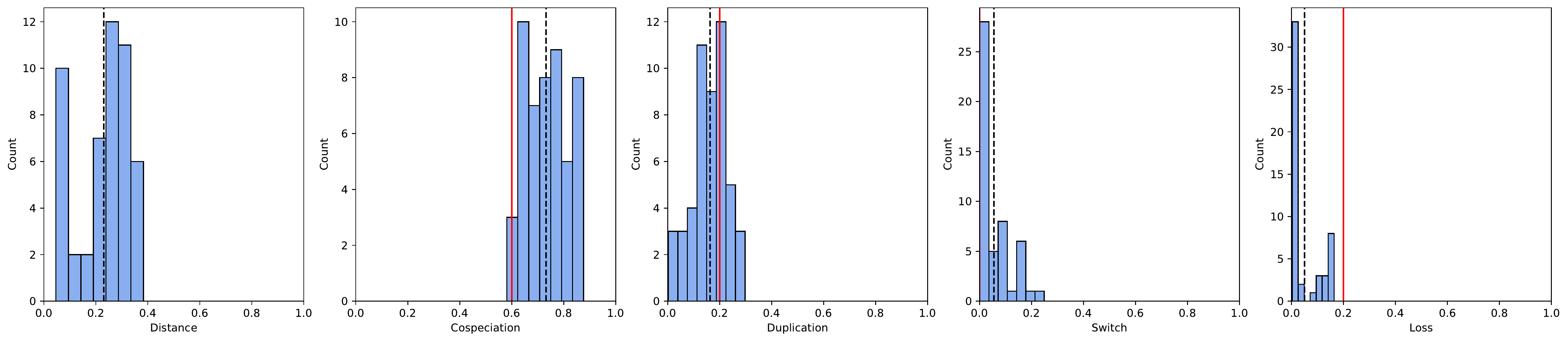} 
    \subcaption{\footnotesize Results for $\theta^\star_5 = \langle 0.60,0.20,0.00,0.20\rangle$}
\label{fig:theta5}
\end{minipage}

\begin{minipage}{\textwidth}
\includegraphics[scale=0.27]{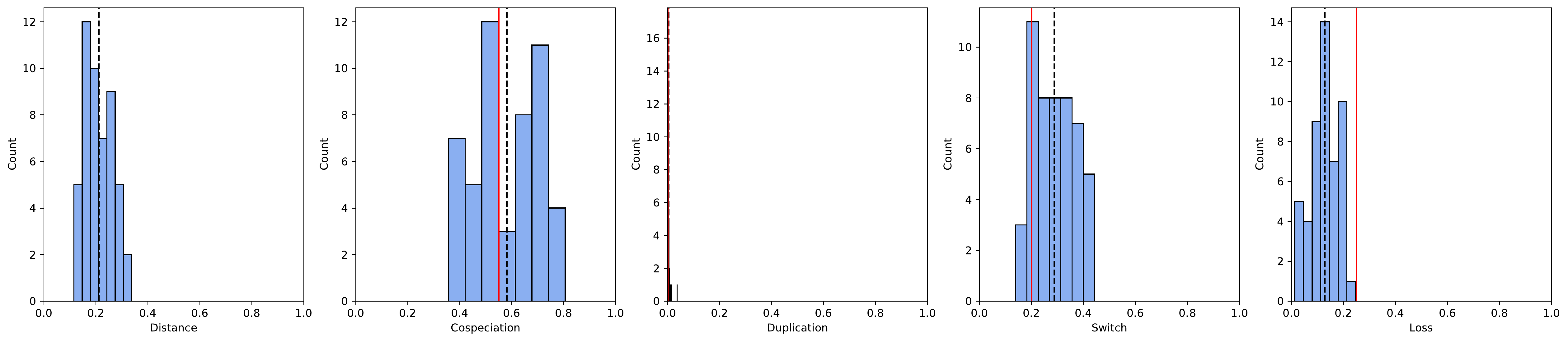} 
    \subcaption{\footnotesize Results for $\theta^\star_6 = \langle 0.55,0.00,0.20,0.25\rangle$.}
\label{fig:theta6}
\end{minipage}

\begin{minipage}{\textwidth}
\includegraphics[scale=0.27]{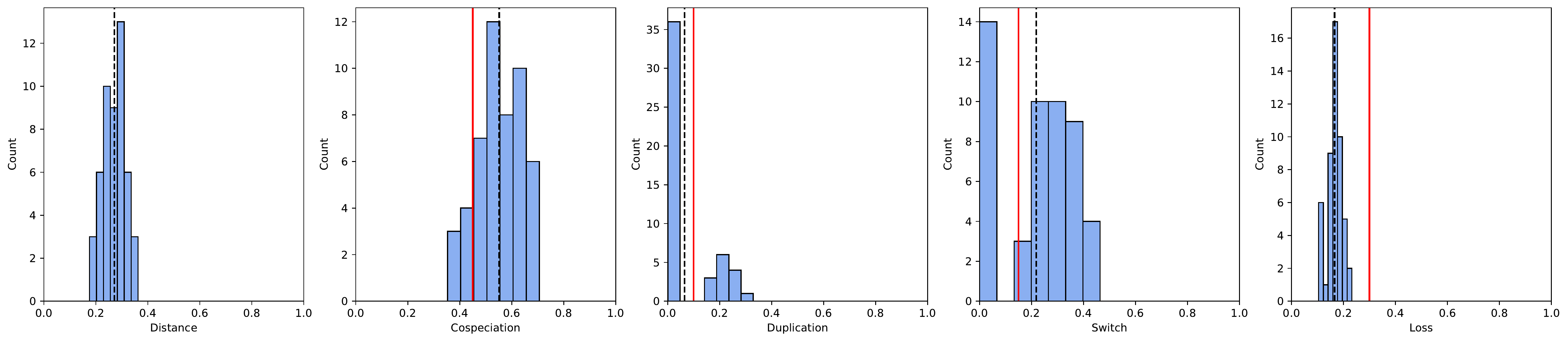} 
    \subcaption{\footnotesize Results for $\theta^\star_7 = \langle 0.45,0.10,0.15,0.30\rangle$}
\label{fig:theta7}
\end{minipage}

\begin{minipage}{\textwidth}
\includegraphics[scale=0.27]{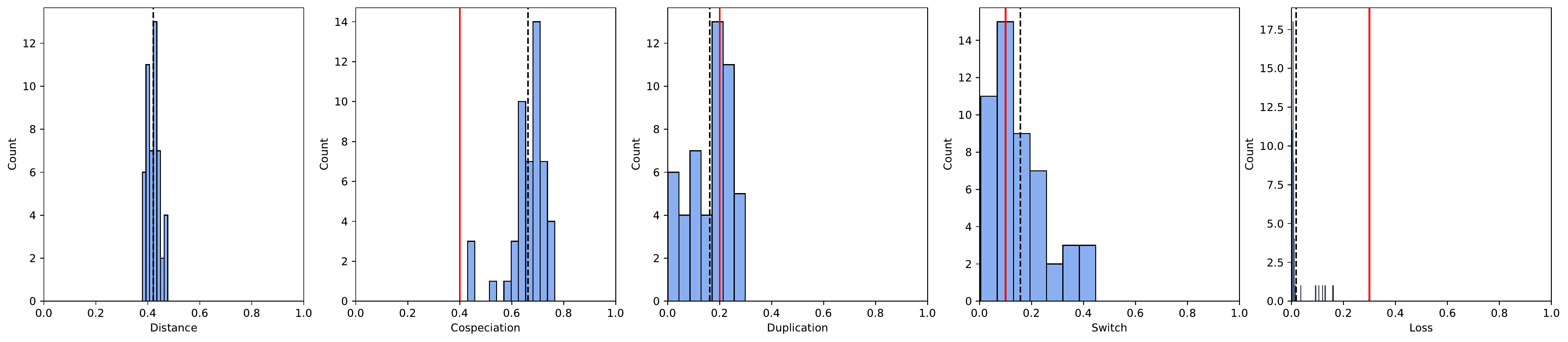} 
    \subcaption{\footnotesize Results for $\theta^\star_8 = \langle 0.40,0.20,0.10,0.30\rangle$.}
\label{fig:theta8}
\end{minipage}

\caption{For each simulated dataset with true parameter value $\theta^\star_i$, we ran \ACoala\ 50 times and, at the end of the third round, we took note of the cluster whose representative parameter vector had the smallest euclidean distance (histograms shown in the first column) to $\theta^\star_i$. Columns 2 to 5 show the histograms of the distributions of the event probabilities in these ``best'' clusters. The dashed vertical black line indicates the mean value. The solid vertical red line indicates the true parameter value.}
\end{figure}

The results for parameter values $\theta^\star_2$ to $\theta^\star_8$ are presented in Figures~\ref{fig:theta2} to~\ref{fig:theta8}. 


\section{Biological datasets}
\begin{landscape}
\begin{figure}
\includegraphics[scale=0.45]{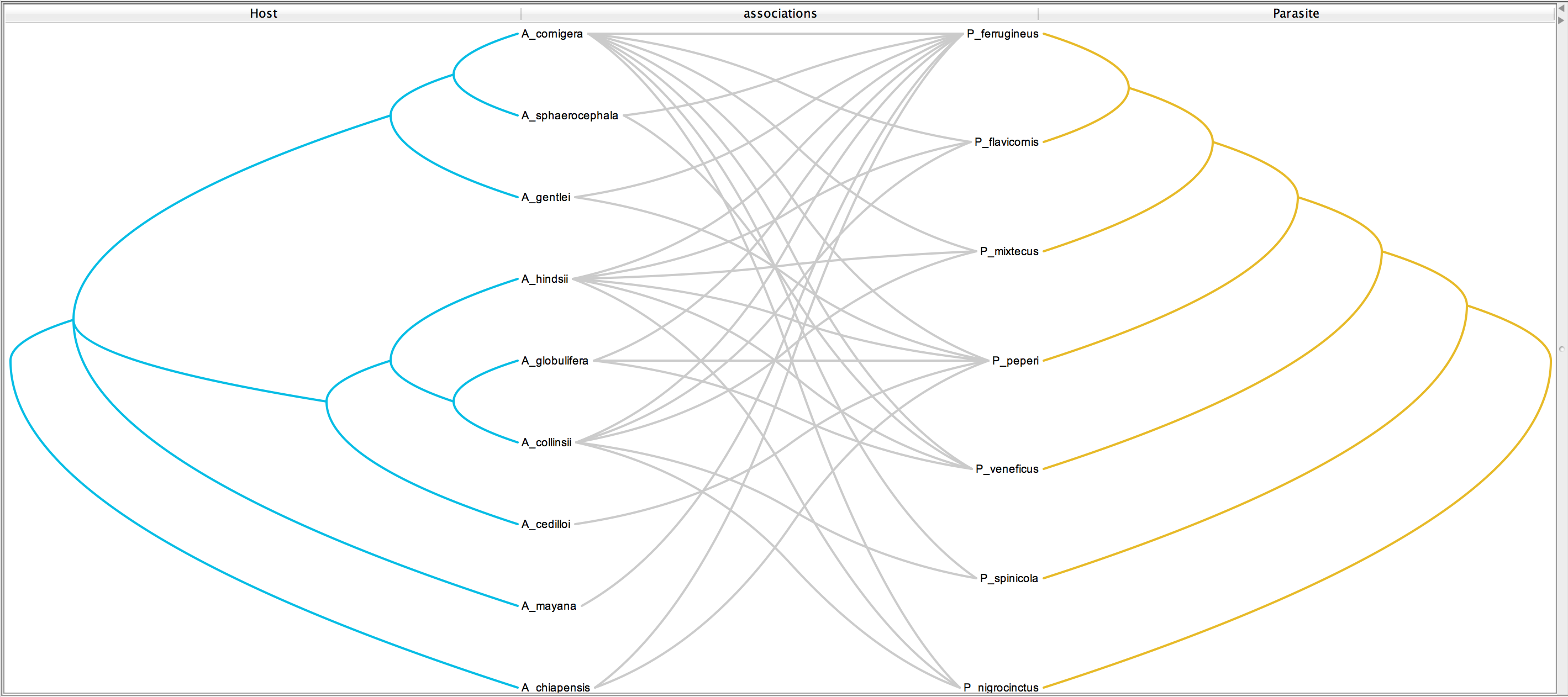}
\caption{AP dataset.}
\label{fig:AP_tree}
\end{figure}
\end{landscape}

\begin{landscape}
\begin{figure}
\includegraphics[scale=0.45]{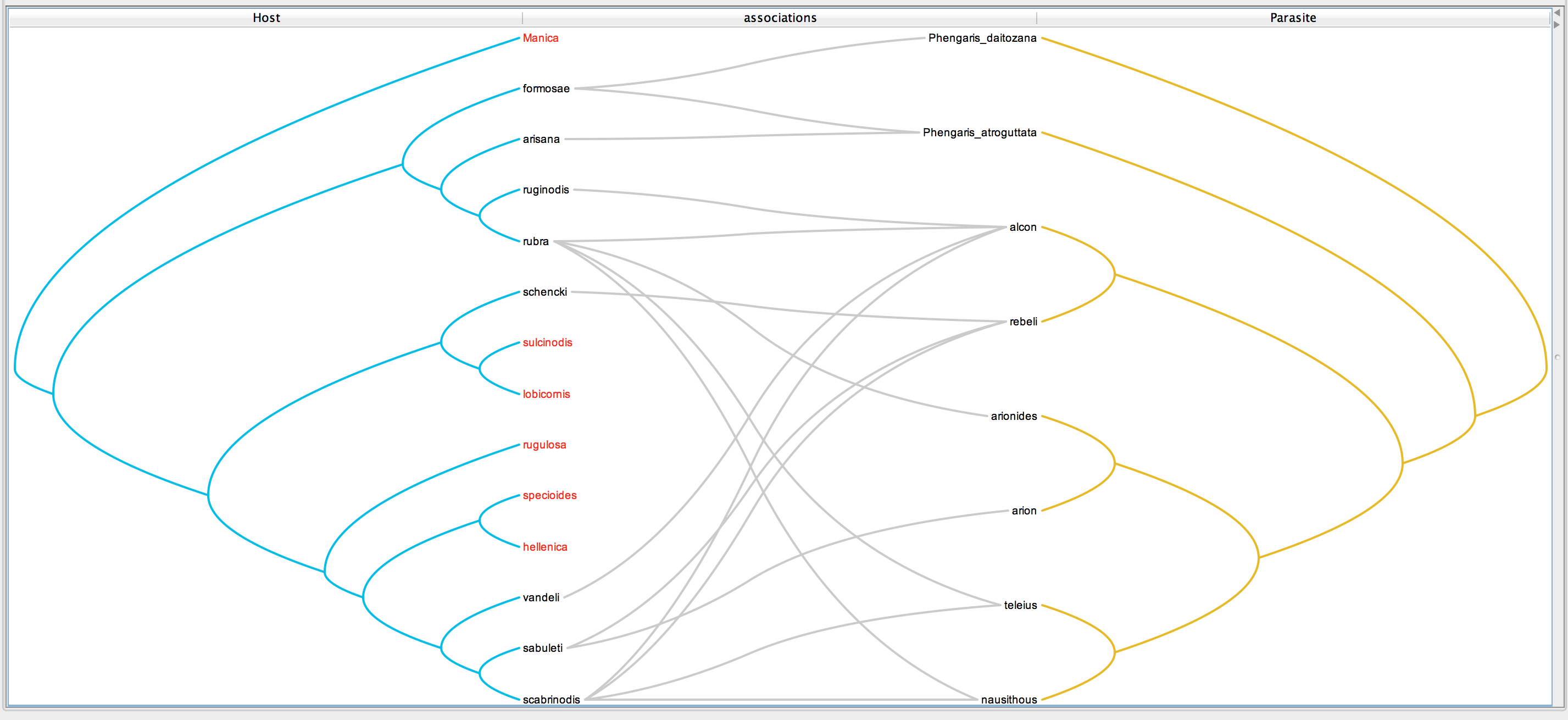}
\caption{MP dataset.}
\label{fig:MP_tree}
\end{figure}
\end{landscape}

\begin{landscape}
\begin{figure}
\includegraphics[scale=0.45]{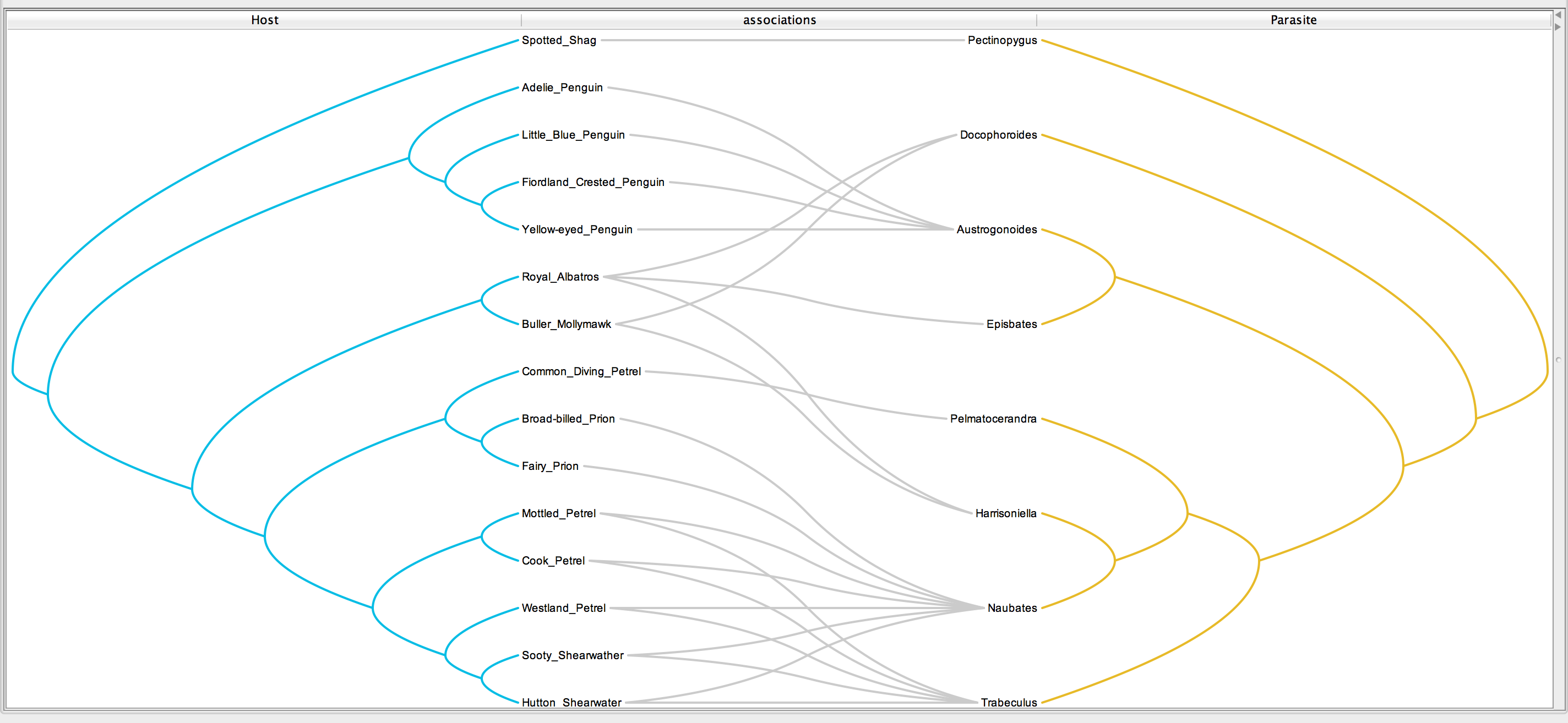}
\caption{SBL dataset.}
\label{fig:SBL_tree}
\end{figure}
\end{landscape}

\begin{landscape}
\begin{figure}
\includegraphics[scale=0.45]{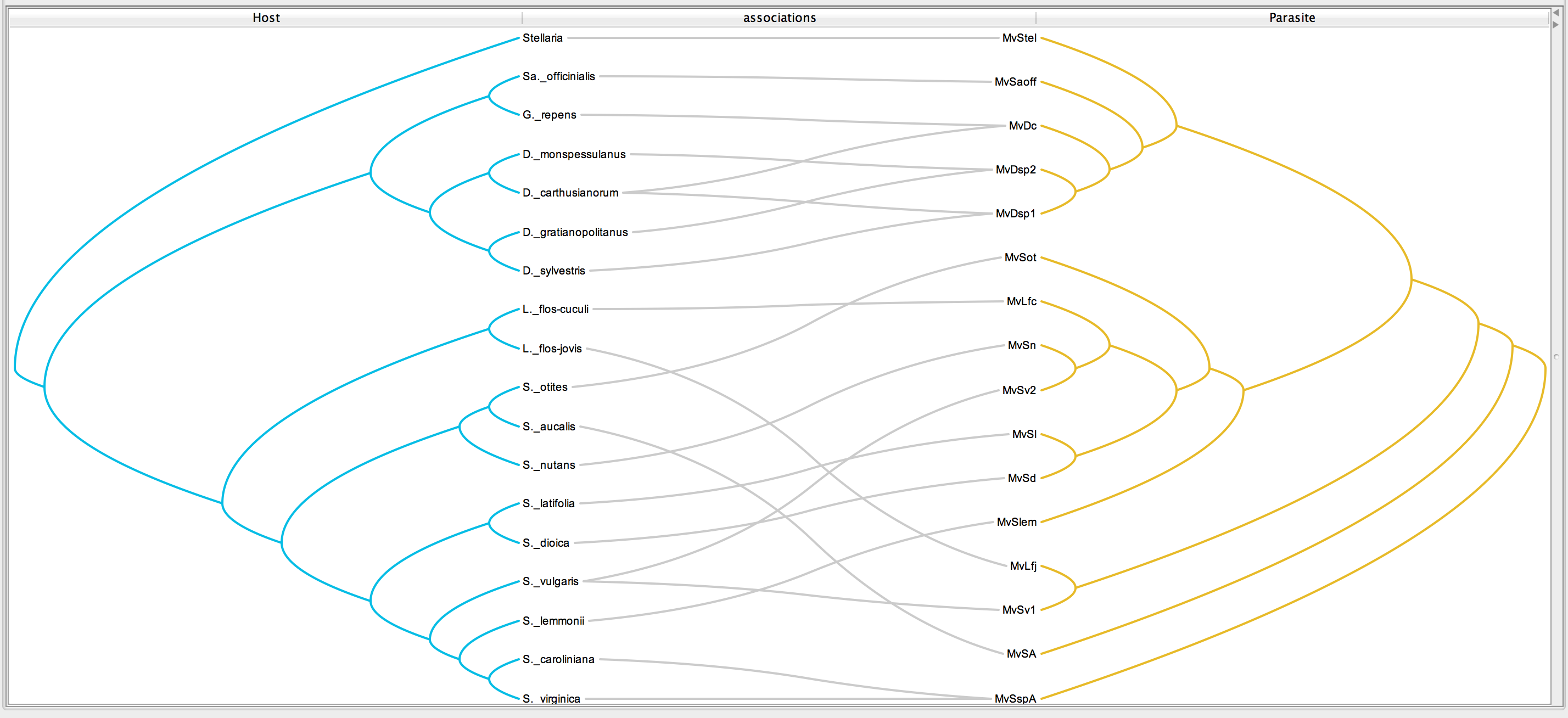}
\caption{SFC dataset.}
\label{fig:SFC_tree}
\end{figure}
\end{landscape}

We provide here a description of the 4 datasets used. The corresponding phylogenetic trees are shown in Figures~\ref{fig:AP_tree} - \ref{fig:SFC_tree}.

\emph{Dataset $1$: AP - \textit{Acacia} \& \textit{Pseudomyrmex}}. This dataset was extracted from 
\cite{gomez2010neotropical} and displays the interaction between \textit{Acacia} plants and \textit{Pseudomyrmex} species of ants. The host and symbiont trees include 9 and 7 leaves, respectively. The dataset has 22 multiple-associations. 

\emph{Dataset $2$: MP - Myrmica \& Phengaris}. This dataset was extracted from \cite{MP2011} and is composed of a pair of host and symbiont trees which have each 8 leaves. The dataset has 8 multiple-associations.

\emph{Dataset $3$: SBL - Seabirds \& Lice}. This dataset was extracted from \cite{SBL1997}. The host and symbiont trees include 15 and 8 leaves, respectively. The dataset has 15 multiple-associations.

\emph{Dataset $4$: SFC - Smut Fungi \& Caryophillaceus plants}. This dataset was extracted from \cite{RGJ2008}. The host and symbiont trees include 15 and 16 leaves, respectively. The dataset has 4 multiple-associations.

\subsection{Results on biological datasets}
\label{sec:SM-real_data}
\begin{landscape}
\begin{figure}
\includegraphics[scale=0.65]{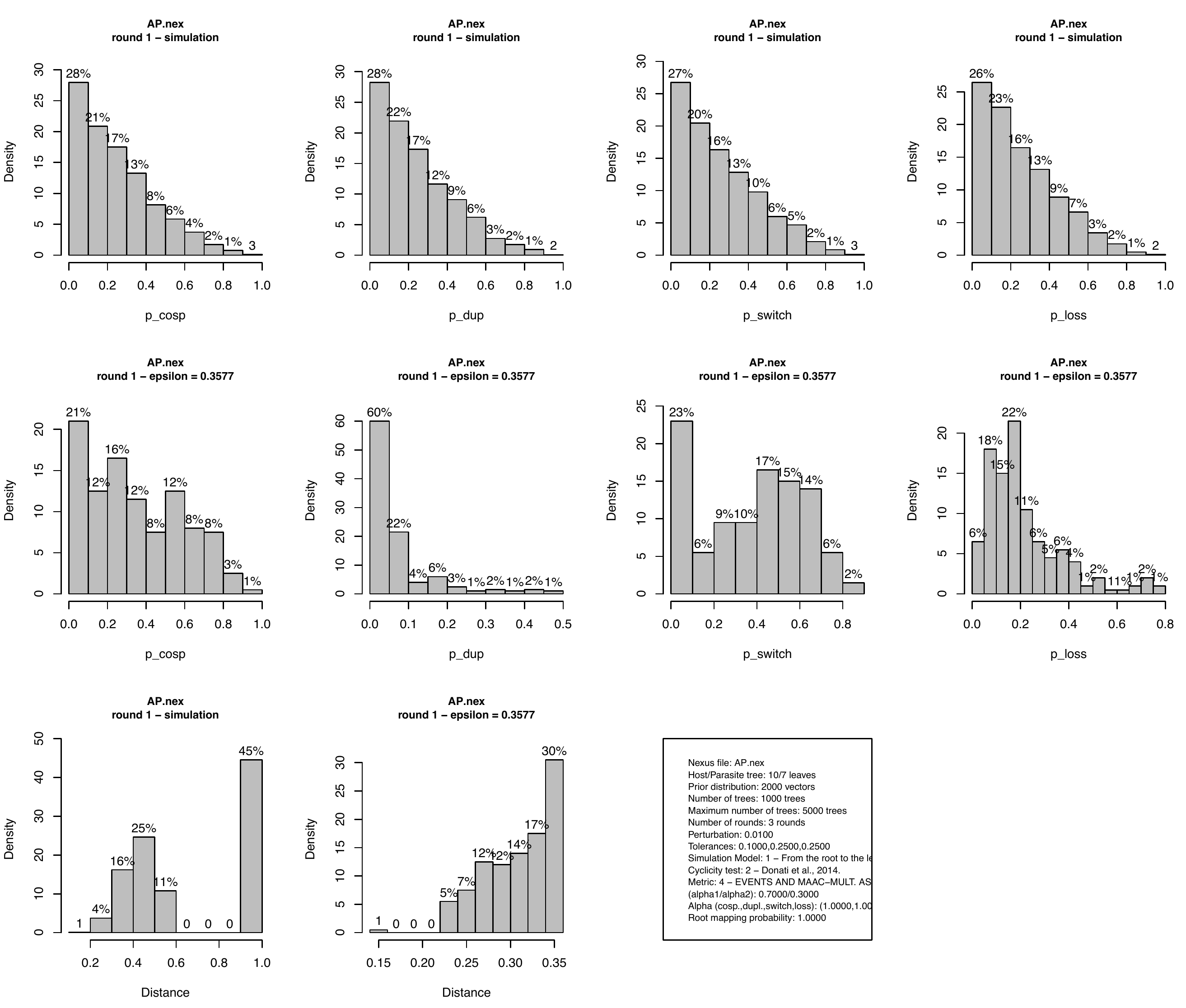}
\caption{AP dataset. First row: histograms of 
the input parameters. Second row: histograms of 
the parameters after round 1. Third row: summary discrepancies of 
the input parameters and of 
the parameters after round 1.}
\label{fig:AP_R1}
\end{figure}
\end{landscape}

\begin{landscape}
\begin{figure}
\includegraphics[scale=0.65]{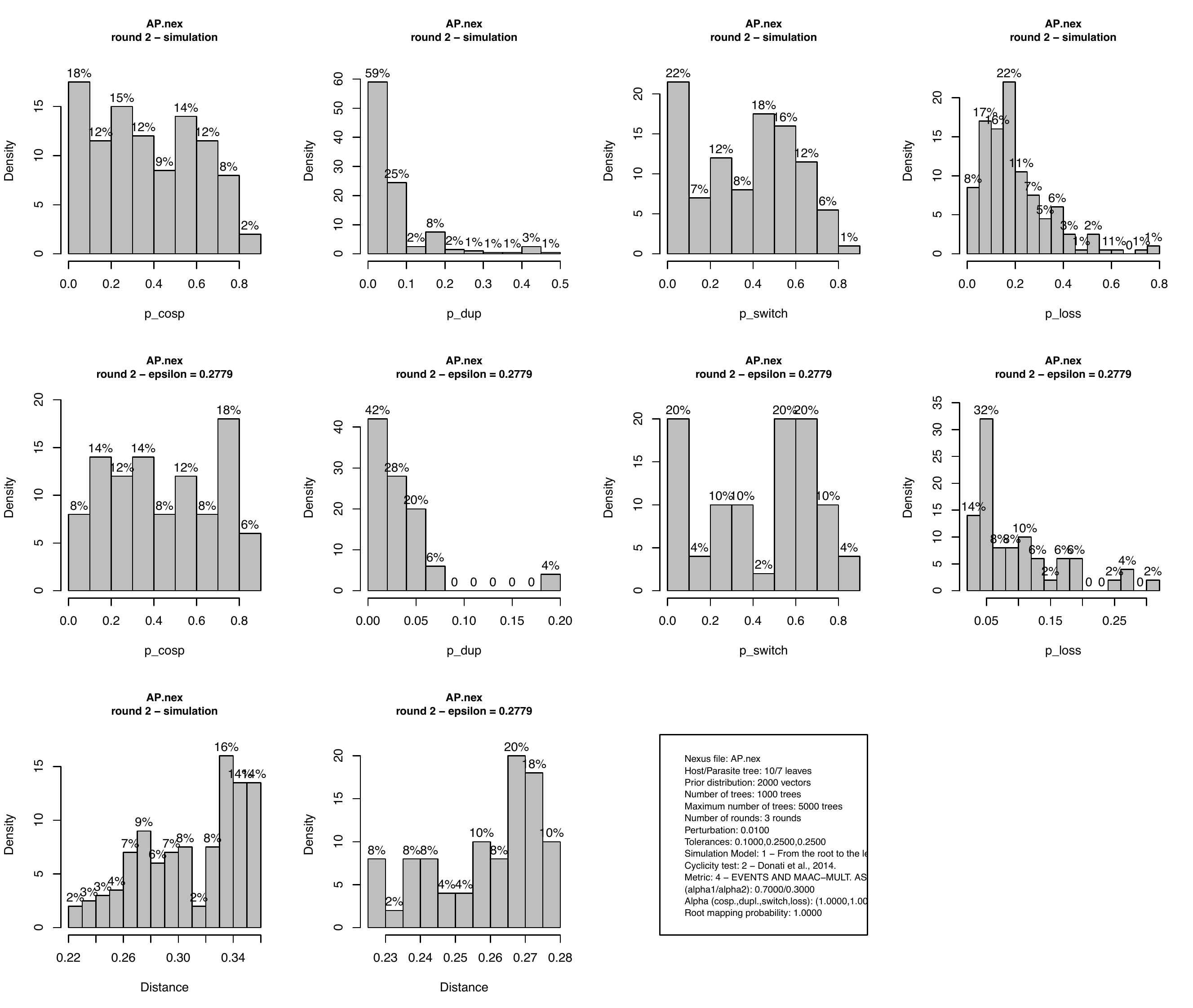}
\caption{AP dataset. First row: histograms of 
the input parameters. Second row: histograms of 
the parameters after round 2. Third row: summary discrepancies of 
the input parameters and of 
the parameters after round 2.}
\label{fig:AP_R2}
\end{figure}
\end{landscape}

\begin{landscape}
\begin{figure}
\includegraphics[scale=0.65]{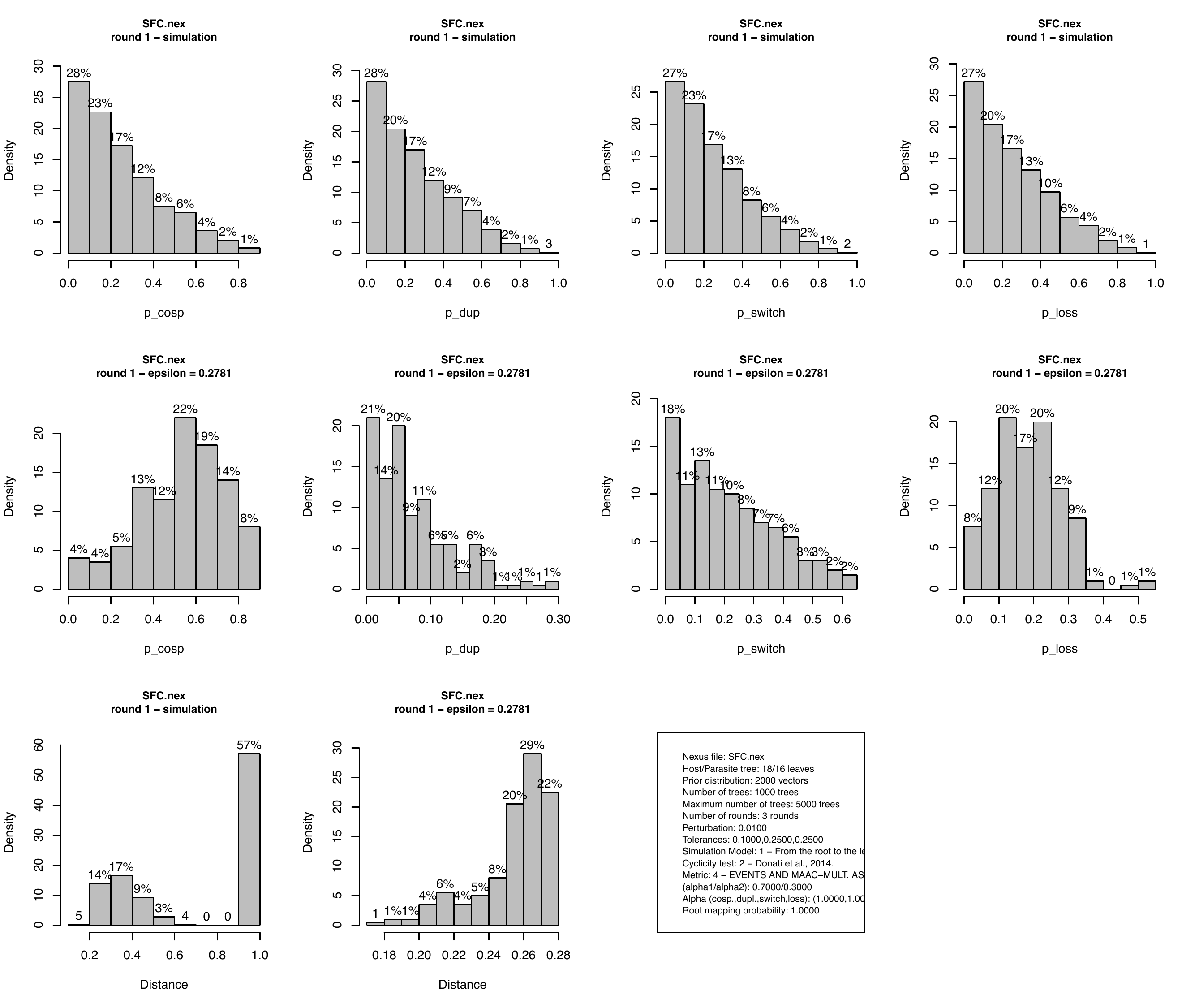}
\caption{SFC dataset. First row: histograms of 
the input parameters. Second row: histograms of 
the parameters after round 1. Third row: summary discrepancies of 
the input parameters and of
the parameters after round 1.}
\label{fig:SFC_R1}
\end{figure}
\end{landscape}

\begin{landscape}
\begin{figure}
\includegraphics[scale=0.65]{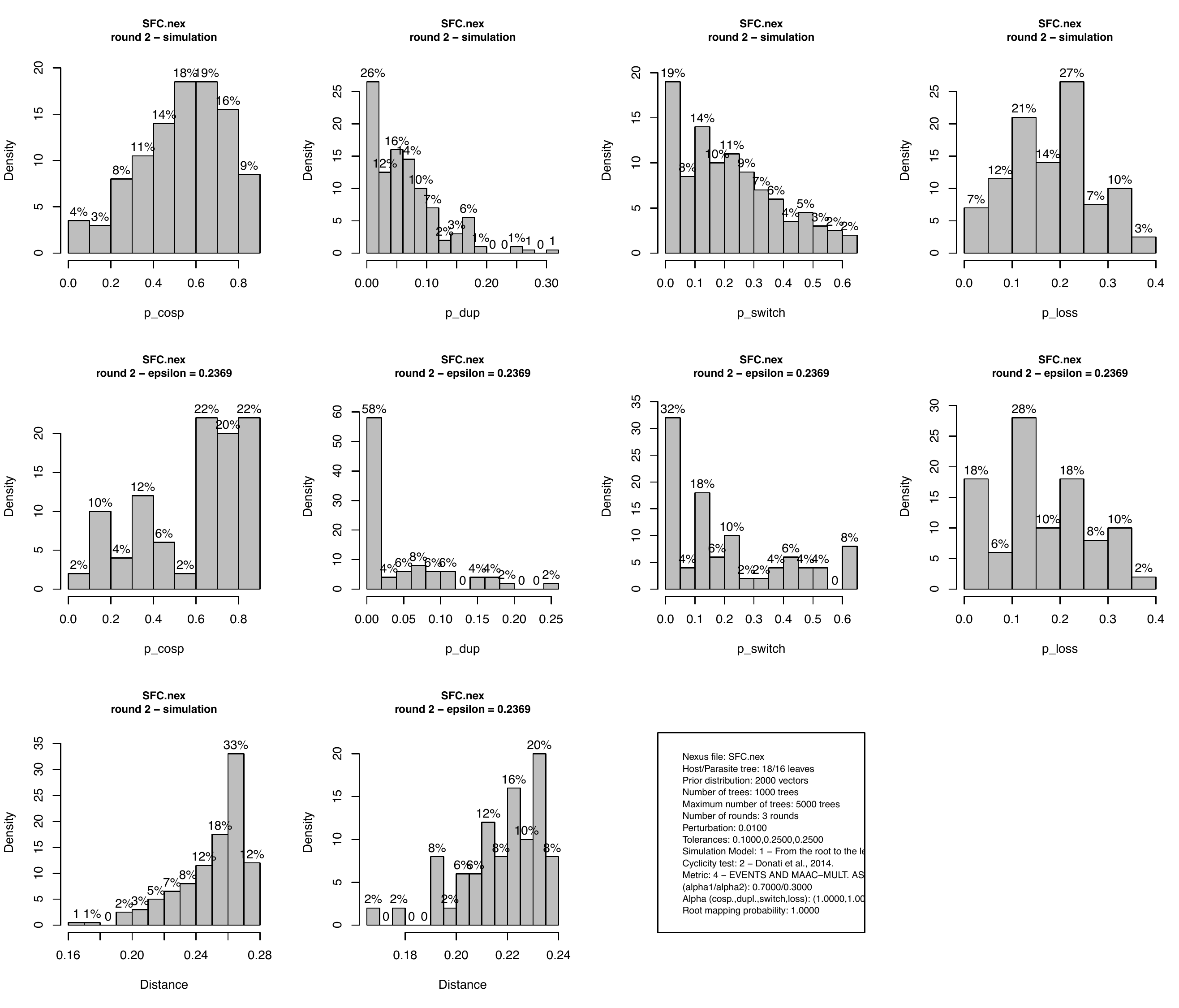}
\caption{SFC dataset. First row: histograms of 
the input parameters. Second row: histograms of
the parameters after round 2. Third row: summary discrepancies of 
the input parameters and of 
the parameters after round 2.}
\label{fig:SFC_R2}
\end{figure}
\end{landscape}

We ran \ACoala\ on all the real datasets and plotted in Figures~\ref{fig:AP_R1} to~\ref{fig:SFC_R3} the histograms of the summary  discrepancies and event probabilities (except for 
the spread probabilities which are not inferred) obtained at the end of each one of the 3 rounds, for each of the 4 datasets. 
We see on the histograms that the summary  discrepancies for the accepted parameter vectors decrease after each round. We recall that the summary discrepancy measures the similarity between the simulated trees and the original symbiont tree, and hence is related to the quality of the vectors. Thus, our result shows that the set of accepted vectors is refined at each round, leading to vectors which can generate trees that are increasingly more similar to the original symbiont tree (and its host associations).

\subsection{Running times}
\label{sec:SM-time}
Table~\ref{tab:time} shows the running times obtained on the 4 biological datasets, together with their sizes (as expressed by the number of leaves in the host and symbiont trees) and the number of multiple associations. The results have been obtained on a computer with a  
AMD EPYC 7542 32-Core processor and 128 CPU (2 sockets of 32 double threads cores) and 675Gb RAM. We used just one core ('nthreads 1', though \ACoala\ has a parallelized version) and  \ACoala\ was run with default values on these datasets.

We also performed an artificial experiment on a host tree with 204 leaves, a symbiont tree with 128 leaves, and six multiple associations. Relying on the above machine and using now 60 threads (which might not have been fully used during the entire computation), the running  time of \ACoala\ (used with default options except for the number of initial vectors $N$ that was set to 1000) was approximately 27.5 hours.

\begin{table}[h]
    \centering
    \begin{tabular}{c|c|c|c}
    Dataset     & (Host,Symbiont) leaves & Multiple associations & Running time \\
    \hline
      AP   &(9,7) & 22 & 23m20.859s\\
      MP & (8,8) & 8 & 21m25.631s\\
      SBL & (15,8)&15 & 28m53.597s\\
      SFC & (15,16) & 4& 117m45.919s 
    \end{tabular}
    \caption{For each of the 4 biological datasets, we indicate the pairs of numbers of host and symbiont trees leaves (2nd column), the number of multiple associations (3rd column) and the running time of \ACoala\ on this dataset (4th column).}
    \label{tab:time}
\end{table}

\subsection{Robustness analysis wrt the pre-estimated spread probabilities}
\label{sec:SM-robust}
\begin{landscape}
\begin{figure}
\includegraphics[scale=0.65]{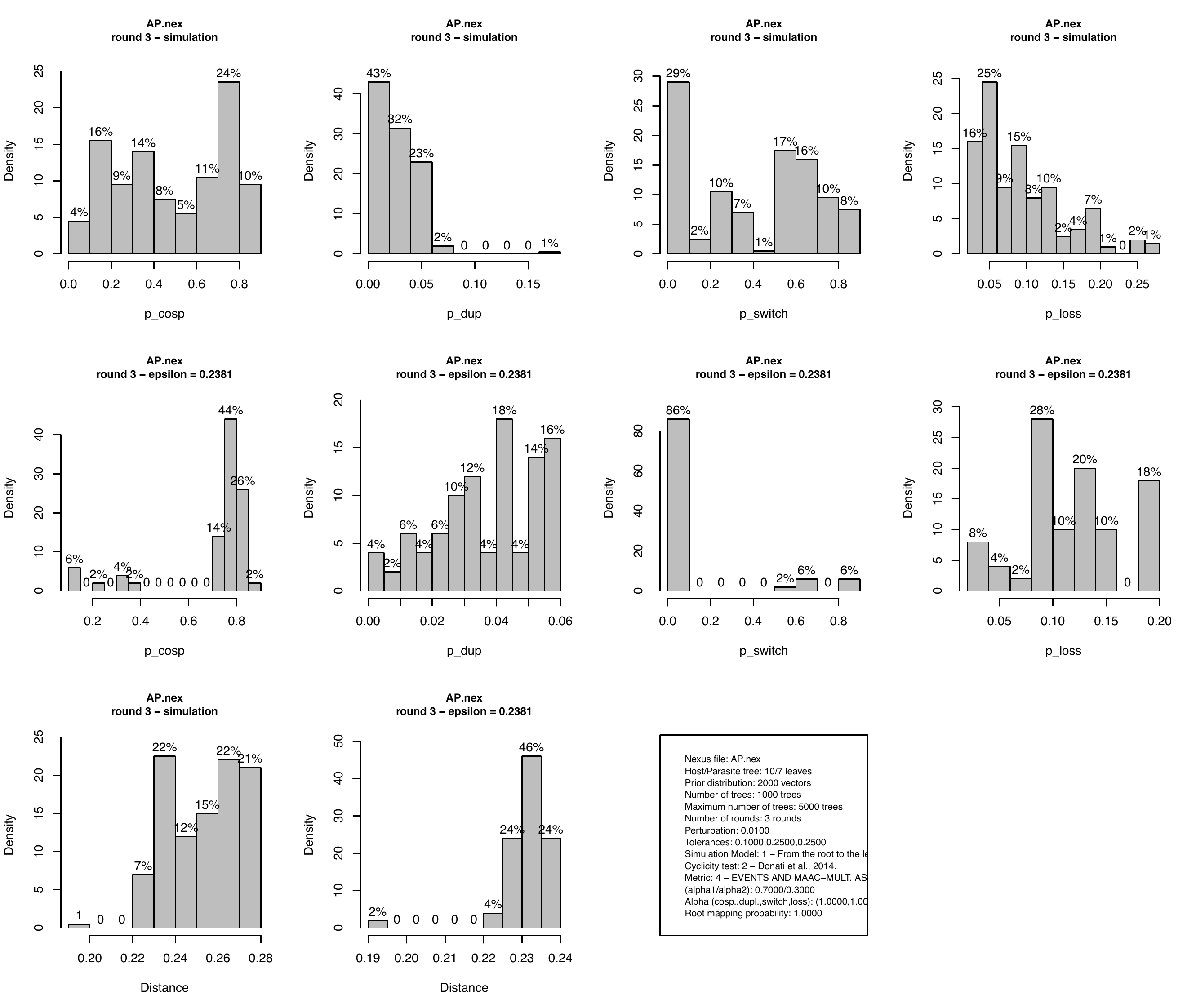}
\caption{AP dataset. First row: histograms of 
the input parameters. Second row: histograms of 
the parameters after round 3. Third row: summary discrepancies of 
the input parameters and of 
the parameters after round 3.}
\label{fig:AP_R3}
\end{figure}
\end{landscape}

\begin{landscape}
\begin{figure}
\includegraphics[scale=0.65]{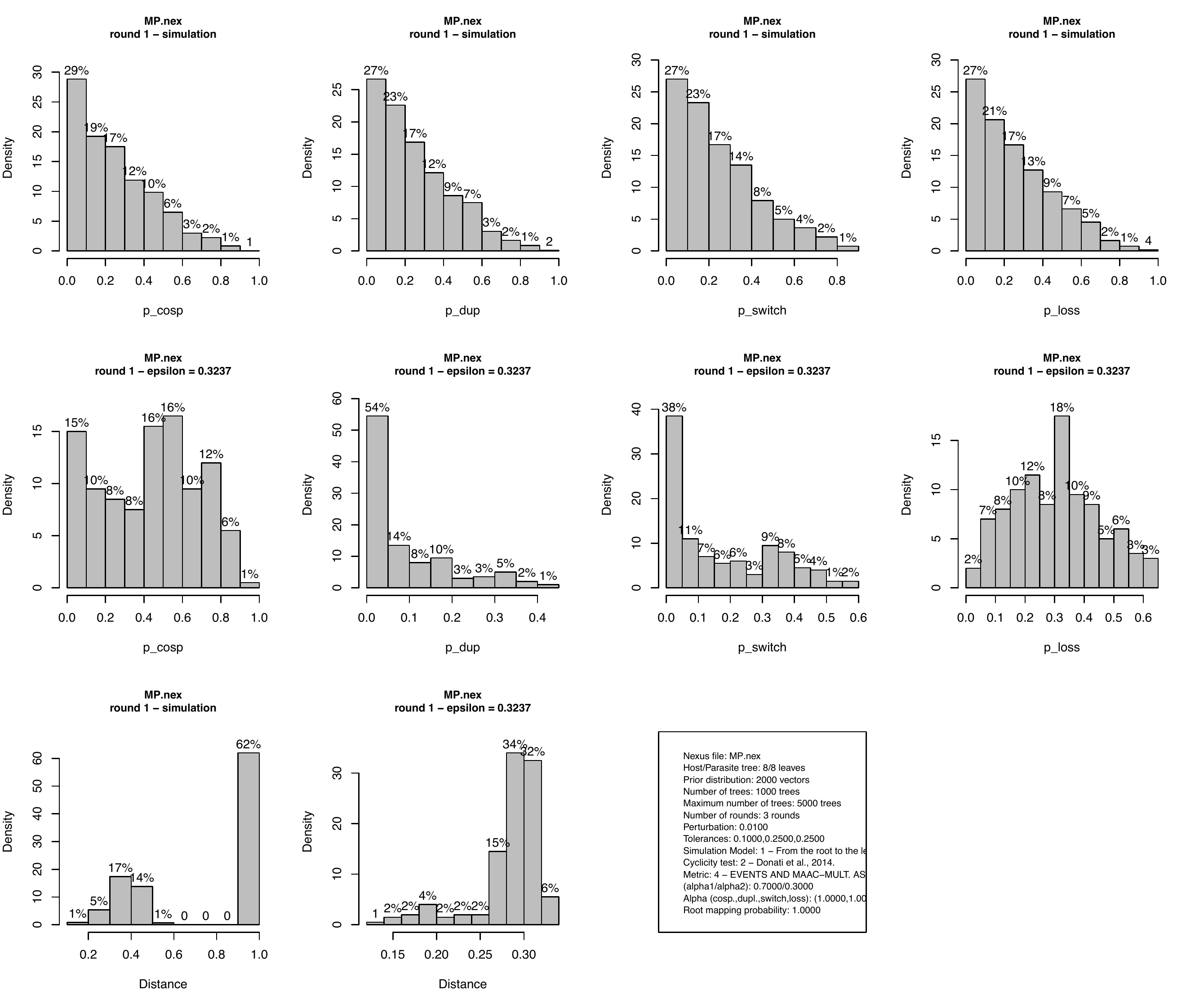}
\caption{MP dataset. First row: histograms of 
the input parameters. Second row: histograms of 
the parameters after round 1. Third row: summary discrepancies of 
the input parameters and of 
the parameters after round 1.}
\label{fig:MP_R1}
\end{figure}
\end{landscape}

\begin{landscape}
\begin{figure}
\includegraphics[scale=0.65]{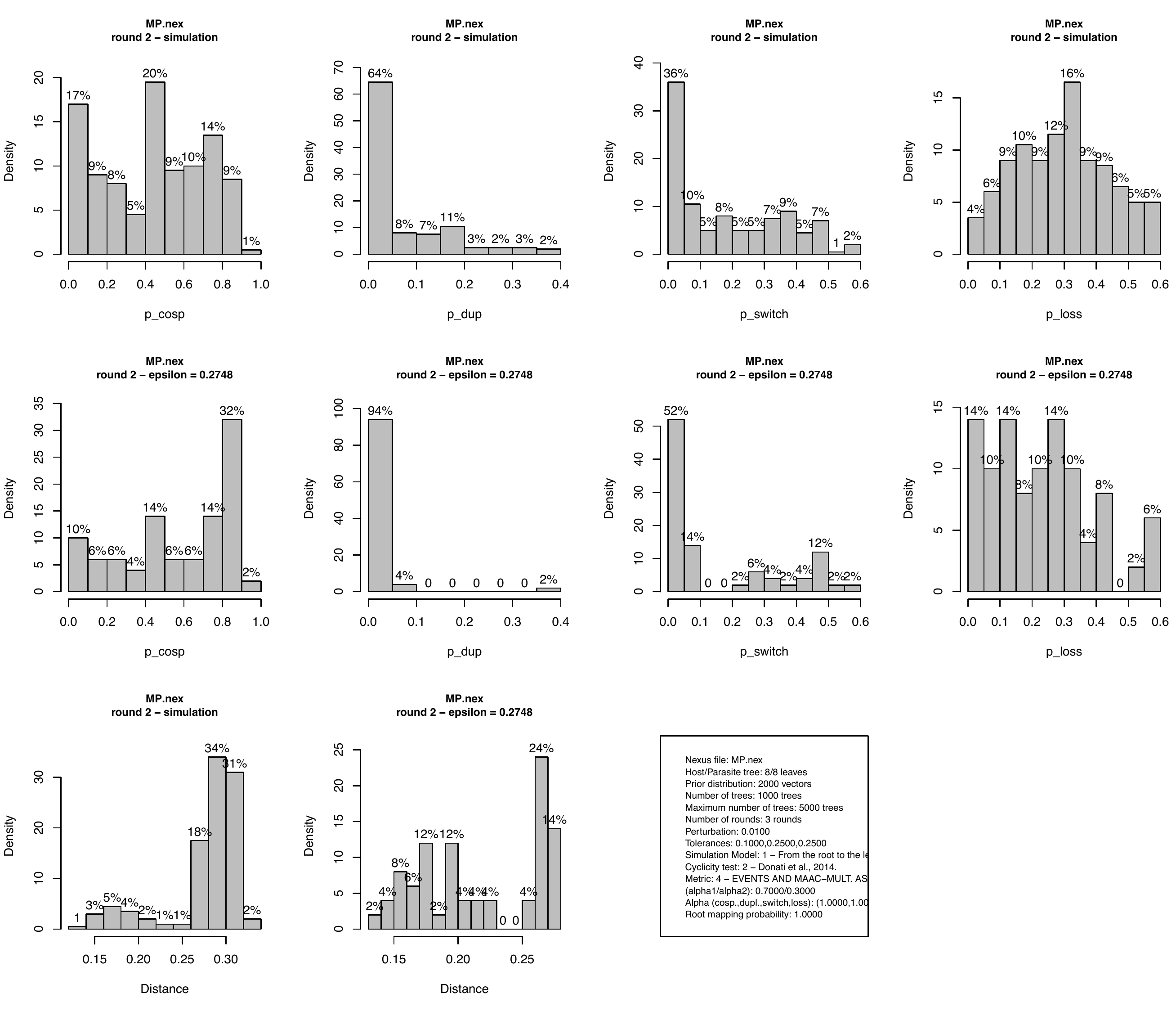}
\caption{MP dataset. First row: histograms of 
the input parameters. Second row: histograms of 
the parameters after round 2. Third row: summary discrepancies of 
the input parameters and of 
the parameters after round 2.}
\label{fig:MP_R2}
\end{figure}
\end{landscape}

\begin{landscape}
\begin{figure}
\includegraphics[scale=0.65]{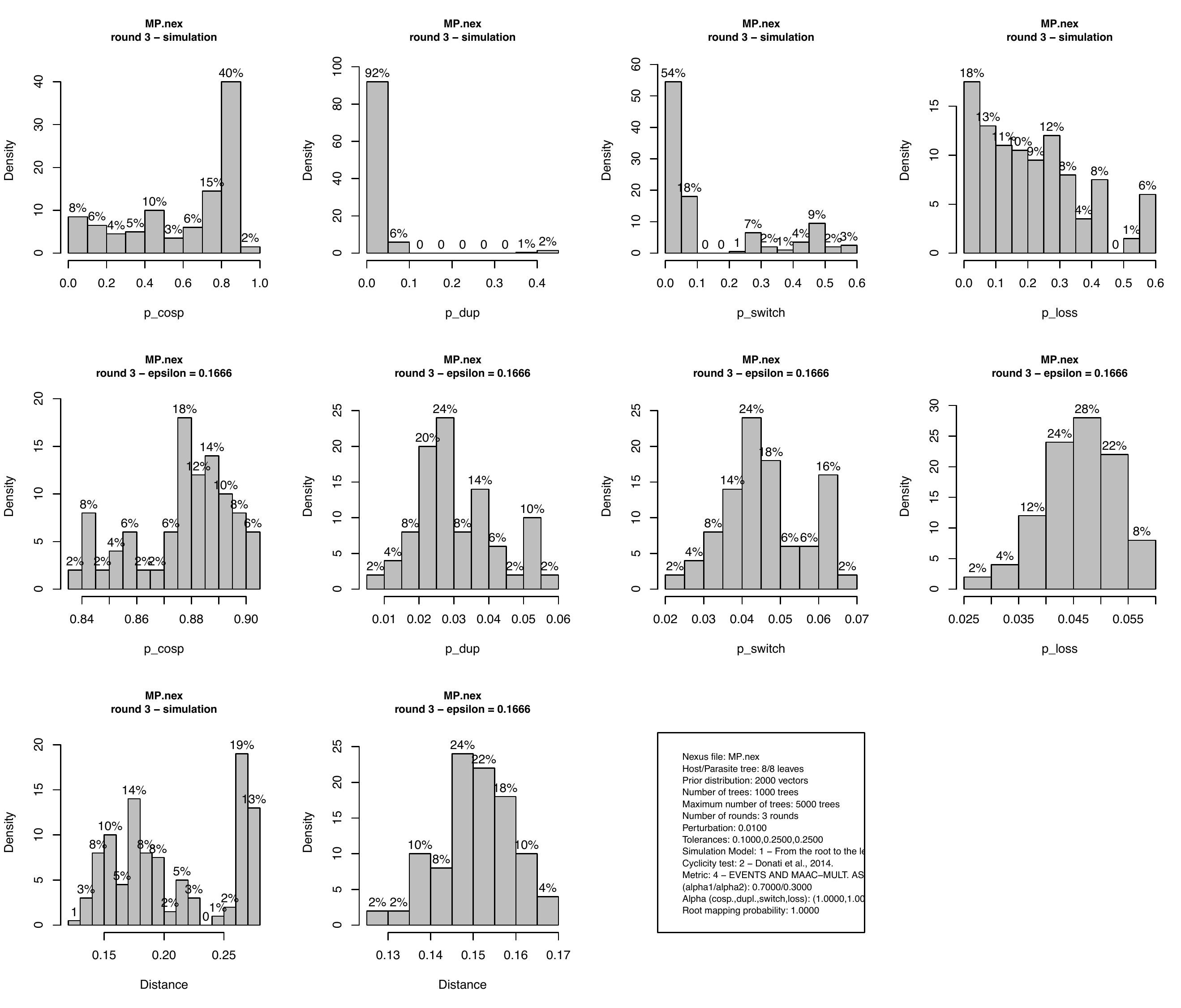}
\caption{MP dataset. First row: histograms of 
the input parameters. Second row: histograms of 
the parameters after round 3. Third row: summary discrepancies of 
the input parameters and of 
the parameters after round 3.}
\label{fig:MP_R3}
\end{figure}
\end{landscape}

\begin{landscape}
\begin{figure}
\includegraphics[scale=0.65]{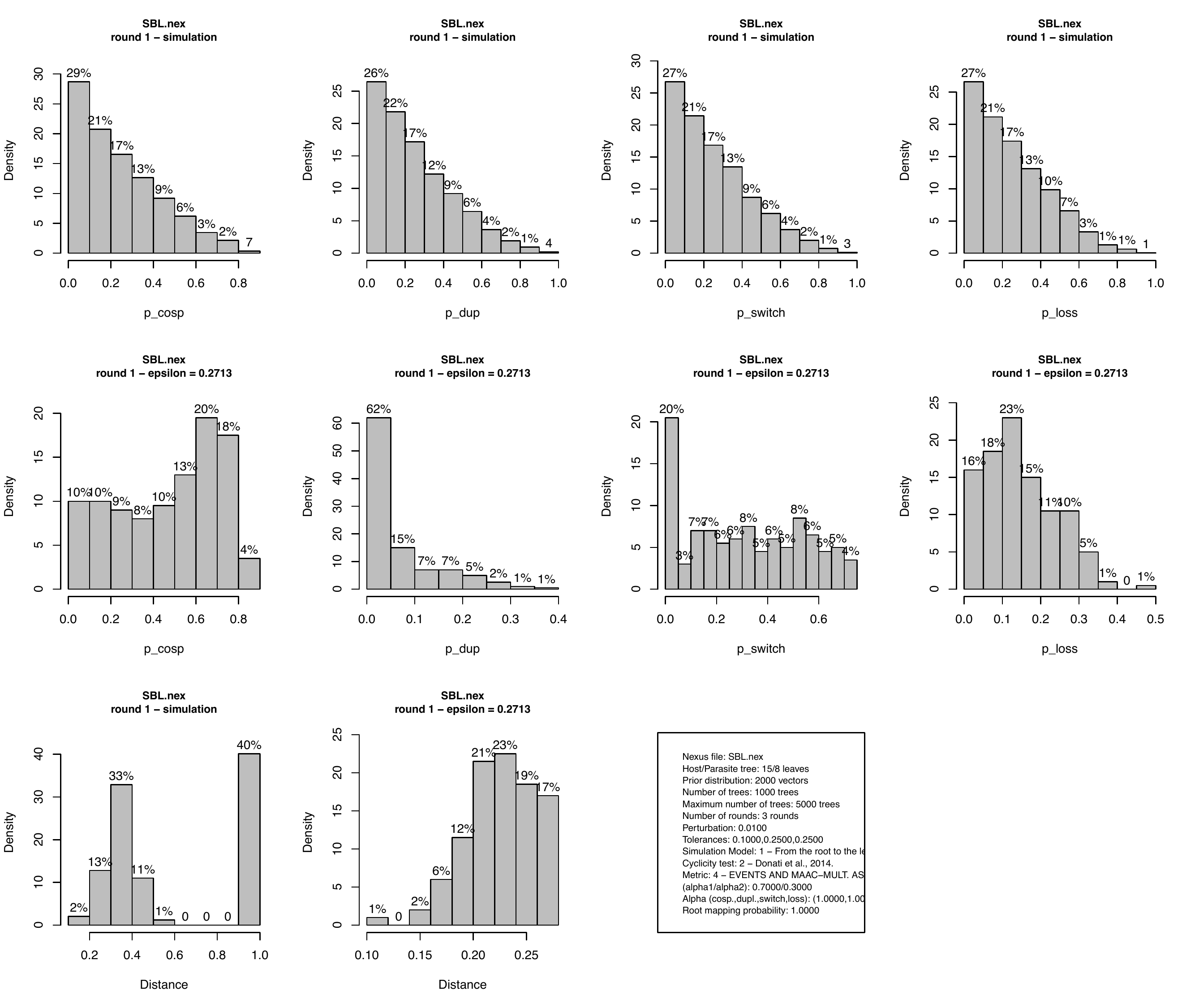}
\caption{SBL dataset. First row: histograms of 
the input parameters. Second row: histograms of 
the parameters after round 1. Third row: summary discrepancies of 
the input parameters and of 
the parameters after round 1.}
\label{fig:SBL_R1}
\end{figure}
\end{landscape}

\begin{landscape}
\begin{figure}
\includegraphics[scale=0.65]{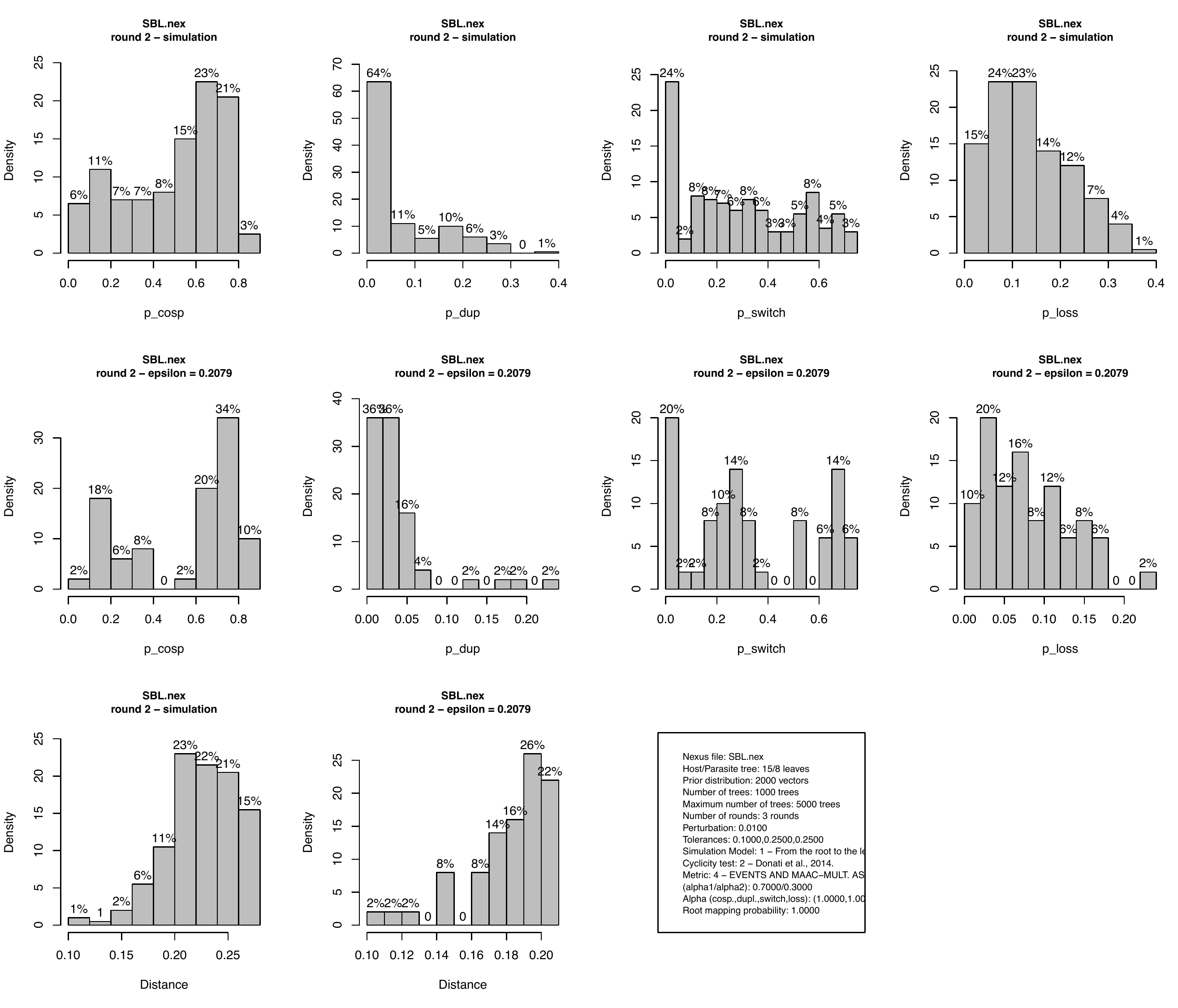}
\caption{SBL dataset. First row: histograms of 
the input parameters. Second row: histograms of 
the parameters after round 2. Third row: summary discrepancies of 
the input parameters and of 
the parameters after round 2.}
\label{fig:SBL_R2}
\end{figure}
\end{landscape}

\begin{landscape}
\begin{figure}
\includegraphics[scale=0.65]{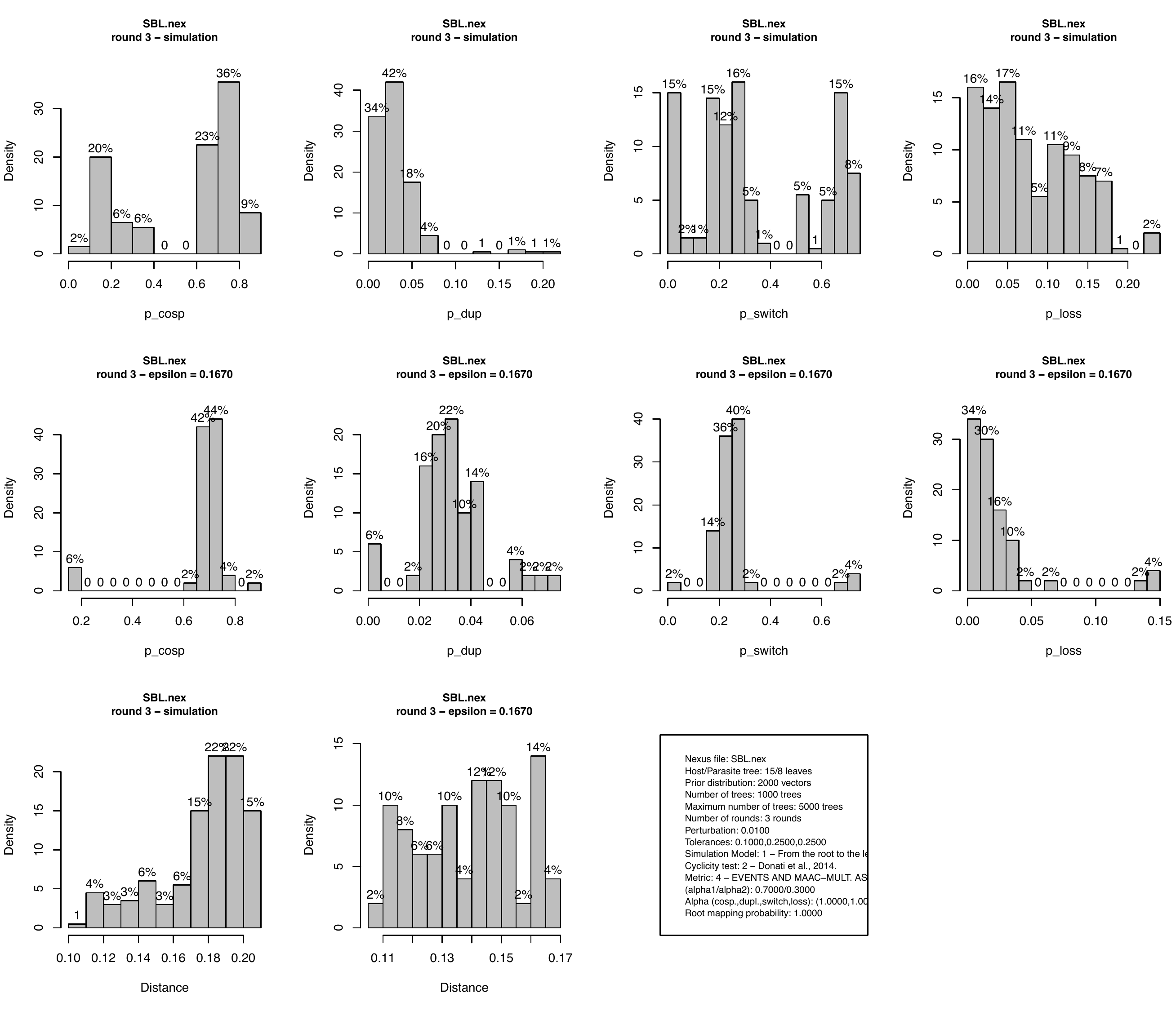}
\caption{SBL dataset. First row: histograms of 
the input parameters. Second row: histograms of 
the parameters after round 3. Third row: summary discrepancies of 
the input parameters and of
the parameters after round 3.}
\label{fig:SBL_R3}
\end{figure}
\end{landscape}

\begin{landscape}
\begin{figure}
\includegraphics[scale=0.65]{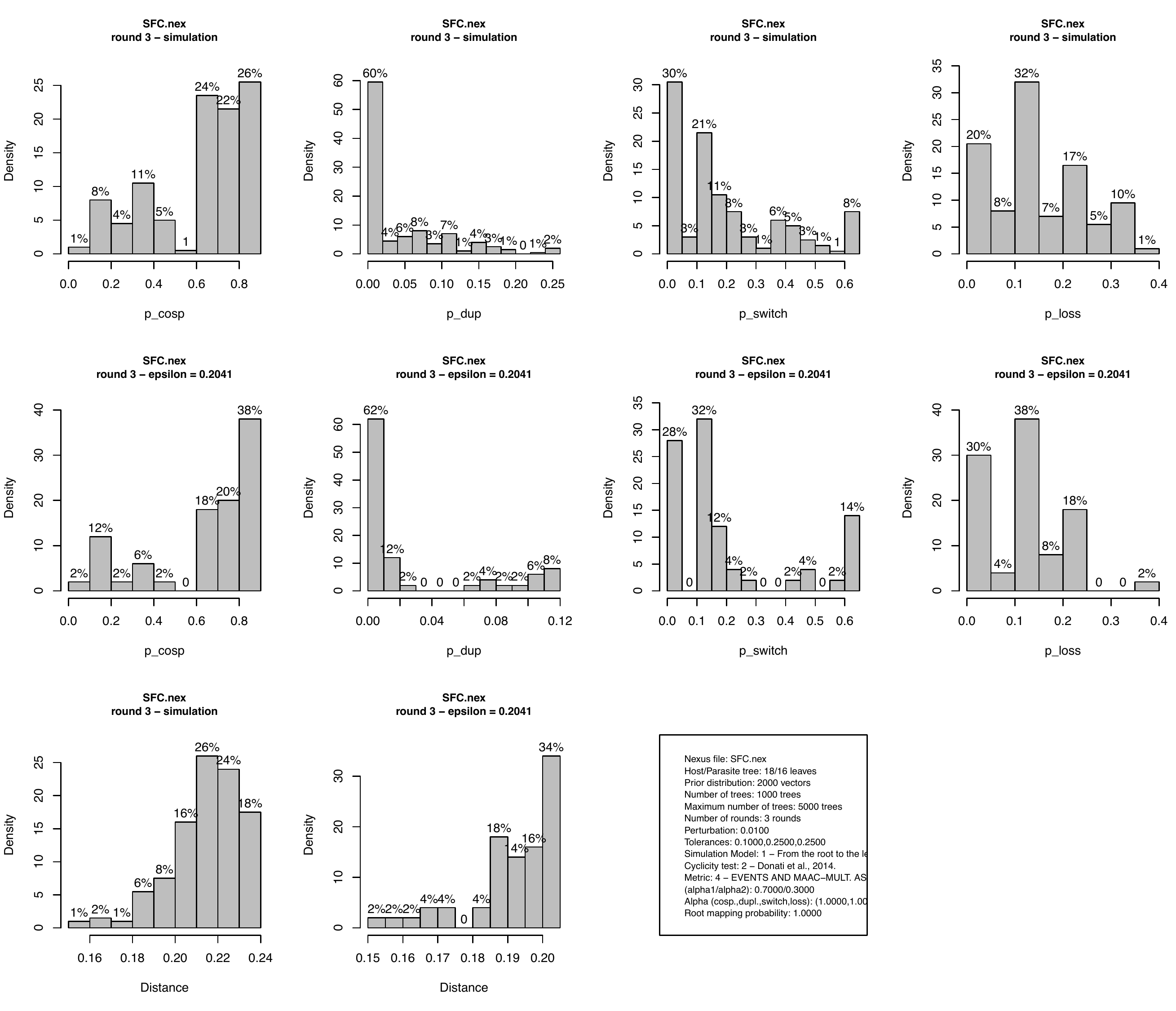}
\caption{SFC dataset. First row: histograms of 
the input parameters. Second row: histograms of 
the parameters after round 3. Third row: summary discrepancies of 
the input parameters and of
the parameters after round 3.}
\label{fig:SFC_R3}
\end{figure}
\end{landscape}

\begin{landscape}
\begin{figure}
\includegraphics[scale=0.65]{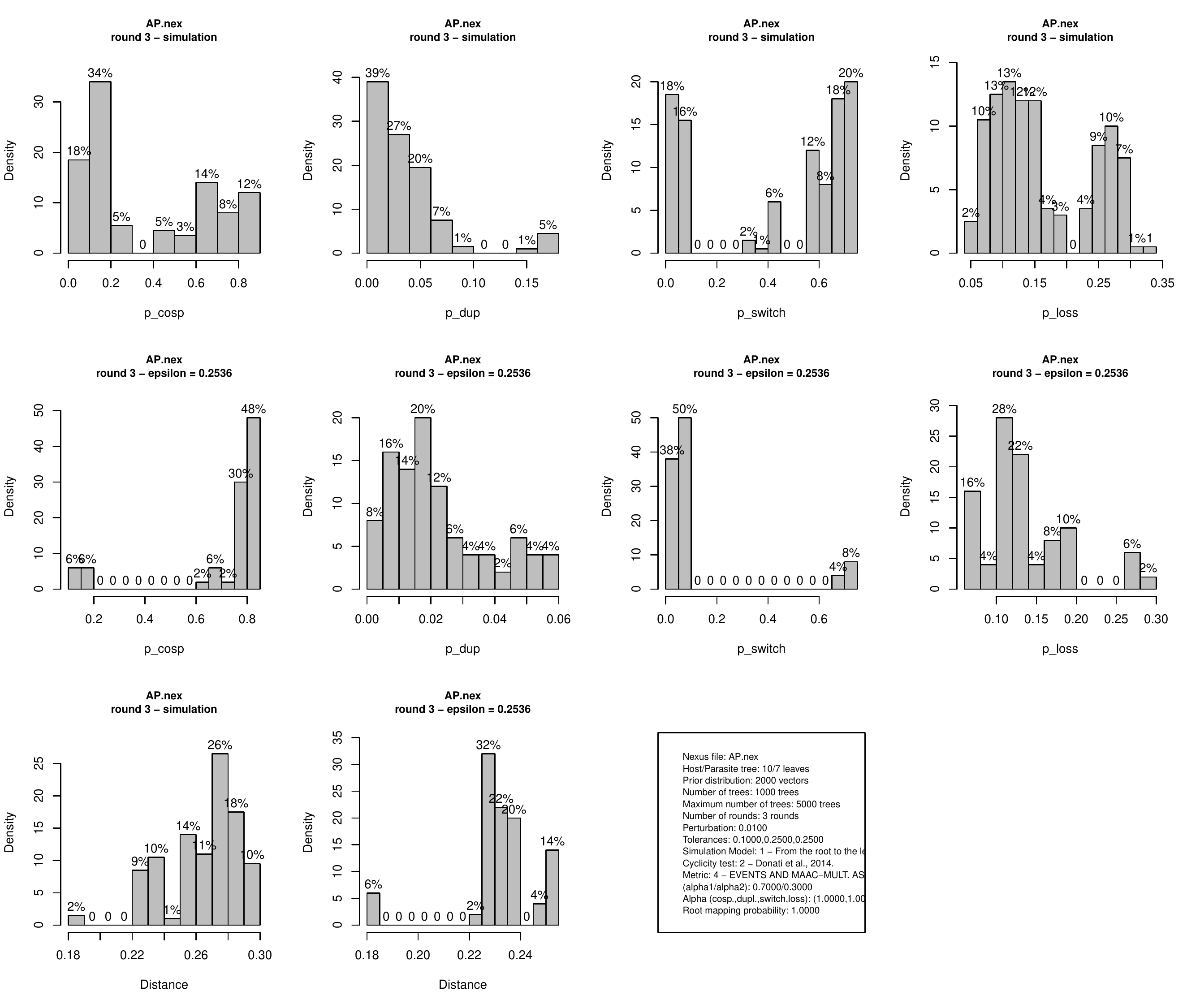}
\caption{AP dataset with perturbated spread probabilities. First row: histograms of 
the input parameters. Second row: histograms of 
the parameters after round 1. Third row: summary discrepancies of 
the input parameters and of 
the parameters after round 1.}
\label{fig:AP_perturbated}
\end{figure}
\end{landscape}

\begin{landscape}
\begin{figure}
\includegraphics[scale=0.65]{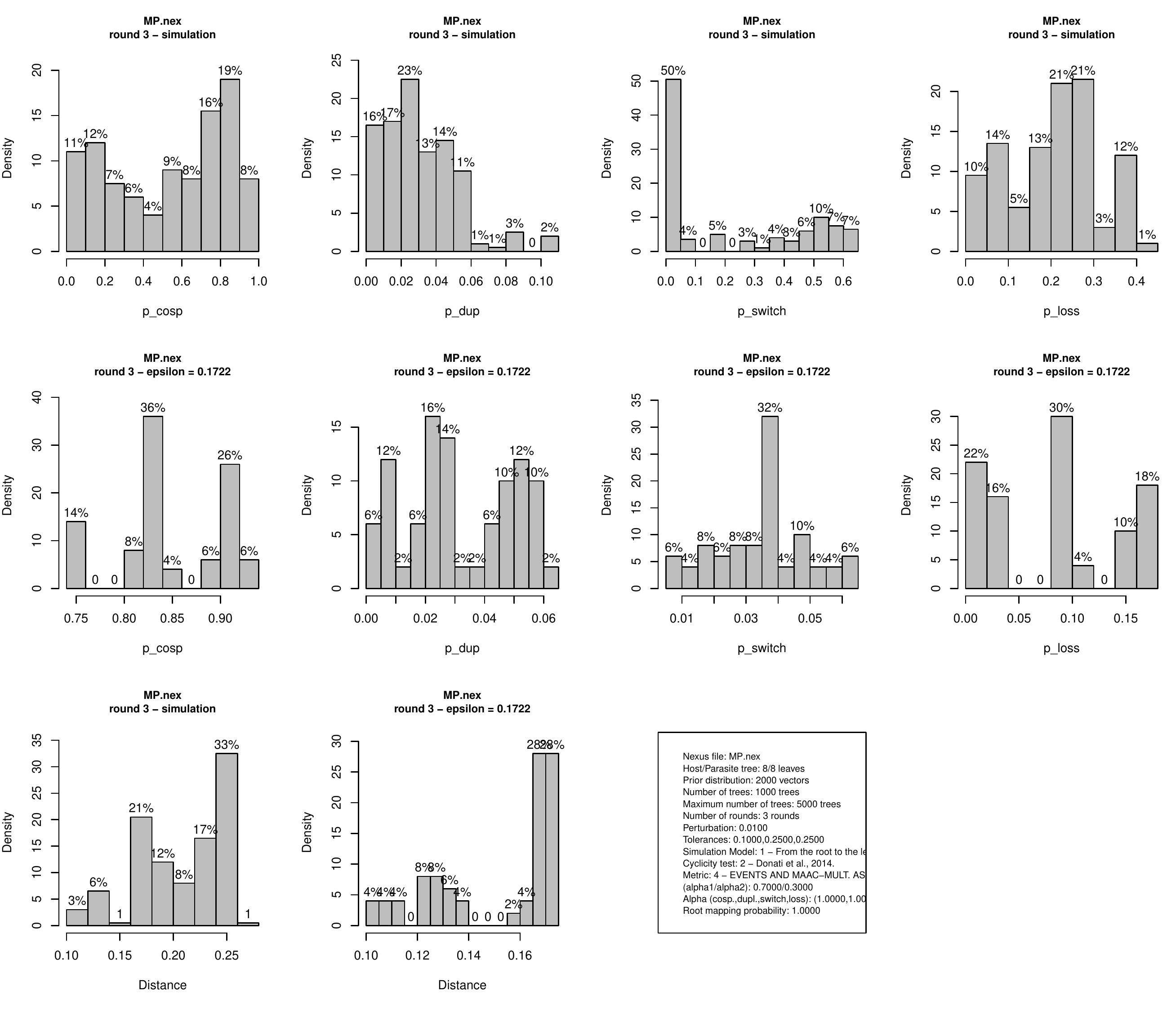}
\caption{MP dataset with perturbated spread probabilities. First row: histograms of 
the input parameters. Second row: histograms of 
the parameters after round 1. Third row: summary discrepancies of 
the input parameters and of 
the parameters after round 1.}
\label{fig:MP_perturbated}
\end{figure}
\end{landscape}

\begin{landscape}
\begin{figure}
\includegraphics[scale=0.65]{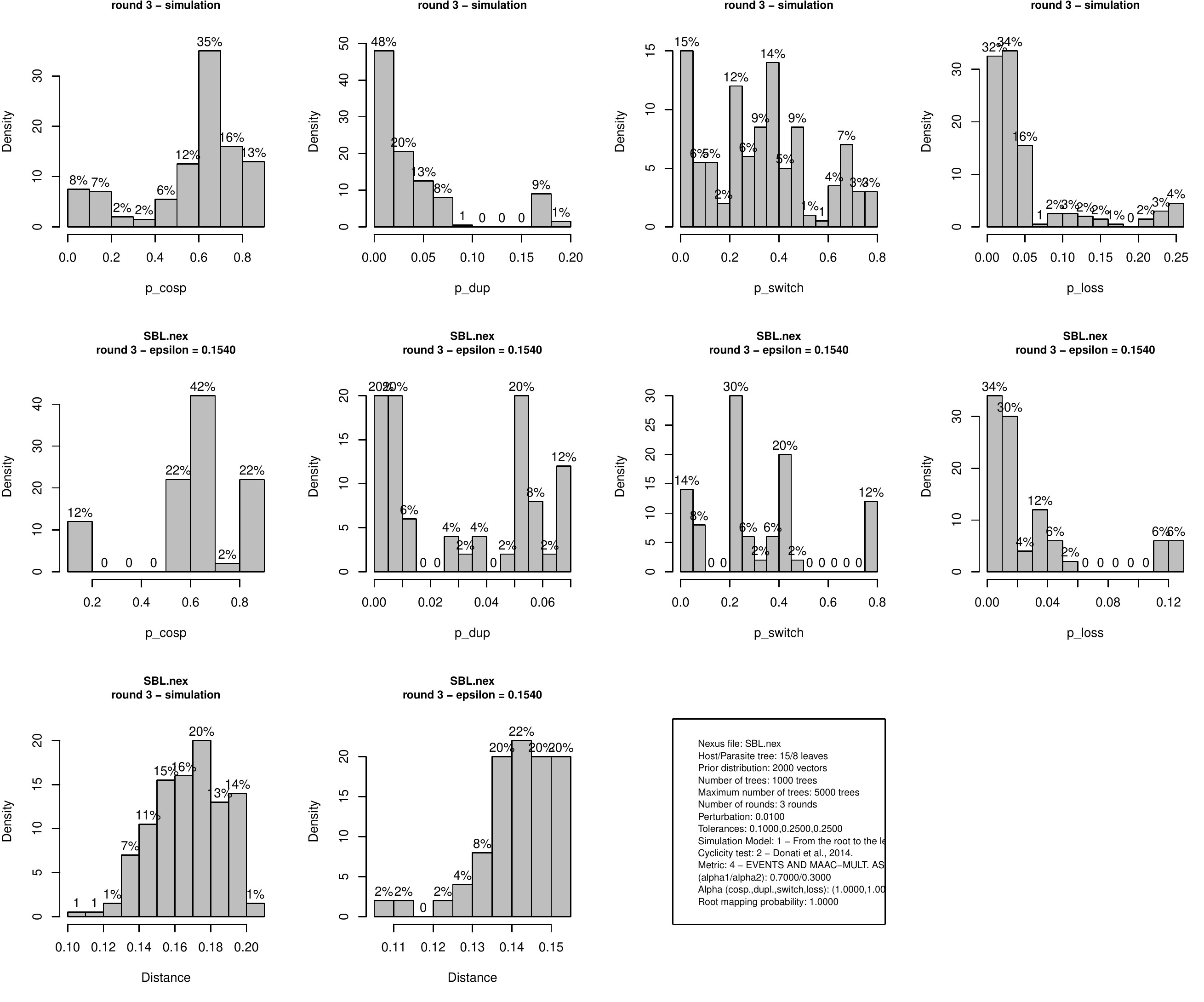}
\caption{SBL dataset with perturbated spread probabilities. First row: histograms of 
the input parameters. Second row: histograms of 
the parameters after round 1. Third row: summary discrepancies of 
the input parameters and of 
the parameters after round 1.}
\label{fig:SBL_perturbated}
\end{figure}
\end{landscape}

\begin{landscape}
\begin{figure}
\includegraphics[scale=0.65]{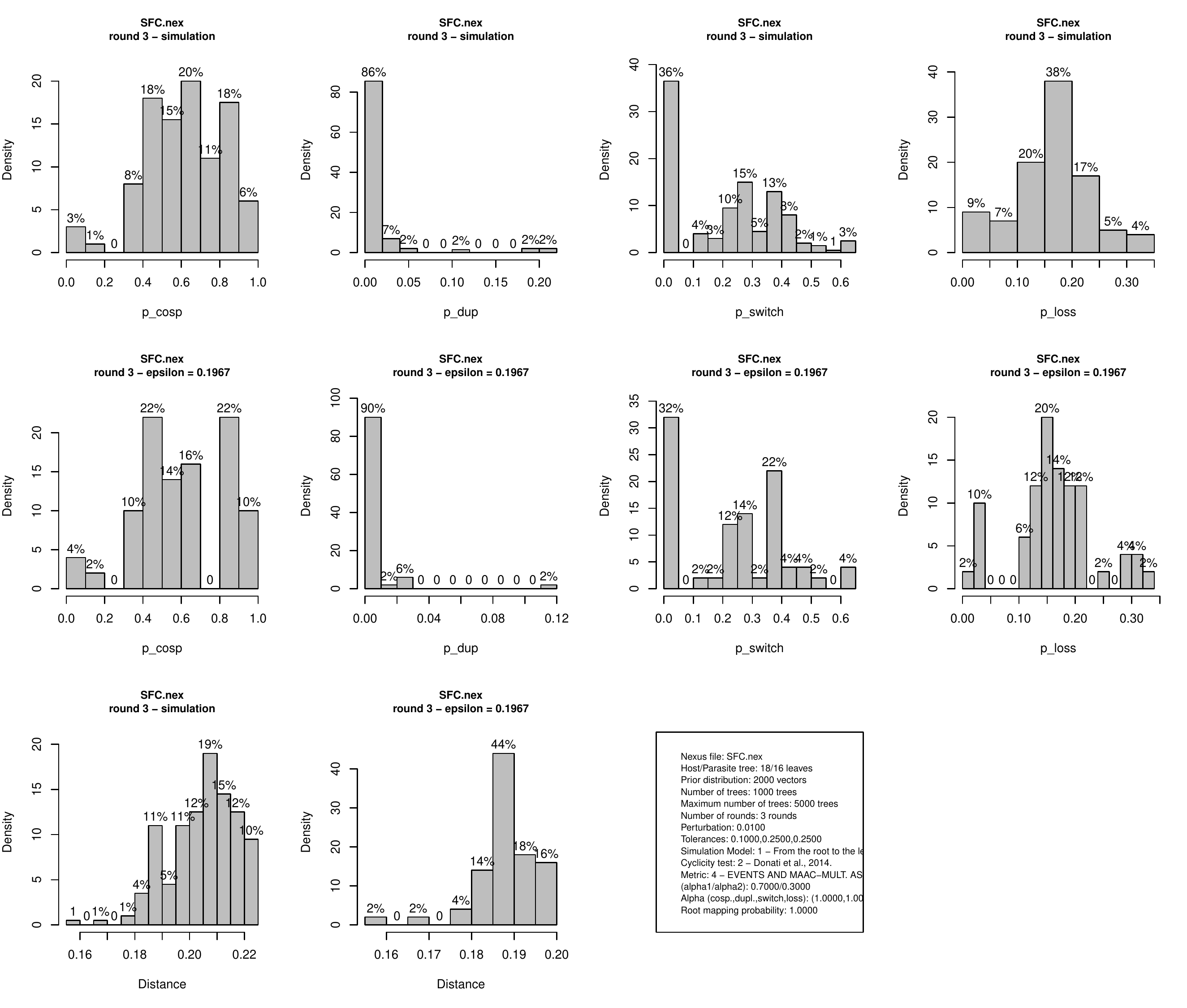}
\caption{SFC dataset with perturbated spread probabilities. First row: histograms of 
the input parameters. Second row: histograms of 
the parameters after round 1. Third row: summary discrepancies of 
the input parameters and of 
the parameters after round 1.}
\label{fig:SFC_perturbated}
\end{figure}
\end{landscape}

In this section, we explore the robustness of our results with respect to the pre-estimated values of the spread events probabilities. On each of the 4 biological datasets, we ran \ACoala\ with perturbated values of $\phs(h),\pvs(h)$. More precisely, to each non zero probability $\phs(h)$ or $\pvs(h)$, we added a noise value uniformly drawn in $[-0.1;0.1]$ (and then took the infimum with 1 and the supremum with 0, in order to ensure the modified probabilities remain in $[0,1]$). 
With these perturbed values, we ran \ACoala\ and output (after 3 rounds) 50 accepted vectors $\theta=\langle p_c,p_d,p_s,p_l \rangle$. 
The results are presented in Figures~\ref{fig:AP_perturbated} to~\ref{fig:SFC_perturbated}. Let us recall that \ACoala\ is a stochastic algorithm and any two runs will give similar but not identical results. The results obtained adding these perturbations are qualitatively the same for the first 3 datasets (namely AP, MP and SBL) as the ones without perturbations (see Figures~\ref{fig:AP_R3} to~\ref{fig:SBL_R3}). 
The results for dataset SFC show more variability wrt those of the unperturbed version (Figure~\ref{fig:SFC_R3}). Thus we also looked at the clusters output by \ACoala\ in this case in Table~\ref{tab:SFC_perturbed_round3}. 
We recall that in \cite{RGJ2008}, the different analyses performed indicated that the most plausible reconciliations presented for the SFC dataset have from 0 to 3 cospeciations, no duplication, 12 to 15 host switches and 0 to 2 losses. 
Here we find that the first main cluster (31 vectors out of 50) has a representative vector with around 50\% of cospeciations (about 7 or 8 events), almost no duplication (about 0 or 1 event), 31\% of host switches (about 4 or 5 events) and 18\% of losses (about 2 or 3 losses).
The second main cluster has a higher probability of cospeciation and less switches. 
Only the third cluster could correspond to \cite{RGJ2008}'s scenario, with 1 or 2 cospeciations, no duplication, 8 or 9 host switches and 4 to 5 losses; though it is supported by only 3 selected vectors out of 50.  
Thus for the SFC dataset, the detection of the biological scenario presented in \cite{RGJ2008} is more difficult to detect with perturbed values of the spread probabilities. 
To conclude, our results are overall robust with respect to potential errors in the  estimation of the spread events probabilities.

\renewcommand\refname{References for Supplementary Material}

\newpage

\clearpage

\clearpage

\clearpage

\clearpage

\clearpage

\clearpage

\clearpage

\clearpage

\clearpage

\clearpage

\clearpage

\clearpage

\clearpage

\clearpage

\clearpage

\clearpage

\clearpage

\clearpage

\clearpage

\clearpage


\setlength{\tabcolsep}{8pt}
\setlength{\extrarowheight}{2pt}
\begin{table}[htbp]
\begin{center}
\caption{Representative vectors of the clusters produced by \ACoala\ with perturbations for the SFC dataset. The column $\# vectors$ indicates the number of vectors in the cluster. }
\label{tab:SFC_perturbed_round3}
\begin{tabular}{ccccccc} \hline
$Dataset$ & $Cluster$ & $p_c$ & $p_d$ & $p_s$ & $p_l$ & $\# vectors$ \\ \hline
\multirow{3}{*}{ SFC} &  1 & 0.4985&	0.0024&	0.3162&	0.1829 & 31  \\ 
\cline{2-7} &  2 & 	0.8738&	0.0147&	0.0180&	0.0935 &16 \\ 
\cline{2-7} & 3 & 0.1087&	0.0012&	0.5770	&0.3131	  & 3\\
\end{tabular}
\end{center}
\end{table}

\end{document}